\documentclass{aa}  
\bibpunct{(}{)}{;}{a}{}{,} 
\usepackage{graphicx}
\usepackage{txfonts}
\usepackage{csquotes}
\usepackage{booktabs}
\usepackage[linktocpage=true, colorlinks=true, linkcolor=blue, citecolor=blue, urlcolor=blue]{hyperref}
\usepackage{comment}
\usepackage{xcolor}
\usepackage{siunitx}

\begin{document} 

\title{Probing the rest-frame wavelength dependence of quasar variability}
\subtitle{Insights from the \textit{Zwicky} Transient Facility Survey}
\titlerunning{Wavelength dependence of quasar variability}

\authorrunning{P. Patel et al.}
\author{P. Patel \inst{1,2}, P. Lira\inst{1,2}, P. Arévalo\inst{3,2}, Mouyuan Sun\inst{4}, S. Bernal\inst{3,2}, M. L. Mart\'inez-Aldama\inst{2,5,6}}
\institute{Departamento de Astronom\'ia, Universidad de Chile, Casilla 36D, Santiago, Chile\\
\email{priya@das.uchile.cl}
\and Millennium Nucleus on Transversal Research and Technology to Explore Supermassive Black Holes (TITANS)
\and Instituto de F\'isica y Astronom\'ia, Universidad de Valpara\'iso, Gran Breta\~na 1111, Valpara\'iso, Chile\\
\email{patricia.arevalo@uv.cl}
\and Department of Astronomy, Xiamen University, Xiamen, Fujian 361005, People’s Republic of China
\and Astronomy Department, Universidad de Concepci\'on, Barrio Universitario S/N, Concepci\'on 4030000, Chile
\and Millennium Institute of Astrophysics MAS, Nuncio Monsenor Sotero Sanz 100, Of. 104, Providencia, Santiago, Chile}
 
\abstract
{Quasar variability can potentially unlock crucial insights into the accretion process. Understanding how this variability is influenced by wavelength is crucial for validating and refining quasar variability models.}
{This paper aims to enhance the understanding of the dependence of variability on the rest-frame wavelength ($\lambda_{RF}$) by isolating the variance in different timescales in well-defined wavelength bins and examining the corona-heated accretion disk (CHAR) model.}
{We investigated the relation between variance and rest-frame wavelength ($\lambda_{RF}$) using optical $g$- and $r$-band light curves from the \textit{Zwicky} Transient Facility (ZTF) Data Release 15 (DR15) for $\sim 5000$ quasars within narrow ranges of the black hole mass ($M_{BH}$) and Eddington ratio ($R_{Edd}$). A spectral model taking into account disk continuum emission, Balmer transitions, Fe II pseudo-continuum emission, and other emission lines is necessary to best interpret the variance spectrum.
}
{Our analysis indicates a strong anticorrelation between median variance and $\lambda_{RF}$ for quasars with $M_{BH} = 10^{8} M_{\odot}$ and $R_{Edd} = 10^{-1}$ at different timescales. This anticorrelation is more pronounced at shorter timescales. The results align well with a bending power-law power spectral density (PSD) model with both the damping timescale and the high-frequency slope of the PSD depending on the wavelength. The predictions provided by the CHAR model on the variance spectrum across most timescales studied showcase its potential in constraining temperature variations within the accretion disk.}
{}

 \keywords{accretion, accretion discs; galaxies:active; quasars: supermassive black holes}

   \maketitle
%
\section{Introduction} 
\label{sec:intro}

Active galactic nuclei (AGNs) are galaxies whose compact nuclear region is much brighter than a region of the same size in a normal galaxy. The AGNs also show time-variable emissions in every waveband, and the characteristic variability timescales range from hours to years, with the shortest timescales associated with shorter emission wavelengths.

The current understanding of AGN continuum and emission properties is based on the standard thin accretion disk theory and reprocessing model of the accretion disc. The standard thin accretion disk theory (e.g., Shakura \& Sunyaev~\citeyear{1973A&A....24..337S}) proposes that, through viscosity, the accretion disk feeds the central object, while angular momentum is transferred outward. As material falls toward the black hole, it releases gravitational energy in the form of electromagnetic radiation. The azimuthal velocity is approximately Keplerian at any particular radius, except for the innermost region near the innermost stable circular orbit. The temperature profile is determined by balancing gravitational energy released during accretion with cooling due to black-body radiation, resulting in a radial temperature profile of $T\propto R^{-3/4}$. While this model explains observed AGN characteristics, such as luminosity and a peak of the spectral energy distribution in the UV, it predicts luminosity variations occurring over viscous timescales, spanning thousands of years, which is inconsistent with observations showing variability over much shorter timescales of weeks to months in quasars. 

The idea of X-ray reprocessing solved these limitations in the 1990s (Clavel et al.~\citeyear{1992ApJ...393..113C}; Collin-Souffrin ~\citeyear{1991A&A...249..344C}; Krolik et al.~\citeyear{1991ApJ...371..541K}; Panagiotou et al.~\citeyear{2022ApJ...935...93P}). The X-ray reprocessing model explains the short-term variability observed in the UV to near-infrared bands. The cold disk surface absorbs the rapidly fluctuating X-ray emission and reprocesses it, resulting in highly correlated rapid changes in the UV to near-infrared flux with a light-travel time delay. The time delay increases as the wavelength increases, which enables the study of the temperature profile and fundamental physical processes of an accretion disk (e.g., Lawrence ~\citeyear{2018NatAs...2..102L}). However, it is important to note that many AGNs do not exhibit the predicted strong correlations between X-ray and UV/optical emissions (Edelson et al.~\citeyear{2019ApJ...870..123E}; Zhu et al.~\citeyear{2018ApJ...860...29Z}; Morales et al.~\citeyear{2019ApJ...870...54M}; Hagen et al.~\citeyear{2023MNRAS.521..251H}). Moreover, this model has an energy-budget problem for luminous AGNs (see Clavel et al.~\citeyear{1992ApJ...393..113C}; Dexter et al.~\citeyear{2019ApJ...885...44D}; Just et al.~\citeyear{2007ApJ...665.1004J}; Strateva et al.~\citeyear{2005AJ....130..387S}; Lusso et al.~\citeyear{2010A&A...512A..34L}), where the required X-ray luminosity to produce the observed fractional variability in the UV/optical bands is too high to be consistent with X-ray observations. This issue is even more critical for highly variable AGNs (e.g., Dexter et al.~\citeyear{2019ApJ...885...44D}). Therefore, it is challenging to explain the entire disk variability using only X-ray reprocessing (Gaskell~\citeyear{2007ASPC..373..596G}).

The CHAR model was proposed by Sun et al.~(\citeyear{2020ApJ...891..178S}) to overcome the limitations of the X-ray reprocessing model. This model predicts a tight coupling between the accretion disk ($\gtrsim$ 10 Schwarzschild radii, $R_s = 2GM/c^{2}$, where $G$ is the gravitational constant and $c$ is the speed of light) and the corona due to the magnetic fields, where magnetic turbulence in the corona produces X-ray fluctuations. Similarly, this process magnetically drives the variations in the heating rate of the accretion disk and results in temperature fluctuations. Although a strong correlation between X-ray and UV/optical emission is expected, the significant advective cooling and corona surface density variations can weaken the correlation. The CHAR model not only resolves the energy-budget problem of the X-ray reprocessing model but also explains the larger-than-expected time lags observed in some AGNs, although this problem might have appeared as a result of the models employed (Kammoun et al.~\citeyear{2021ApJ...907...20K}). Additionally, it has a broad scope for explaining quasar UV/optical variability and its connection to physical properties. 

Previous studies have looked into the dependence of variability parameters on the rest-frame wavelength and some have reported a correlation between the variability amplitude and the rest-frame wavelength (Cutri et al.~\citeyear{1985ApJ...296..423C}; Paltani \& Courvoisier~\citeyear{1994A&A...291...74P}; Vanden Berk et al.~\citeyear{2004AAS...20512002V}; MacLeod et al.~\citeyear{2010ApJ...721.1014M},~\citeyear{2012ApJ...753..106M}; Meusinger et al.~\citeyear{2011A&A...525A..37M}; Zuo et al.~\citeyear{2012ApJ...758..104Z}; Morganson et al.~\citeyear{2014ApJ...784...92M}; Li et al.~\citeyear{2018ApJ...861....6L}; Sánchez-Sáez et al.~\citeyear{2018ApJ...864...87S}). However, there is no consensus regarding the dependence of the logarithmic gradient of variability ($\gamma$) on the rest-frame wavelength, with contradictory results reported, including positive (Li et al.~\citeyear{2018ApJ...861....6L}) or no correlation (Sánchez-Sáez et al.~\citeyear{2018ApJ...864...87S} and Morganson et al.~\citeyear{2014ApJ...784...92M}). Besides, these studies have primarily examined the relationship between variability and rest-frame wavelengths on a single timescale. To gain a comprehensive understanding of the processes driving the observed variability, it is necessary to investigate this relationship across different timescales.

Variance also correlates with the physical properties of black holes — that is, their masses and Eddington ratios (see Ar\'evalo et al. \citeyear{2023MNRAS.526.6078A} and references therein) — making it challenging to establish a strong statistical link between AGN variability and rest-frame wavelength. To solve this problem, a comprehensive analysis of well-sampled individual AGN light curves with known physical properties is needed to unveil the relationship between variability and rest-frame wavelength, as we show here.

In this work, \textit{Zwicky} Transient Facility (ZTF; Masci et al.~\citeyear{2019PASP..131a8003M}) data are used to quantify the variance as a function of wavelength between 2700 Å and 5200 Å for a sample of 2533 quasars in the $g$-band and 2795 quasars in the $r$-band, selected to have approximately the same $M_{BH}$ and $R_{Edd}$. We then compare our results with CHAR model predictions, offering valuable insights into the underlying mechanisms driving AGN optical variability.

The paper is structured as follows. Section~\ref{sec:method} provides a detailed explanation of the methodology employed and presents the results of the variability analysis. Section~\ref{sec:Var_vs_wavelength} presents the statistical analysis connecting variability and rest-frame wavelength. Section~\ref{sec:spectral study} examines if the emission lines and Balmer continuum can account for the variability wiggles observed in the variance--rest-frame wavelength spectrum. Section~\ref{sec: CHAR model} compares the observational results with the predictions of the CHAR model. Section~\ref{sec:discussion} focuses on how this correlation modifies the power spectrum, and the insights obtained from the comparison between the CHAR model and observed data. Section~\ref{sec:conclusion} summarizes key findings and discusses their physical implications.

\section{Methods} \label{sec:method}
\subsection{Sample selection} \label{subsec:sample}

Our sample is based on the crossmatch between the ZTF Data Release 15 (DR15) (Masci et al.~\citeyear{2019PASP..131a8003M}) and the catalog of Rakshit et al.~\citeyear{2020yCat..22490017R}, which consists of 526,265 confirmed quasars from the Sloan Digital Sky Survey Data Release 14 (SDSS-DR14). This catalog provides black hole masses and Eddington ratios, among other quantities, through careful and homogeneous analysis of the SDSS spectra. Virial black hole mass values were calculated based on the H$\beta$ line (\(z<0.8\)) or the Mg II line (\(0.8<z<1.9\)). Since the H$\beta$ line is a more confident single epoch mass estimator, we selected quasars for the range of $z=0.1$ to 0.8 and with mass measurement errors of less than 0.2 dex. We used redshift to probe different rest-frame wavelengths by dividing the observed effective wavelengths, 4722.74 Å and 6339.61 Å for the $g$ and $r$-bands, respectively, by $(1 + z)$.\footnote{Effective wavelengths in the $g$ and $r$ bands are 4722.74 Å and 6339.61 Å, respectively, as is listed on the SVO Filter Profile Service website (Rodrigo et al.~\citeyear{2012ivoa.rept.1015R}; Rodrigo \& Solano~\citeyear{2020sea..confE.182R}) on May 4, 2022. These values were updated on June 19, 2023.} Thus, the sample covers the rest-frame wavelength from 2624.18 Å to 5686.25 Å.

To control for the dependency of variability on the black hole mass and Eddington ratio, we defined a narrow range in these two properties corresponding to the peak values of the quasar population found in Rakshit et al.~\citeyear{2020yCat..22490017R}. This selected sample consists of 7392 quasars, covering a range in black hole mass ($\log{M_{BH}}$ in units of $M_{\odot}$) between $8.0$ to $8.5$, Eddington ratio ($\log{R_{Edd}}$) between $-1.3$ to $-0.8$, and bolometric luminosity ($\log{L_{bol}}$ in units of erg s$^{-1}$) between 44.85 and 45.83. This selection allows us to measure the variability amplitude as a function of rest-frame wavelength largely independently of other physical parameters. As PSF photometry, performed to generate data release ZTF light curves, can yield an inaccurate flux for extended objects, we narrowed our selection to point-like sources classified by Tachibana \& Miller~\citeyear{2018PASP..130l8001T}. We also excluded radio-loud sources from our analysis. Radio-loud quasars typically possess strong jets, which may introduce a substantial contribution to the overall emission. This approach allows us to concentrate solely on analyzing the emission from the accretion disc. Notably, radio-loud quasars exhibit approximately 30\% larger variability amplitudes than radio-quiet objects (MacLeod et al.~\citeyear{2010ApJ...721.1014M}, Vanden Berk et al.~\citeyear{2004ApJ...601..692V}). We obtained 20 cm continuum measurements for all the sources from FIRST and NVSS, prioritizing NVSS values for radio flux when available in the unified radio catalog of Kimball \& Ivezić (\citeyear{2008AJ....136..684K}). We calculated the radio-loudness parameter, ${\rm R}_{i}$, using the formula by Ivezić et al.~(\citeyear{2002AJ....124.2364I}), defined as the ratio of radio flux density to optical flux density without K-correction. Thus,

\begin{equation} \label{1}
\mathrm{R_{i} = log(F_{radio}/F_{optical}) = 0.4 (i - t)}
.\end{equation}
 
In the equation, i represents the SDSS magnitude, while t represents the AB radio magnitude, which is calculated as 
\begin{equation} \label{2}
\mathrm{t = -2.5 \: log(\frac{F_{int}}{3631 Jy})}
,\end{equation}
where F$_{\rm int}$ is the flux density.

\begin{figure}[!ht]%
\centering{
\includegraphics[width=0.47\textwidth]{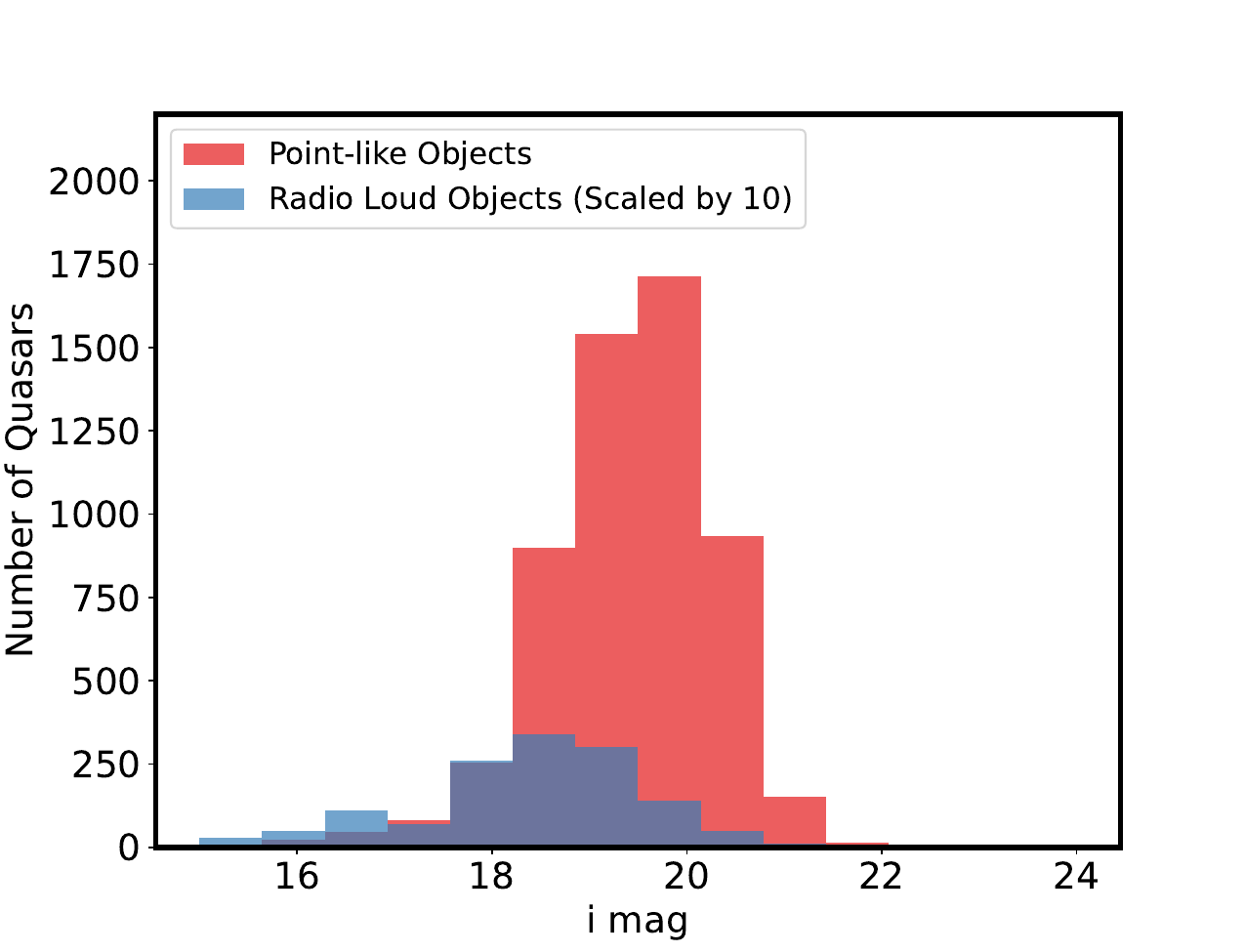}}
\caption{Comparison of the number of radio-loud quasars (adjusted by a scaling factor of ten), and the overall population of quasars in our sample as a function of the $i$-band magnitude.}
\label{fig : radio loud}
\end{figure}

We identified radio-loud quasars with R$_{\rm i} \geq 1 $ using the criterion of Ivezić et al. (\citeyear{2002AJ....124.2364I}), resulting in 136 radio-loud objects. Figure \ref{fig : radio loud} presents the proportion of radio-loud quasars in the sample as a function of magnitude. 

Finally, to minimize the effect of host galaxy emission on the normalized variance, we excluded 553 quasars with host contributions exceeding 0.0 at 4200\AA\ and 5100\AA, reported by Rakshit et al.~(\citeyear{2020yCat..22490017R}). In their study, Rakshit et al. (\citeyear{2020yCat..22490017R}) applied principal component analysis to decompose the host galaxy and quasar contributions to the spectra of low-redshift ($z < 0.8$) systems, based on the methodologies presented by Vanden Berk et al.~(\citeyear{2006AJ....131...84V}) and Shen et al.~(\citeyear{2008AJ....135..928S}, \citeyear{2015ApJ...805...96S}). The dominant quasar emission at 4200 \AA\ could result in significant negative host flux during the fitting process. In such cases, host decomposition was not performed, leading to the assignment of a value $-999$. This resulted in 553 quasars also excluded from our sample.

We extracted ZTF DR15 light curves for 4975 quasars in our final sample in the $g$ band and $r$ I band. The ZTF DR15 survey has light curves with a cadence of approximately four days and is up to 4 years and 8 months in length, with yearly gaps. We also removed bad measurements for which the catflags parameter is not zero. The ZTF observations use a camera consisting of 16 detectors,with known cross-calibration offsets. To minimize cross-calibration uncertainties, we constructed separate light curves for each CCD (see below). This approach can generate multiple light curves for certain objects. We introduced further quality filtering of each light curve by (1) excluding observations with limitmag $< 20$ (essentially an image quality cut), (2) restricting the sample to include only light curves that are at least 550.0 days long in the rest frame, (3) keeping observations captured by a single CCD, and (4), considering only light curves that contain at least 90 data points. After quality filtering, we determined the mean flux of the filtered ZTF light curves and excluded objects with a mean $g$-band and/or $r$-band magnitude of more than 20 mag to have good signal-to-noise ratios ($5-\sigma$). 

In the $g$ band, there are a total of 2533 objects, of which 314 have two acceptable light curves and 14 have three. In the $r$ band, there are a total of 2795 objects, of which 353 have two acceptable light curves and 17 have three. Hence, there are 2875 and 3182 valid light curves in the g and r bands, respectively. These valid light curves have in the rest frame a mean (median) length of 1059 (1043) days for the $g$ band, with a standard deviation of 120 days, and contain an average (median) of 270 (266) good data points, with a standard deviation of 98 data points. In the $r$ band, the mean (median) length is 1052 (1036) days, with a standard deviation of 122 days, and they show a mean (median) of 293 (295) good data points, with a standard deviation of 95 data points.

\subsection{Calculation of variance}\label{subsec: Calc_var}

Before estimating the variance, we first subtracted the average flux from each light curve. The resulting zero-mean light curves were normalized by their respective (pre-subtraction) means, allowing us to compare the amplitude of variability (or variance) of light curves of different flux levels. Then, the variance at a given timescale was calculated by convolving the light curves with a Mexican hat filter developed to work with data presenting gaps, such as those found in long light curves due to the seasonal observability of the astronomical sources (Ar\'evalo et al.~\citeyear{2012MNRAS.426.1793A}). In this method, the original light curve is convolved using two Gaussian filters of slightly different widths. The difference found between these convolved light curves is similar to introducing low- and high-pass filtering. The method is equivalent to multiplying the original power spectrum of the light curve by the power spectrum of the Mexican hat filter. This filter power spectrum can be defined as the square of the difference between the Fourier transform of the two Gaussian kernels of slightly different widths. The power spectrum of the filter peaks at $k_{p} = 0.225/\sigma$, and its width is $\sim 1.16\ k_{p}$, where $\sigma$ is the Gaussian width. The filtered variances, which have days as their units, estimate the normalized power density. After filtering, the normalized variance estimates are multiplied by the peak frequency, $k_{p}$, of each frequency filter to obtain dimensionless variance estimates. This can recover the broadband shape and normalization of the power spectrum. For our work, the variance of the filtered light curve was calculated for four rest-frame timescales around 300, 150, 75, and 30 days (see Appendix~\ref{sec:appendixA}). 

In order to measure the real variability of our sources, it is necessary to have a correct estimate of the level of observational noise in the ZTF light curves. We used simulated light curves with additional noise to estimate the observational noise. The variance of the simulated light curve was expressed as the sum of intrinsic variance due to real variability ($\sigma_{intrinsic}$), noise variance derived from the error bars ($\sigma_{noise}$), and an extra noise estimate ($\sigma_{noise\ estimate}$) (for more details, please refer to Sect. 3 of Ar\'evalo et al. \citeyear{2023MNRAS.526.6078A}). The difference between the variance of the simulated light curve and the variance of the original light curve generates an estimate of the variance of the noise; that is, ($\sigma^2_{\rm sim})-(\sigma^2_{\rm original})=(\sigma^2_{\rm intrinsic}+\sigma^2_{\rm noise}+\sigma^2_{\rm noise, estimate}) - (\sigma^2_{\rm intrinsic}+\sigma^2_{\rm noise}) = \sigma^2_{\rm noise, estimate}$.

If the error bars on the flux properly represent the observational noise, then, on average, $\sigma_{noise}^2$ = $\sigma_{noise\ estimate}^2$. Subtracting this noise estimate from the variance of the original light curve provided the intrinsic variance. The procedure was repeated ten times for different timescales (300, 150, 75, and 30 days). We analyzed a sample of non-variable stars observed with ZTF to validate the noise estimates. We chose and extracted 4100 ZTF light curves of individual stars from the SDSS Stripe 82 Standard Star Catalog stars from Thanjavur et al. (\citeyear{2021MNRAS.505.5941T}) with $16< g <20$, and $15.5< r <20$. The quasar sample covers the broad redshift range from 0.1 to 0.8, so we transformed the star light curves using eight different redshift values [0.1, 0.19, 0.28, 0.35, 0.44, 0.53, 0.62, 0.71, 0.8] to transform the time axis in the same way as was done to the quasar light curves. Then, we used the same Mexican hat filter applied to the quasar light curves to estimate the variance of the star light curves for the four timescales of interest, and we computed the variance attributed to observational noise mentioned above. We corrected the quasar variances by accounting for excess noise estimated from the corresponding star bins. Although we anticipated symmetric distributions of the estimated intrinsic variances for the star sample, we observed mostly negative and some positive net variances. Negative values show that the noise is overestimated, a more significant problem for dimmer objects and for shorter timescales.

\begin{figure*}[ht]%
\centering{
\includegraphics[width=8cm]{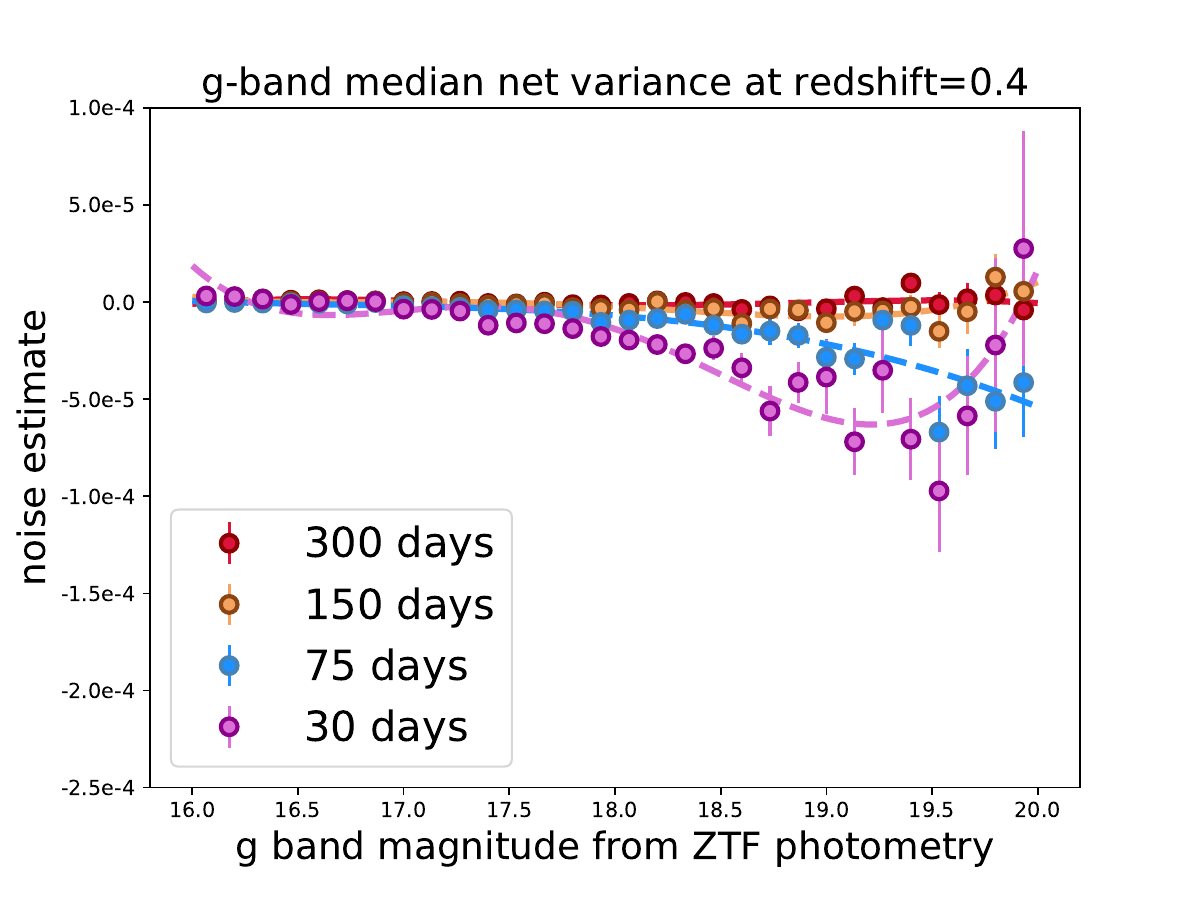}
\includegraphics[width=8cm]{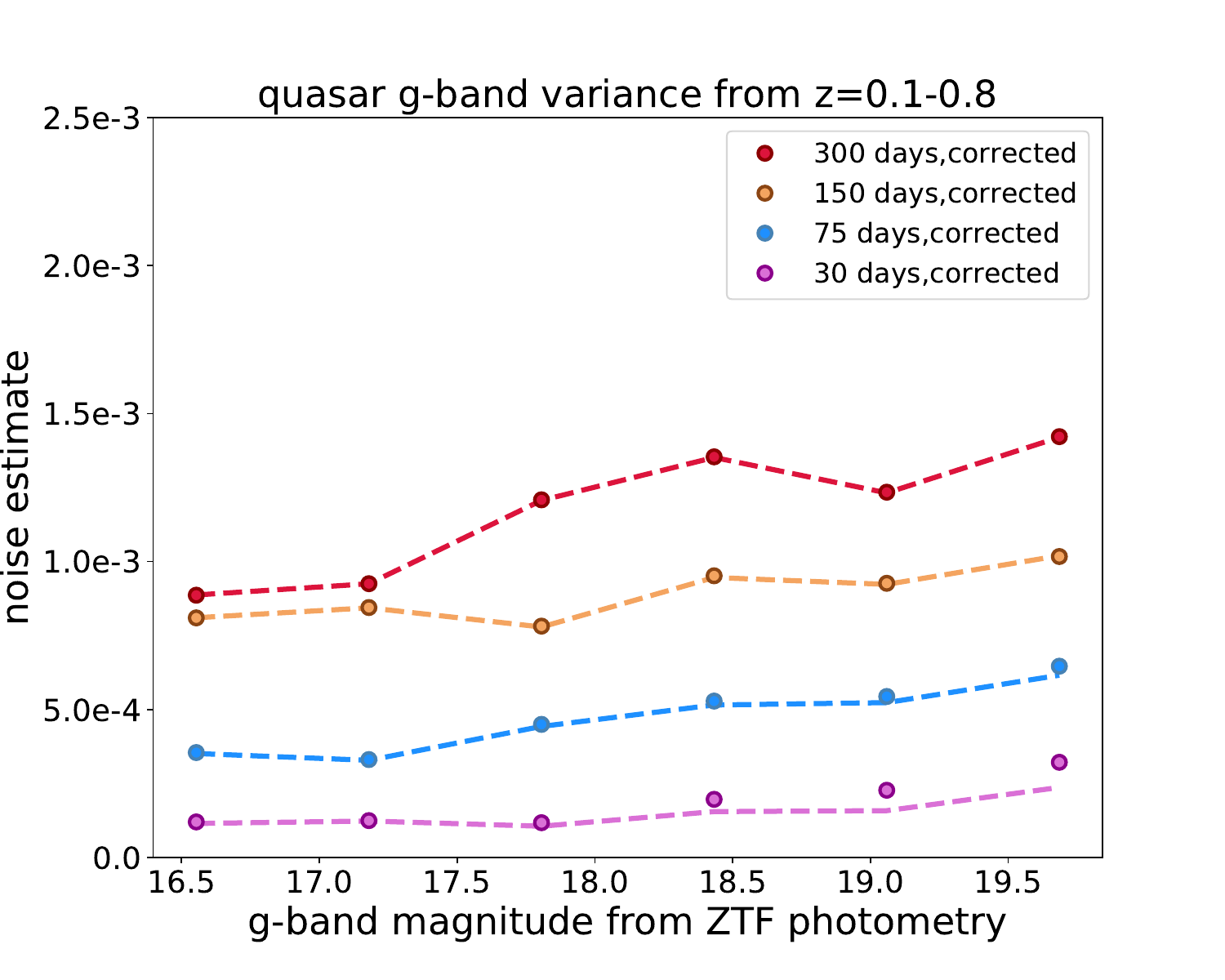}
\includegraphics[width=8cm]{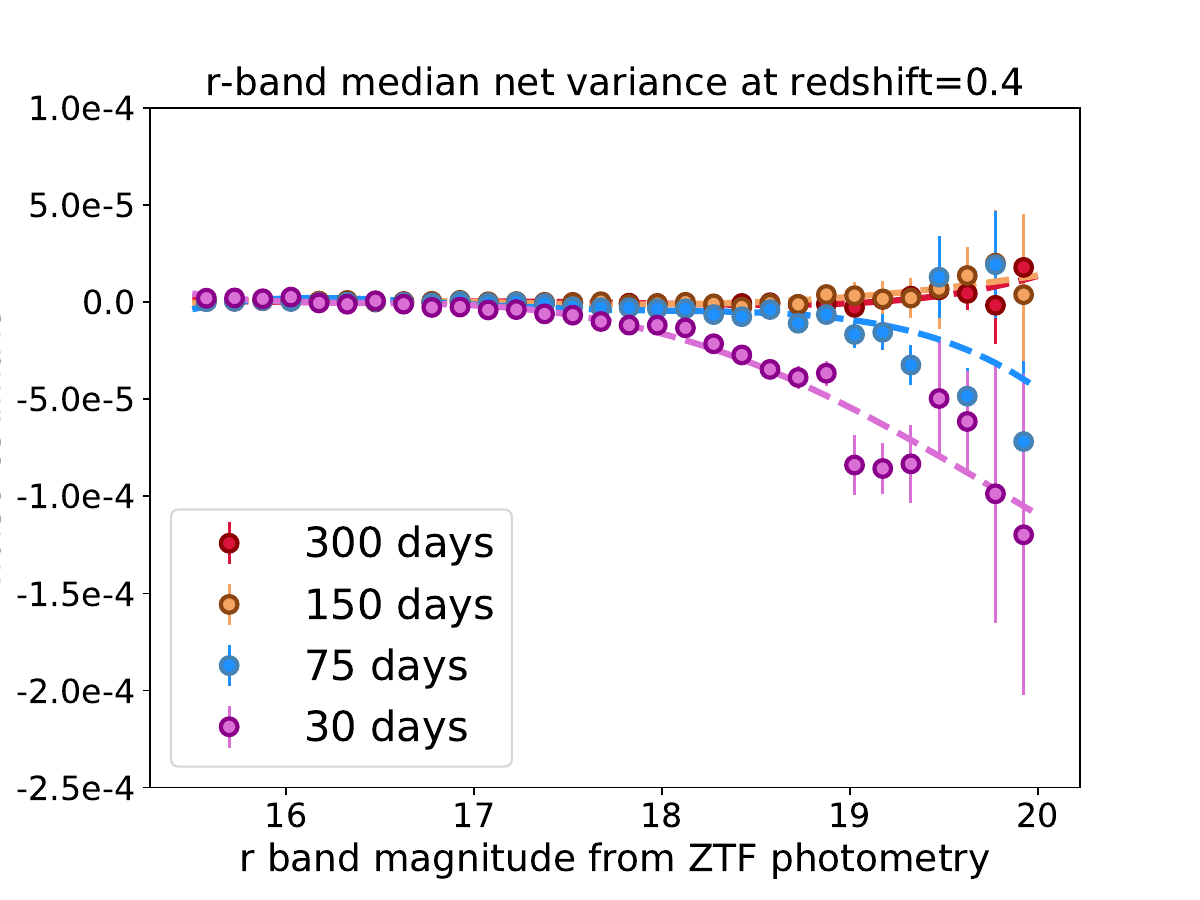}
\includegraphics[width=8cm]{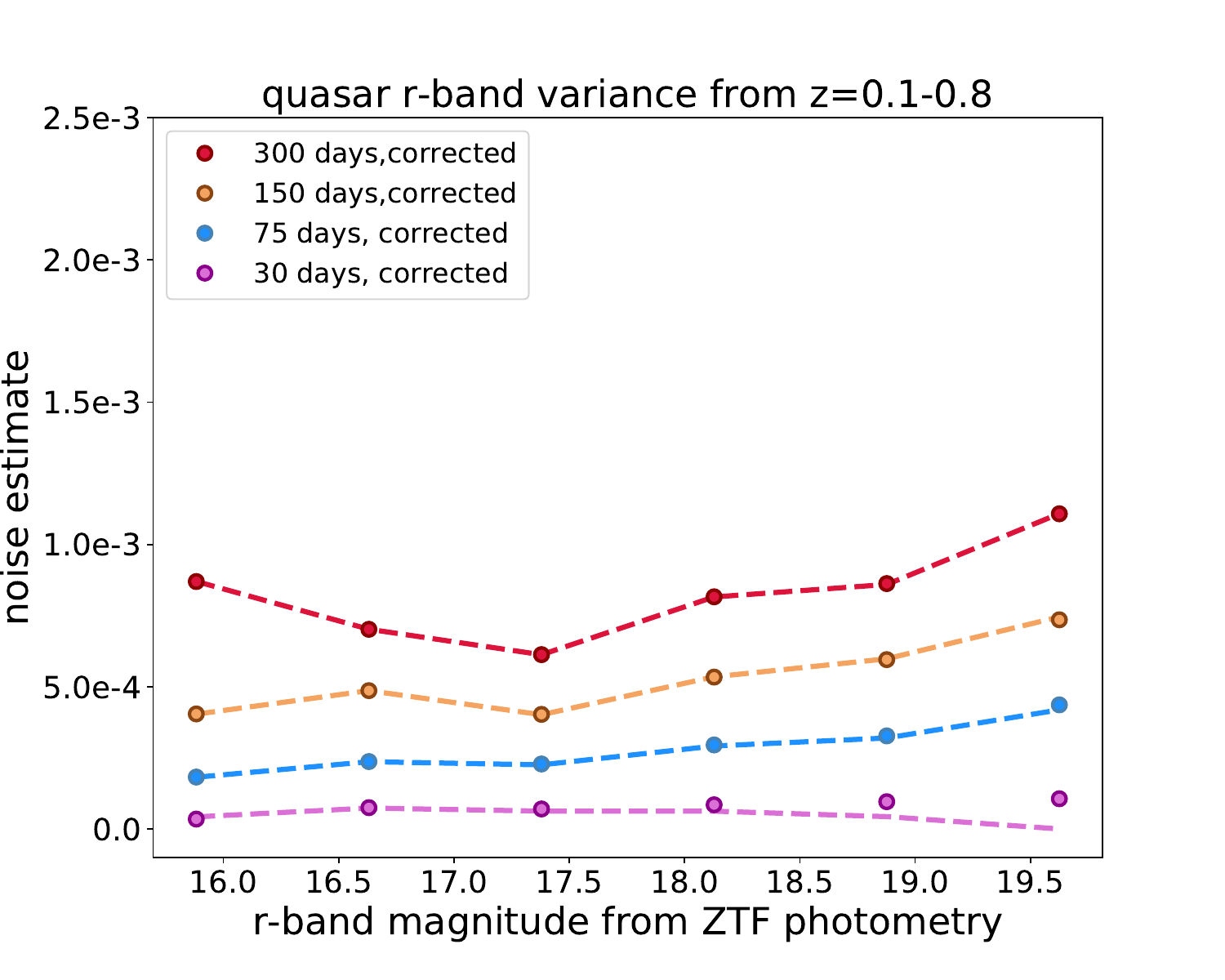}
}
\caption{Correction in the variance for the $g$ and $r$ bands. Left$\colon$ Net variance (total variance - noise estimate) of the standard stars as a function of ZTF magnitude for four different timescales at redshift bin 0.4, as is shown in the legend. The markers show the median net variances for each bin in magnitude, and the polynomial fits are shown as dashed lines. The error bars representing the median variance of the binned data were determined through a bootstrapping approach using 1000 re-samples within each magnitude bin and computing the root-mean-squared scatter of the medians. Right$\colon$ Median net quasar variances, shown with dashed lines, as a function of ZTF magnitude for four different timescales at redshifts $0.1 - 0.8$. Upon implementing the noise estimate correction, the median variances exhibit a shift from the values indicated by the lines to those represented by the solid markers (circles).}\label{fig: var vs mag}
\end{figure*}

Figure \ref{fig: var vs mag} shows the corrections applied to the quasar variances in the $g$ and $r$ bands using the results obtained from the previous analysis. The corrections affect only the shortest timescales of variability; that is, 30 days, and dim objects. From here on, we shall refer to this noise-subtracted, corrected variance simply as the variance.

After obtaining the variance for each quasar, we need to further homogenize the sample, since it covers a non-negligible range in black hole mass ($\log{M_{BH}} = 8.0 - 8.5\ M_{\odot}$) and Eddington ratio ($\log{R_{Edd}} = -1.3 - -0.8$), as the 0.5 dex in the parameter range is sufficient to introduce changes in the variability properties of the quasars. To remove the small differences in variance expected for quasars of slightly different properties, we rescaled our values to $M_{BH}$=$10^{8.0} M_{\odot}$ and $R_{Edd}=0.1$ by means of the scaling relations presented in tables 2 and 3 of Arévalo et al. (\citeyear{2023MNRAS.526.6078A}), which prescribe corrections on a log-log scale. These tables show the relation between variance and $M_{BH}$ and variance and $R_{Edd}$ for the same four timescales used in our work. The linear fit relations between log(variance) and $\log{M_{BH}}$ as well as between log(variance) and $\log{R_{Edd}}$ were used to obtain the following expressions$\colon$
{
\small
\begin{eqnarray}
\text{rescaled var(300 days)} &=& \text{var(300 days)} \times 10^{(0.17 \pm 0.04) \times (\log{M_{BH}} - 8.0)} \nonumber \\
&& \times 10^{(0.50 \pm 0.04) \times (\log{R_{Edd}} + 1)} \label{4} \\
\text{rescaled var(150 days)} &=& \text{var(150 days)} \times 10^{(0.37 \pm 0.05) \times (\log{M_{BH}} - 8.0)} \nonumber \\
&& \times 10^{(0.61 \pm 0.06) \times (\log{R_{Edd}} + 1)} \label{5} \\
\text{rescaled var(75 days)} &=& \text{var(75 days)} \times 10^{(0.66 \pm 0.02) \times (\log{M_{BH}} - 8.0)} \nonumber \\
&& \times 10^{(0.81 \pm 0.07) \times (\log{R_{Edd}} + 1)} \label{6} \\
\text{rescaled var(30 days)} &=& \text{var(30 days)} \times 10^{(1.02 \pm 0.07) \times (\log{M_{BH}} - 8.0)} \nonumber \\
&& \times 10^{(0.96 \pm 0.14) \times (\log{R_{Edd}} + 1)} \label{7}
.\end{eqnarray}
}
\section{Variability parameters as a function of rest-frame wavelength} \label{sec:Var_vs_wavelength}

In this section, we present the different correlations between the rest-frame wavelength ($\lambda_{RF}$) and the variability features measured from the ZTF light curves for the rest-frame timescales of 300, 150, 75, and 30 days before and after median binning. We used redshift as a tool to study different rest-frame wavelengths using  $4722.74/(1+z)$ \AA\ and $6339.61/(1+z)$ \AA\ for the $g$ and $r$ bands, respectively. We considered the final samples defined in Sect.~\ref{subsec:sample} for our analysis. For median binning, we grouped the variance of quasars into 19 equal-width bins of rest-frame wavelength. Each value in the rest-frame wavelength bin was replaced by its bin median value. We excluded bins containing fewer than 40 quasars. For further data analysis, we also obtained variance ratios. A variance ratio is defined as a ratio between variance at a longer timescale and a shorter timescale. We then grouped the variance ratios into 19 equal-width bins of rest-frame wavelength. Given that the errors in ratio determinations are larger, a robust statistical analysis requires a substantial number of quasars. To ensure reliability and minimize error propagation, we restricted our analysis to those wavelength bins that contained more than 50 quasars. 

\subsection{Correlations}

We first analyzed the correlation strength between variance and rest-frame wavelength on the variability timescales of 300, 150, 75, and 30 days before and after median binning, using Spearman’s rank correlation coefficient ($\rho_{s}$), which does not consider measurement errors in the variables. Table \ref{table : corr_varaince} presents the correlations between variance at four different timescales and rest-frame wavelength for 2533 sources in the $g$ band and 2795 sources in the $r$ band.

Prior to binning, variance at timescales of 300, 150, and 75 days shows a weak but significant anticorrelation with the rest-frame wavelength, whereas variance at a timescale of 30 days exhibits a moderate negative correlation. After median binning, the correlations between variance and the rest-frame wavelength become significantly stronger and anticorrelated at all timescales. This finding broadly supports the work of previous research linking the amplitude of variability with the rest-frame wavelength (Cutri et al.~\citeyear{1985ApJ...296..423C}; Paltani \& Courvoisier~\citeyear{1994A&A...291...74P}; Vanden Berk et al.~\citeyear{2004AAS...20512002V}; MacLeod et al.~\citeyear{2010ApJ...721.1014M},~\citeyear{2012ApJ...753..106M}; Meusinger et al.~\citeyear{2011A&A...525A..37M}; Zuo et al.~\citeyear{2012ApJ...758..104Z}; Morganson et al.~\citeyear{2014ApJ...784...92M}; Li et al.~\citeyear{2018ApJ...861....6L}; Sánchez-Sáez et al.~\citeyear{2018ApJ...864...87S}).

As the next step of our analysis, we investigated the relationship between all six variance ratios and the rest-frame wavelength. We computed Spearman's rank correlation between log($\lambda_{RF}$) and log(variance ratio) both before and after median binning, as is shown in Table \ref{table : corr_varaince_ratio}. For the variance ratio (300/150) — that is, the ratio of variance(300 days) to the variance(150 days) — we observed a negligible correlation strength between the two variables both before and after binning, implying that the two variables cannot be represented by a monotonic relationship. However, for four variance ratios (300/30, 150/30, 150/75, and 75/30), we found a weak positive correlation with rest-frame wavelength before median binning, which was statistically significant. After applying median binning, the correlations between the variance ratios and the rest-frame wavelength became significantly stronger, suggesting a clear relationship. For the variance ratio (300/75), the relationship is moderately significant. As the log(variance ratio) is a proxy for the power spectral slope, this finding indicates that the power-law slope depends on the rest-frame wavelength. These results corroborate those obtained by Li et al.~(\citeyear{2018ApJ...861....6L}). In their study, Li et al.~use a power-law model to describe the variability structure function, denoted as $V=A{(t\ /1\ \mathrm{year})}^{\gamma }$. Furthermore, they observe that the logarithmic gradient of variability, represented by $\gamma$, can also be stated as a positive function of the rest-frame wavelength. 

\begin{table}
\centering
\caption{Correlation between log(variance) and log($\lambda_{RF}$).}
\label{variance vs rest-frame wavelength}
\begin{tabular}{@{}ccc@{}}
\toprule
timescale & Unbinned   & Binned    \\ \midrule
300 days  & -0.23 (\textless{1e-8}) & -0.93 (4e-8) \\
150 days  & -0.24 (\textless{1e-8}) & -0.92(1e-7)   \\
75 days   & -0.35 (\textless{1e-8}) & -0.99 (\textless{1e-8}) \\
30 days   & -0.47 (\textless{1e-8}) & -0.97 (\textless{1e-8})\\ \bottomrule
\end{tabular}
\tablefoot{Correlation coefficient $\rho_s$ and p-value (in parenthesis) determined between log(variance) and log($\lambda_{RF}$).}
\label{table : corr_varaince}
\end{table}

\begin{table}
\centering
\caption{Correlation between log(variance ratio) and log($\lambda_{RF}$).}
\label{variance ratio vs rest-frame wavelength}
\begin{tabular}{@{}ccc@{}}
\toprule
timescale & Unbinned & Binned \\ \midrule
300/30    &  0.17 (\textless{1e-8}) &   0.74 (2e-3)     \\ 
300/75    &  0.05 (1e-3)          &   0.58 (0.02)       \\
300/150   &  -0.03 (0.08)         &   0.04 (0.9)        \\
150/30    &  0.23 (\textless{1e-8})  &   0.74 (2e-3)    \\
150/75    &  0.12 (\textless{1e-8})  &   0.90 (6e-6)    \\
75/30     &  0.20 (\textless{1e-8})  &   0.69 (4e-3)    \\
\bottomrule
\end{tabular}
\tablefoot{
Correlation coefficient, $\rho_s$, and p value (in parentheses) determined between log(variance ratio) and log($\lambda_{RF}$).
}
\label{table : corr_varaince_ratio}
\end{table}

\subsection{Linear regression}\label{regression}

\begin{figure*}[!ht]%
\centering
\includegraphics[width=0.33\textwidth]{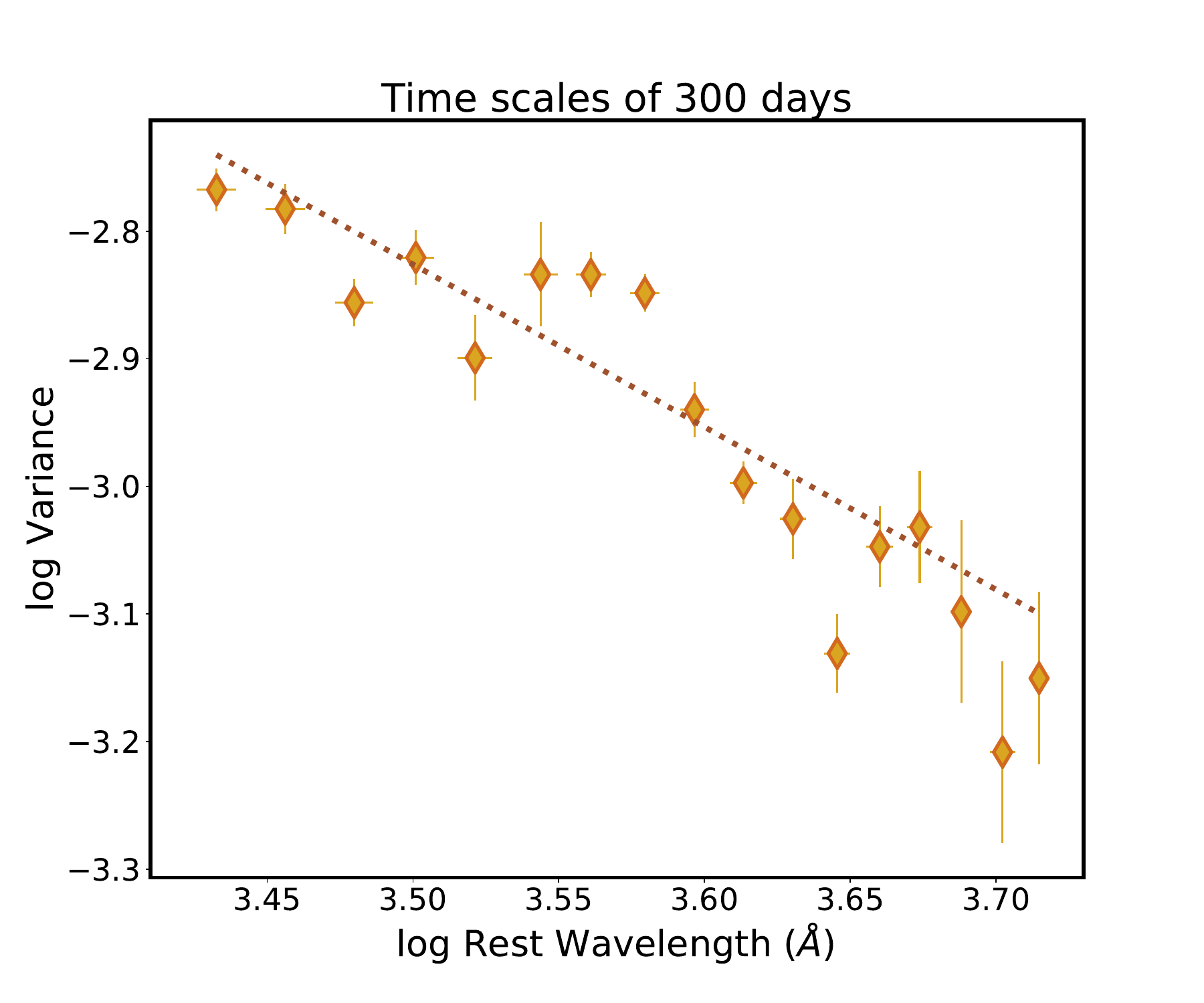}
\includegraphics[width=0.33\textwidth]{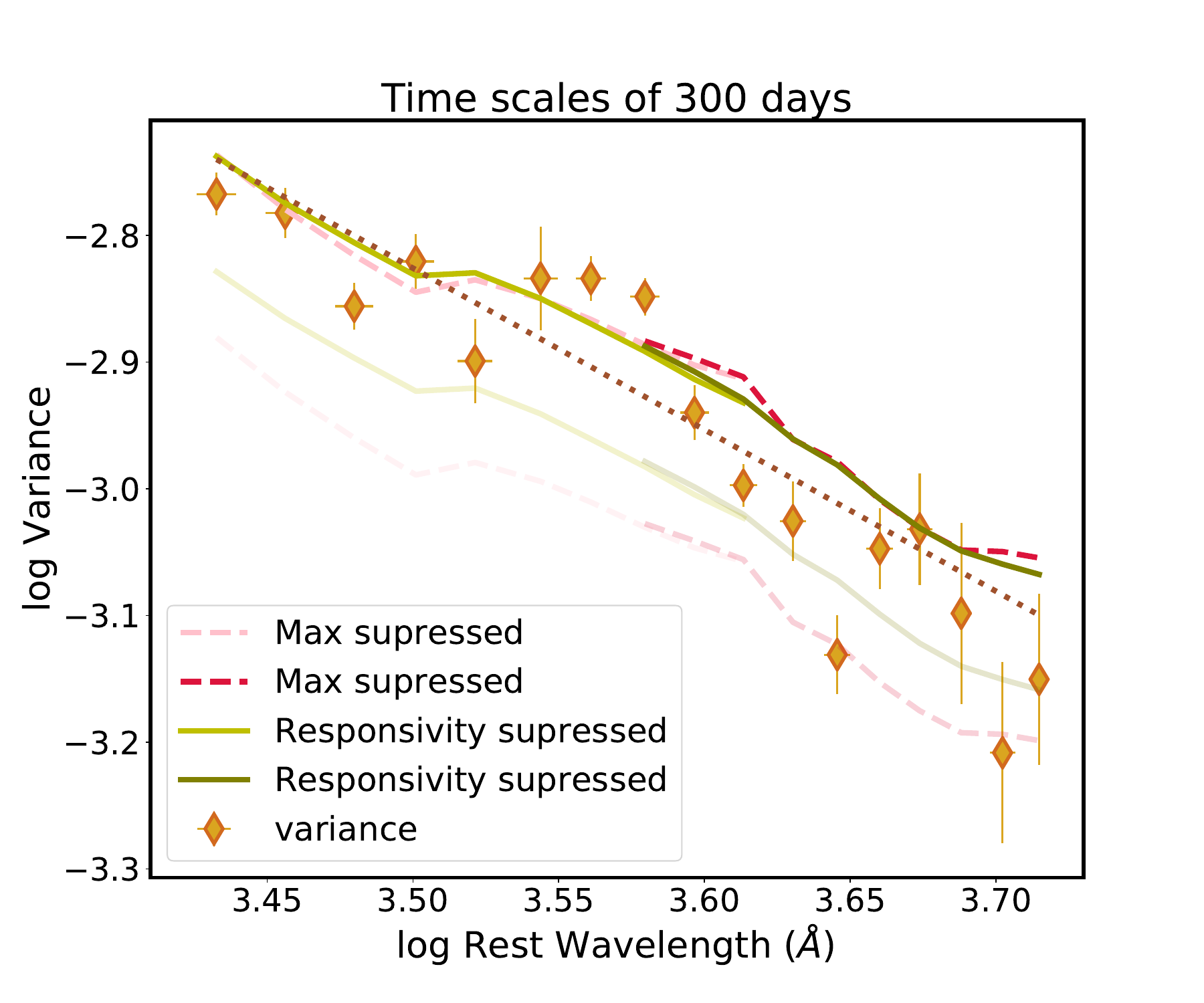}
\includegraphics[width=0.33\textwidth]{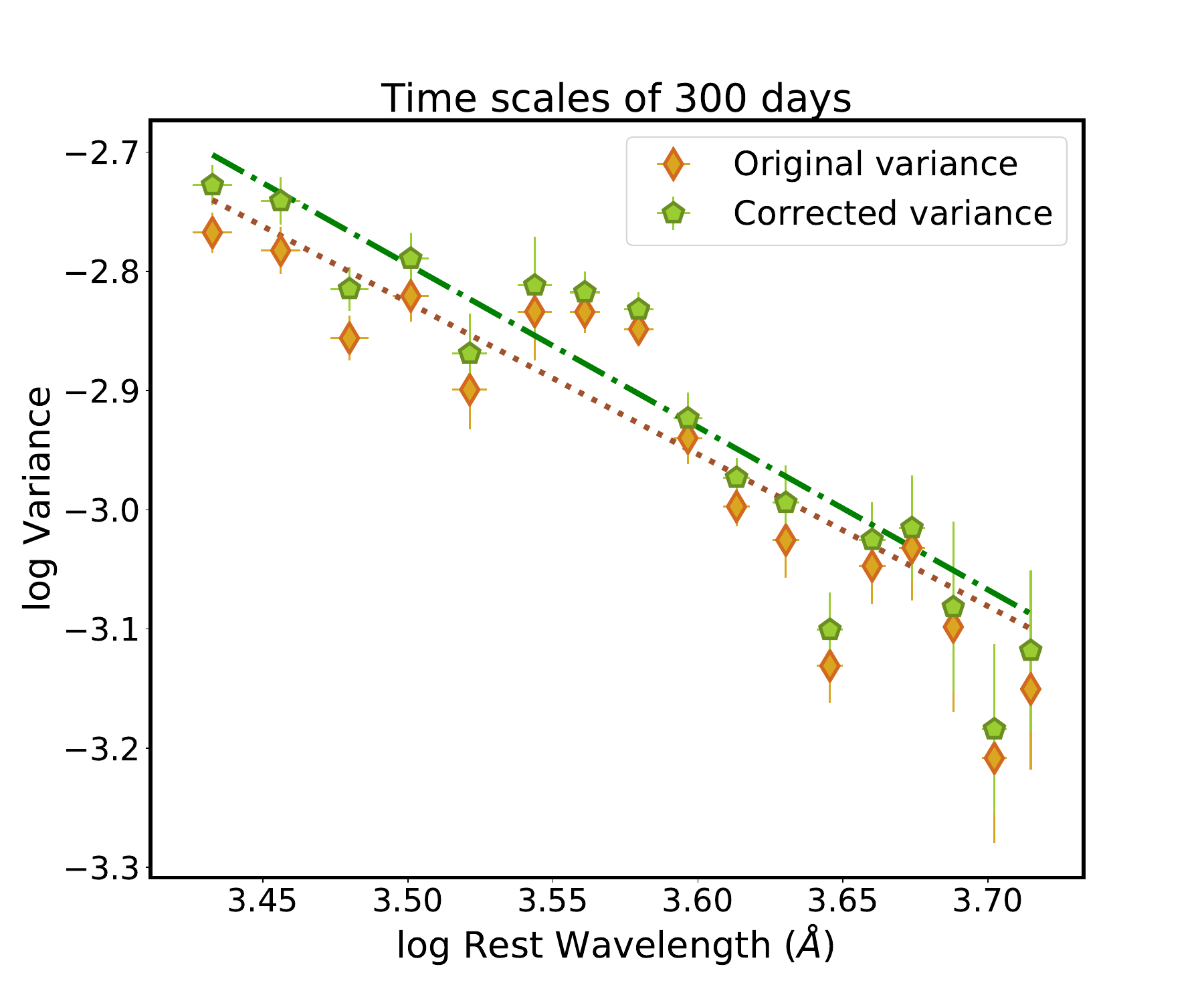}
\includegraphics[width=0.33\textwidth]{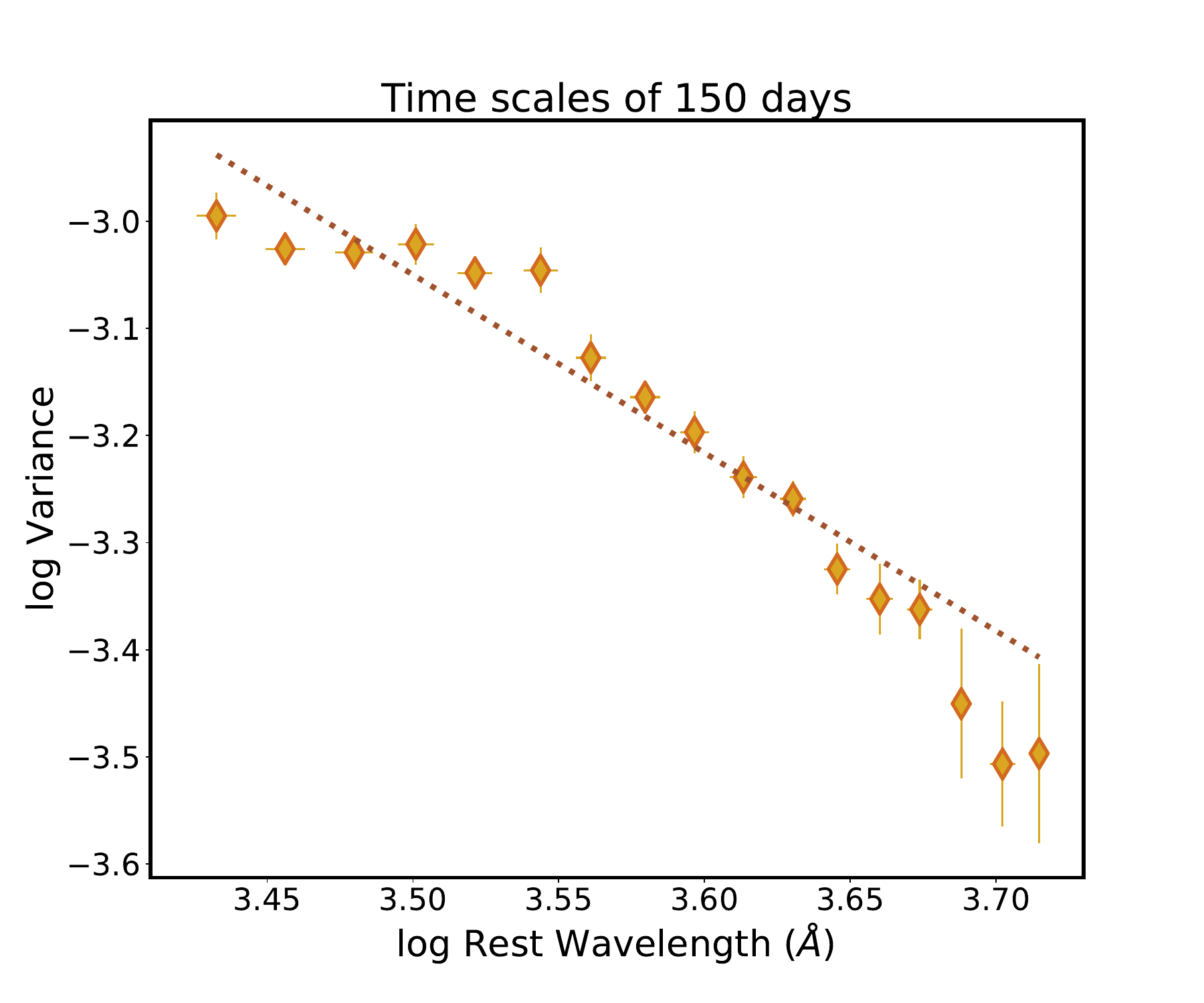}
\includegraphics[width=0.33\textwidth]{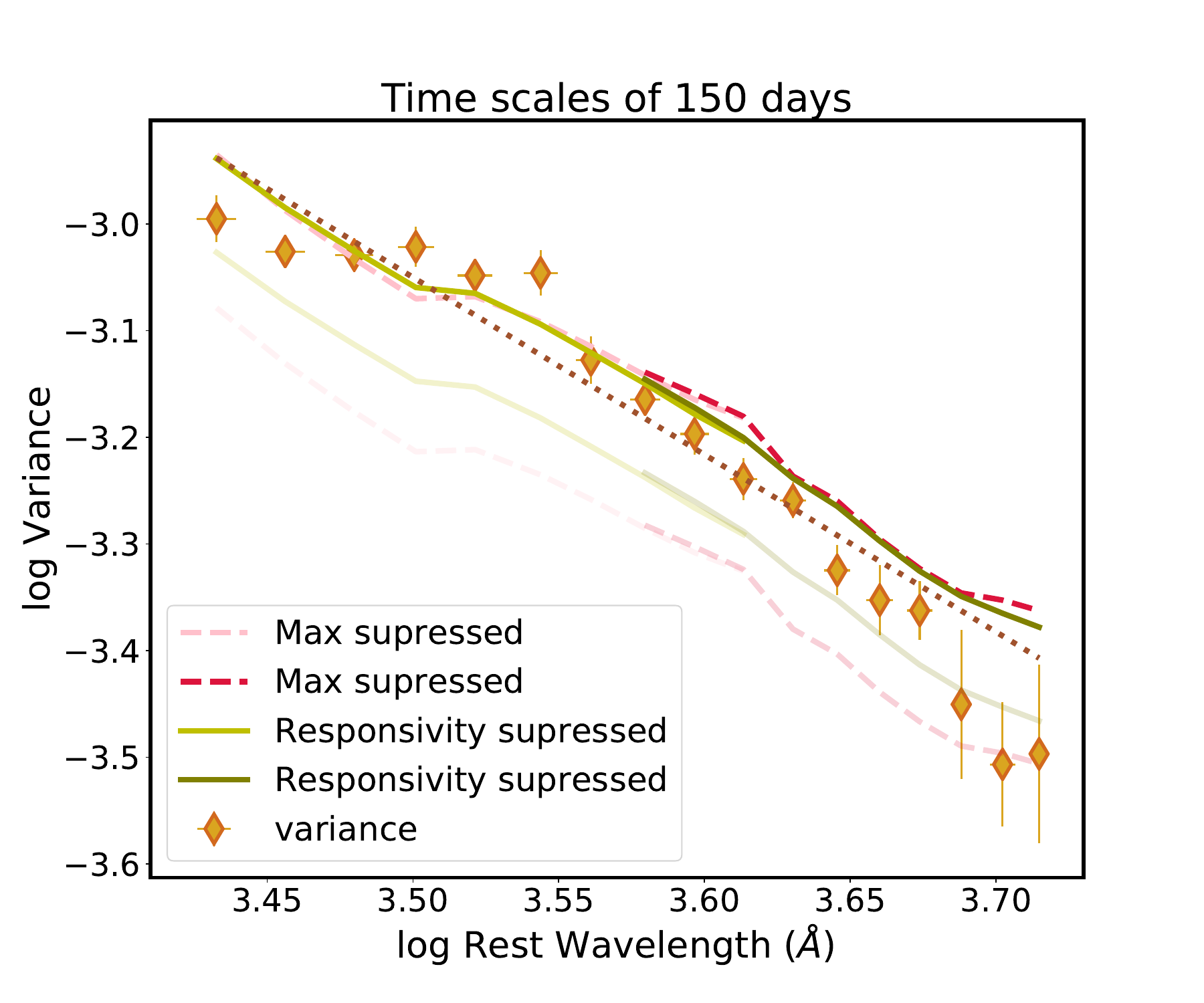}
\includegraphics[width=0.33\textwidth]{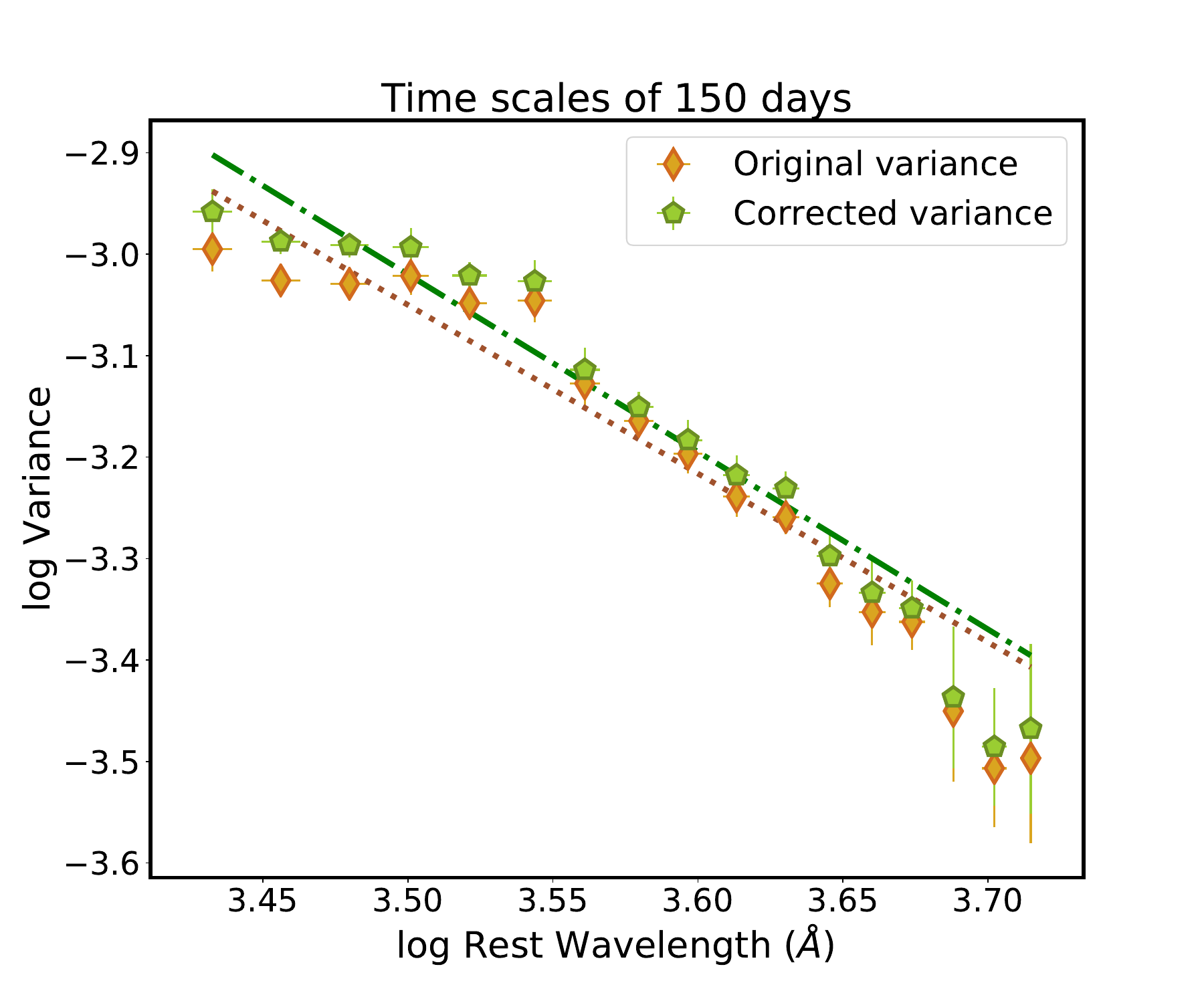}
\includegraphics[width=0.33\textwidth]{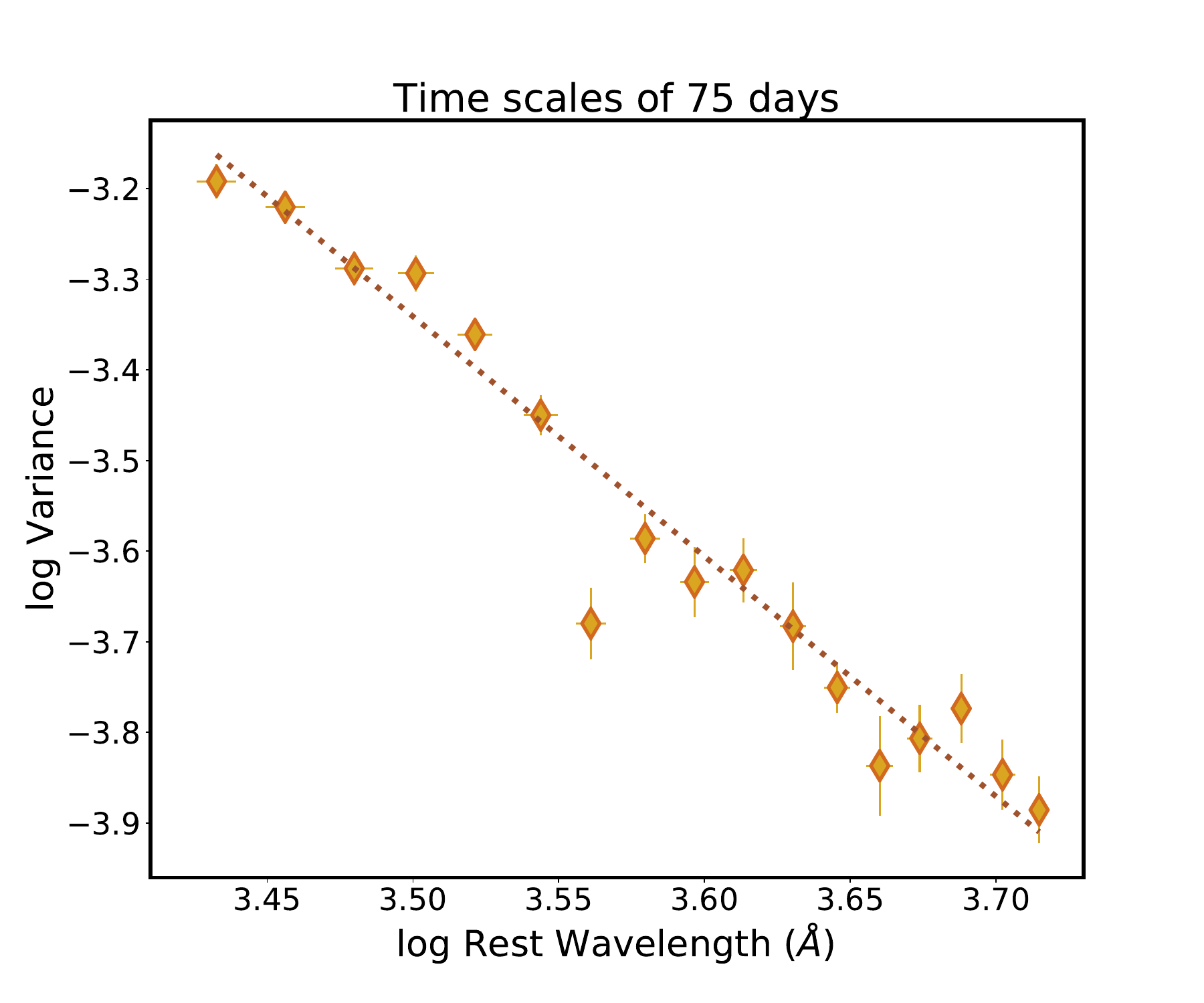}
\includegraphics[width=0.33\textwidth]{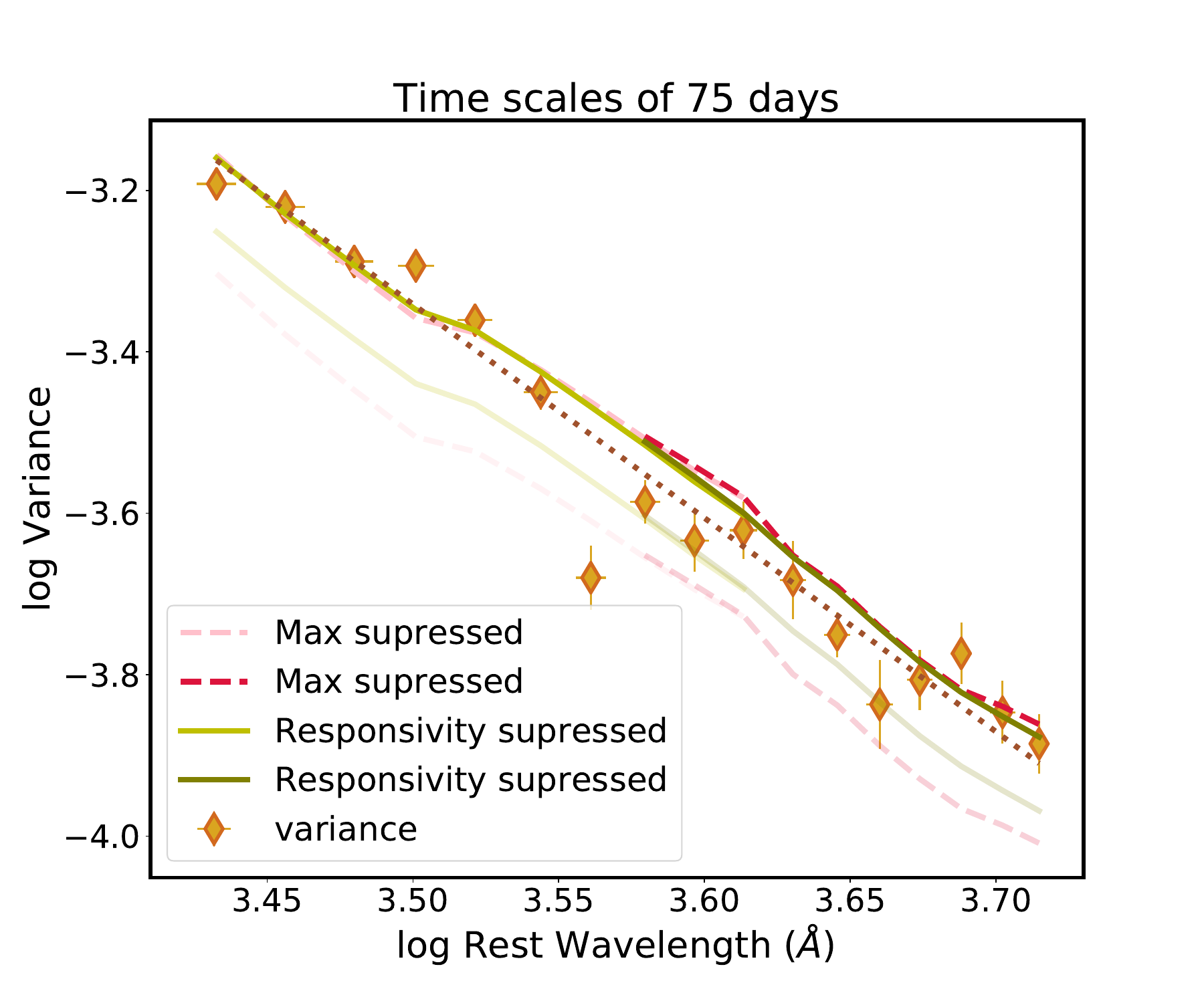}
\includegraphics[width=0.33\textwidth]{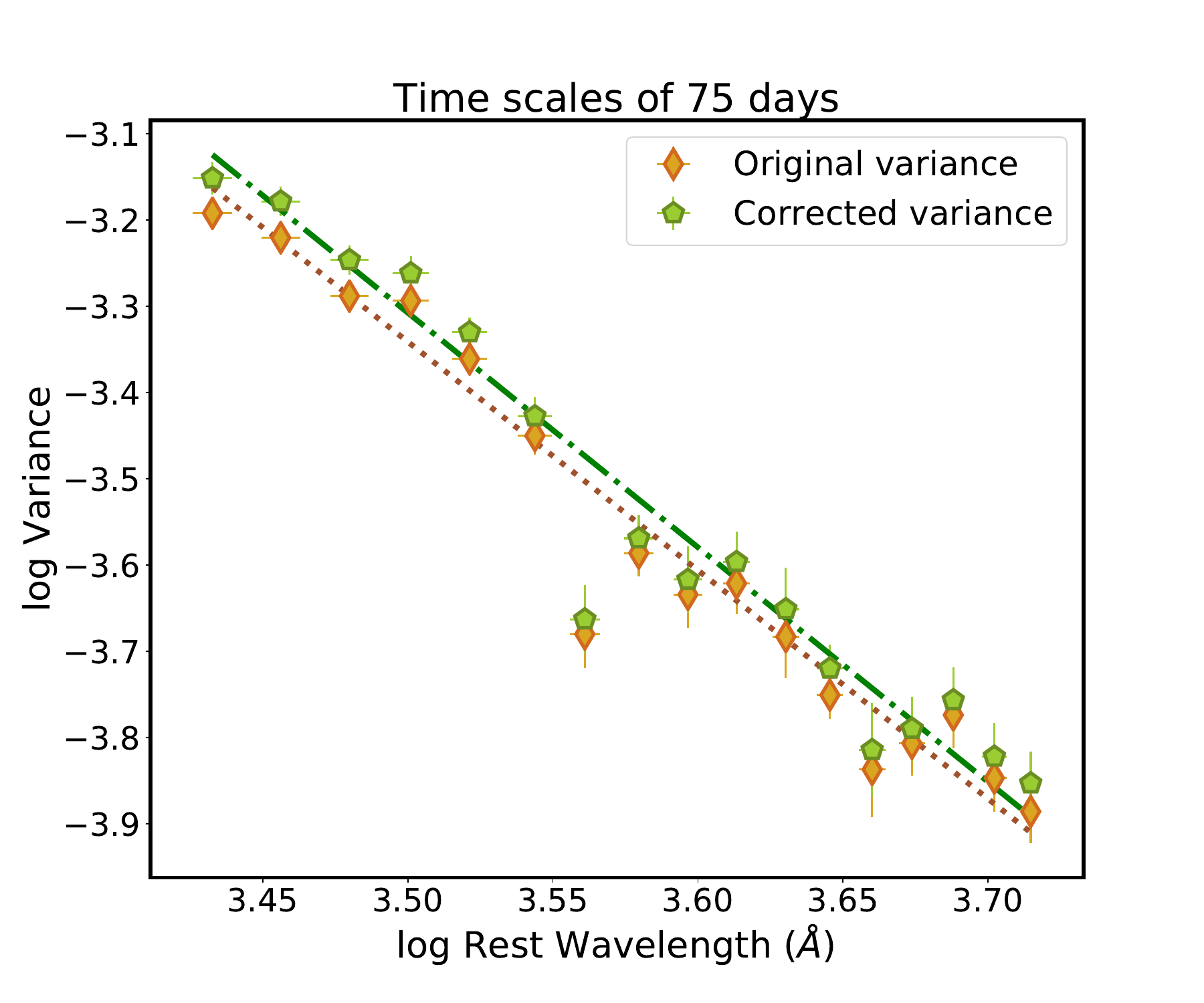}
\includegraphics[width=0.33\textwidth]{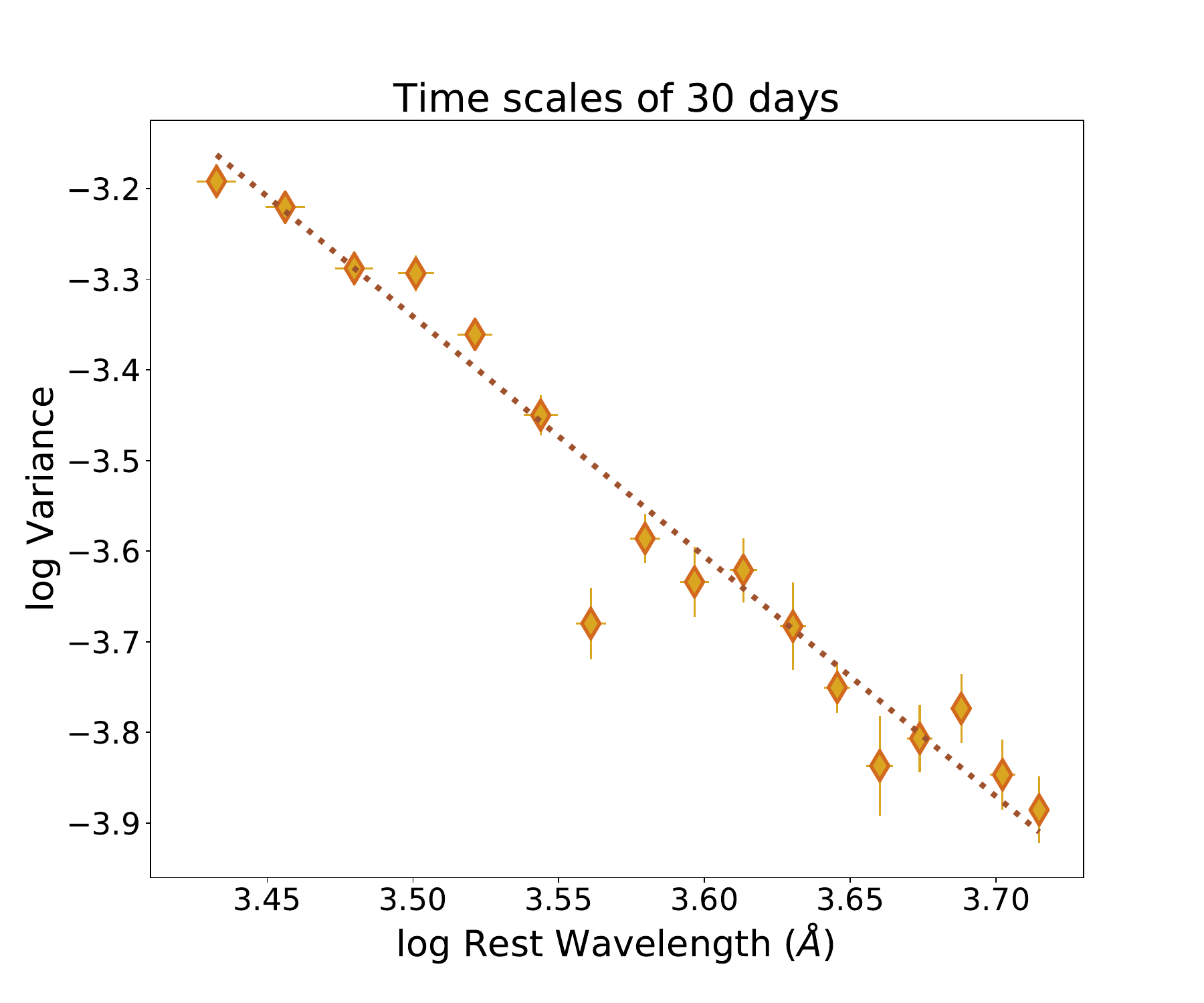}
\includegraphics[width=0.33\textwidth]{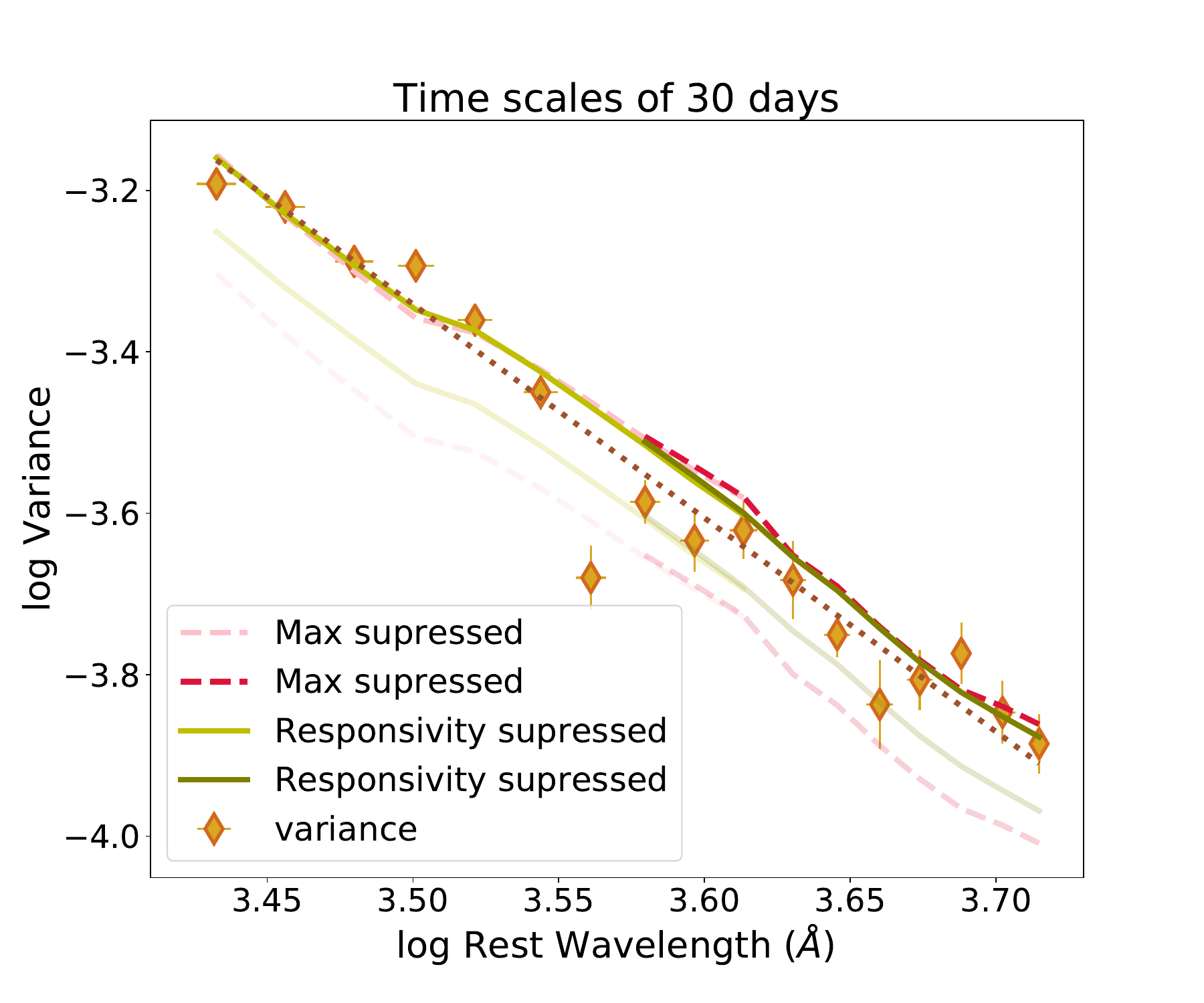}
\includegraphics[width=0.33\textwidth]{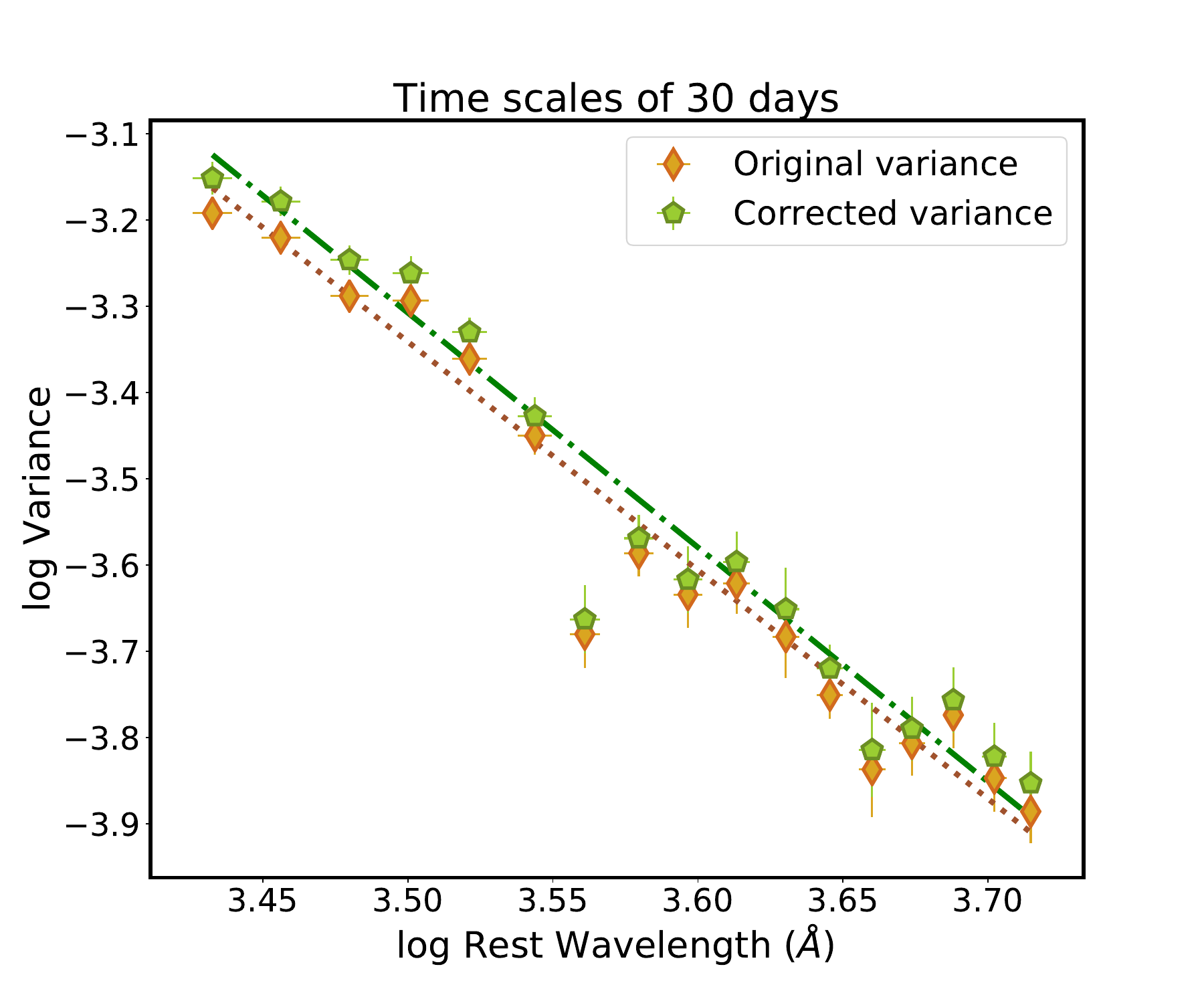}
\caption{Variance vs. rest-frame wavelength for four variability timescales. Each row represents a specific timescale of 300, 150, 75, or 30 days. Left column: Original variance and linear fits. Steeper negative correlations are seen for shorter timescale fluctuations (notice the different dynamical range of the y axis). Middle column: Original variance and linear fits including spectral corrections due to variance suppression caused by emission lines and Balmer and Fe pseudo-continua. Suppression was obtained assuming that emission lines and pseudo-continua are not variable at all or show partial variability given by their responsivities (Kokubo et al.~\citeyear{2014ApJ...783...46K}), as shown in the legend. Lighter lines show the full spectral suppression while darker lines show the suppression after and arbitrary vertical shift is applied. Right column: Original (gold diamonds) and new variance values (green pentagons) after adding the suppressed variance as determined using line and pseudo-continua responsivities, including original and new linear fits. See Section~\ref{sec:spectral study} for details on the spectral suppression correction.}
  
\label{fig:variance plots}
\end{figure*}

We performed a linear regression between variance and the rest-frame wavelength ($\lambda_{RF}$) to further characterize the variance$-\lambda_{RF}$ dependency for our sample of quasars with black hole masses of $10^{8} M_{\odot}$ and Eddington ratios of $10^{-1}$. This analysis only used the median-combined data. As was mentioned above, we divided the variance of quasars into equal-width bins of rest-frame wavelength for median binning. Bootstrapping was used to compute the standard error on these medians using 1000 random samples from the original sample and calculated the standard deviation of the medians for each bin. The median variance of these bootstrapping samples provided the error estimates. We used the orthogonal distance regression method, which considers errors in both axes and perpendicular to the fitted line instead of just vertically or horizontally. As is shown in the left column of Fig. \ref{fig:variance plots}, there is a significant decreasing trend for the variance with increasing rest-frame wavelength for the four variability timescales studied. Although a linear fit provides an acceptable description of the general trend, there is a significant deviation from linearity at all four timescales in the ranges of log$(\lambda_{RF}) = 3.50-3.56$ and $3.68-3.71$ \AA. This deviation could be due to spectral features. Motivated by this, we shall investigate the possibility that spectral features cause the observed wiggles in Sect. ~\ref{sec:spectral study}. 
 
After obtaining the linear correlation between variance and rest-frame wavelength, we need to also examine the relationship between log(variance ratio) and log($\lambda_{RF}$). We find a significant increasing trend for the variance ratio with increasing rest-frame wavelength (see Fig. \ref{fig : variance ratio correction}). 

The first column of Table \ref{tab : linear_fits} (Original data) displays the regression results for the variance and variance ratio. The table includes the best-fitting values of parameters with their $1-\sigma$ errors.

\section{Spectral study} \label{sec:spectral study}

We first examined the spectral components observed in the wavelength range studied to understand the nature of the features seen in Fig. \ref{fig:variance plots}. The rest-frame wavelength range covered by our sources extends from 2700 Å to 5200 Å. This wavelength range includes the continuum from the accretion disk, broad and narrow emission lines, particularly Mg II 2800 Å, H$\gamma$ 4341.68 Å, and H$\beta$ 4862.68 Å, and the emission from the Balmer pseudo-continuum and Fe II pseudo-continuum. We considered these spectral features introduced by different redshifts as they shift in and out of the spectral range of the filters, while calculating the contribution of the emission lines and continuum emission.

The estimated normalized variance, which is variance divided by the total mean flux squared, is dimensionless. If the broad emission lines and pseudo-continua fractionally fluctuate in the same way as the quasar continuum, then they behave as a multiplier for the un-normalized variance in the numerator and the total flux squared in the denominator. Therefore, there will be no change in the normalized variance. In contrast, if the lines and pseudo-continua do not change at all, they will only contribute to the total flux in the denominator, reducing the normalized variance. The intermediate situation should take into account the fraction of these spectral components that vary. It can be shown that the correction term to apply to the observed variance to obtain the true continuum variance is $(1 + c)^2$, where $c = \Sigma_i\, (1-\eta_i)\times f_i\,/\, \Sigma_i\ \eta_i \times f_i$, and $\eta_i$ is the fraction of the flux of the spectral components that responds linearly to changes in the continuum, and $f_i$ is its flux. In particular, the accretion disk continuum has a multiplicative value $\eta = 1$.

To determine whether the amplitude and spectral shape of the emission lines and pseudo-continua can reproduce the wiggles in the variance-rest-frame wavelength plot, we calculated an expression for $c$ as a function of wavelength. First, we selected the SDSS DR14 spec-3767-55214-0738 quasar spectrum, which has signal-to-noise ratios of 10 at 5100 Å and 14 at 3000Å, located at RA = $143.45$ and DEC $= -1.71$. We selected this spectrum at $z = 0.71$ because higher-redshift quasars allow for a better estimate of the Balmer continuum. We then used the penalized pixel-fitting package (pPXF; Cappellari \& Emsellem~\citeyear{2004PASP..116..138C}; Cappellari~\citeyear{10.1093/mnras/stw3020}) to perform the spectral fitting of the source (see Appendix~\ref{sec:appendixB}). After fitting the spectrum, we estimated the flux associated with the different spectral components within the $g$ and $r$ bands for the different redshift bins of the sample.

The $c$ ratio was estimated and used to calculate the new variance using the above equation. We assumed two cases: 1) that the emission lines, and the Balmer and FeII pseudo-continua, do not vary (i.e., $\eta = 0$), and 2) that these spectral components show partial variability. For this later case, and following Kokubo et al.~(\citeyear{2014ApJ...783...46K}), we assumed the responsivity values $\eta = 0.2$ for MgII and the FeII pseudo-continuum, and $\eta = 0.6$ for the more variable Balmer lines and continuum (see also Goad et al.~\citeyear{Goad_1999}).

The middle and right columns of Figure \ref{fig:variance plots} illustrate our findings. In the middle column we incorporated the amount of suppression in the linear correlations previously obtained, while in the right column we corrected the observed variance for these suppression factors to obtain a new corrected variance. Since the spectral components move in and out of all the spectral bins of our determined variance, variability suppression affects the whole variance spectrum. Hence, an arbitrary shift was also applied to the corrected linear fits. This shift was found by a weighted $\chi^2$ minimization between the corrected linear relations and the observed variance values. Figure \ref{fig:variance plots} shows that broad emission lines, Fe II emission, and Balmer pseudo-continuum effectively suppress all the variance (lighter lines in the middle column), and after the described shift, those regions where the spectral modulations of the observed variance look stronger coincide where the regions where the spectral corrections present breaks. Also, the corrections affect more strongly the longer time scales, as the variance spectrum is shallower. That is also seen in the measured variance itself, where wiggles are more pronounced at longer timescales. The difference between adopting $\eta = 0$ for the lines and pseudo-continua or partial variability of these components is very small, as can be seen when comparing the full and segmented lines after the vertical shift is applied.

\begin{figure*}[ht]%
\centering
\includegraphics[width=0.33\textwidth]{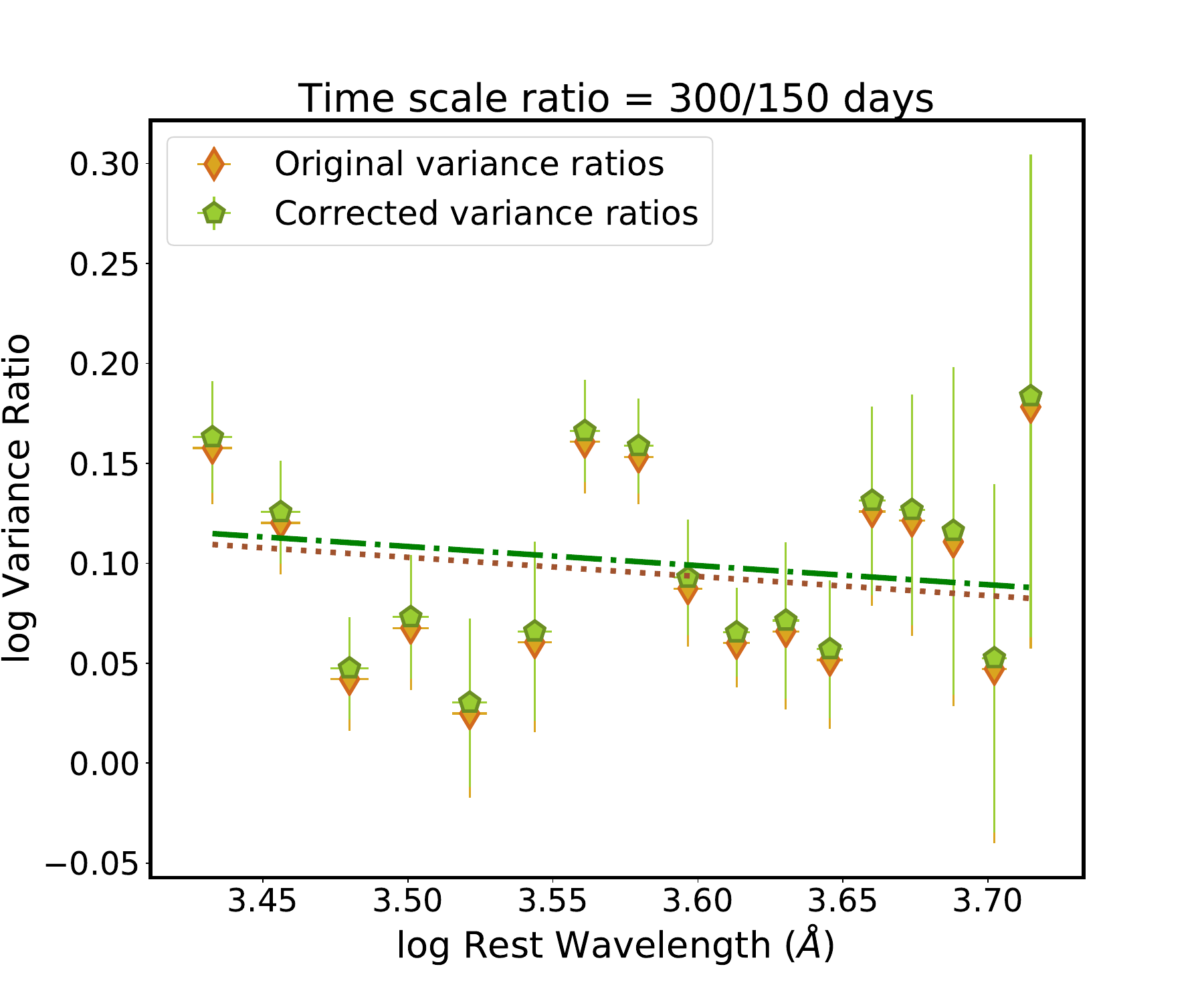}
\includegraphics[width=0.33\textwidth]{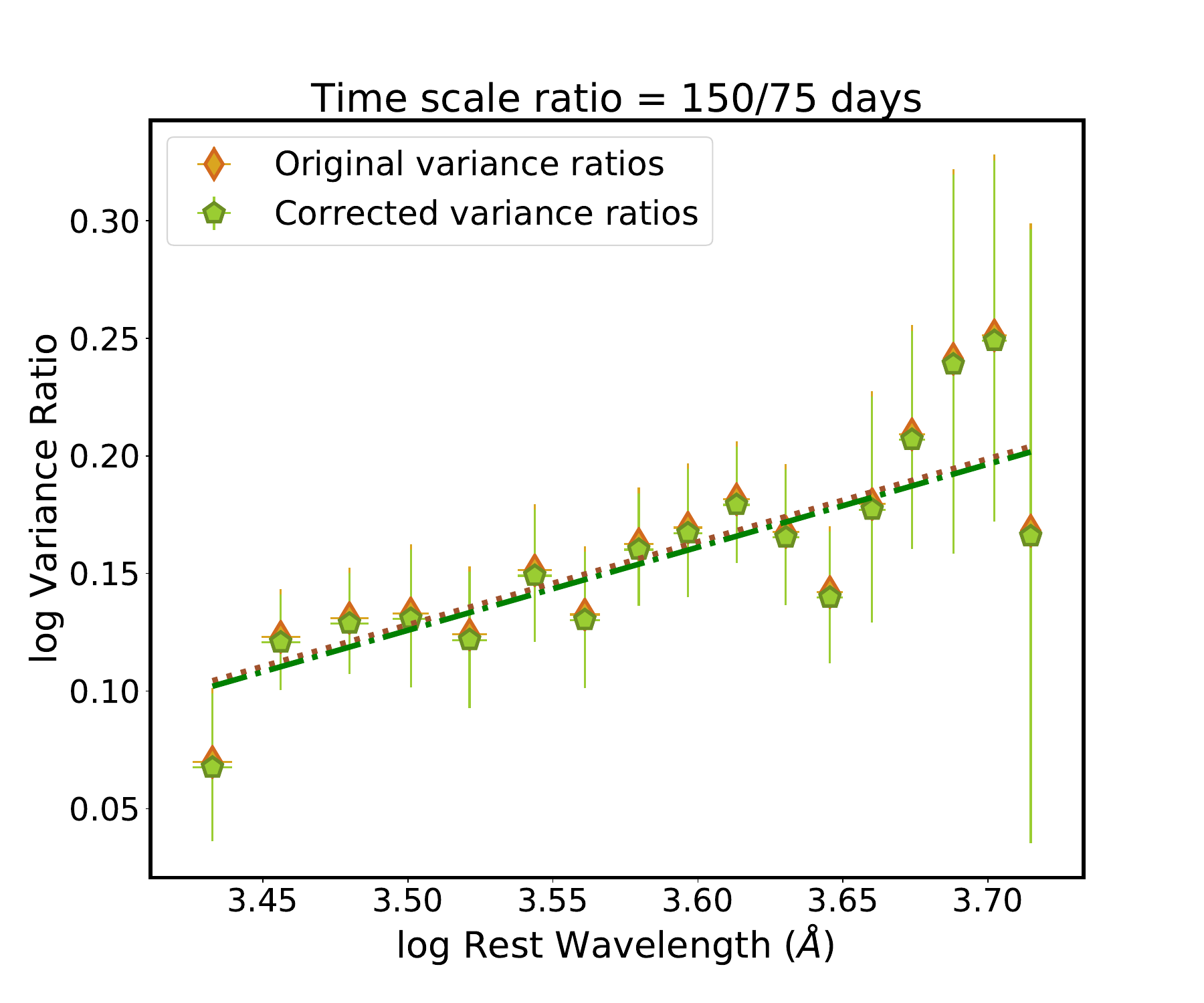}
\includegraphics[width=0.33\textwidth]{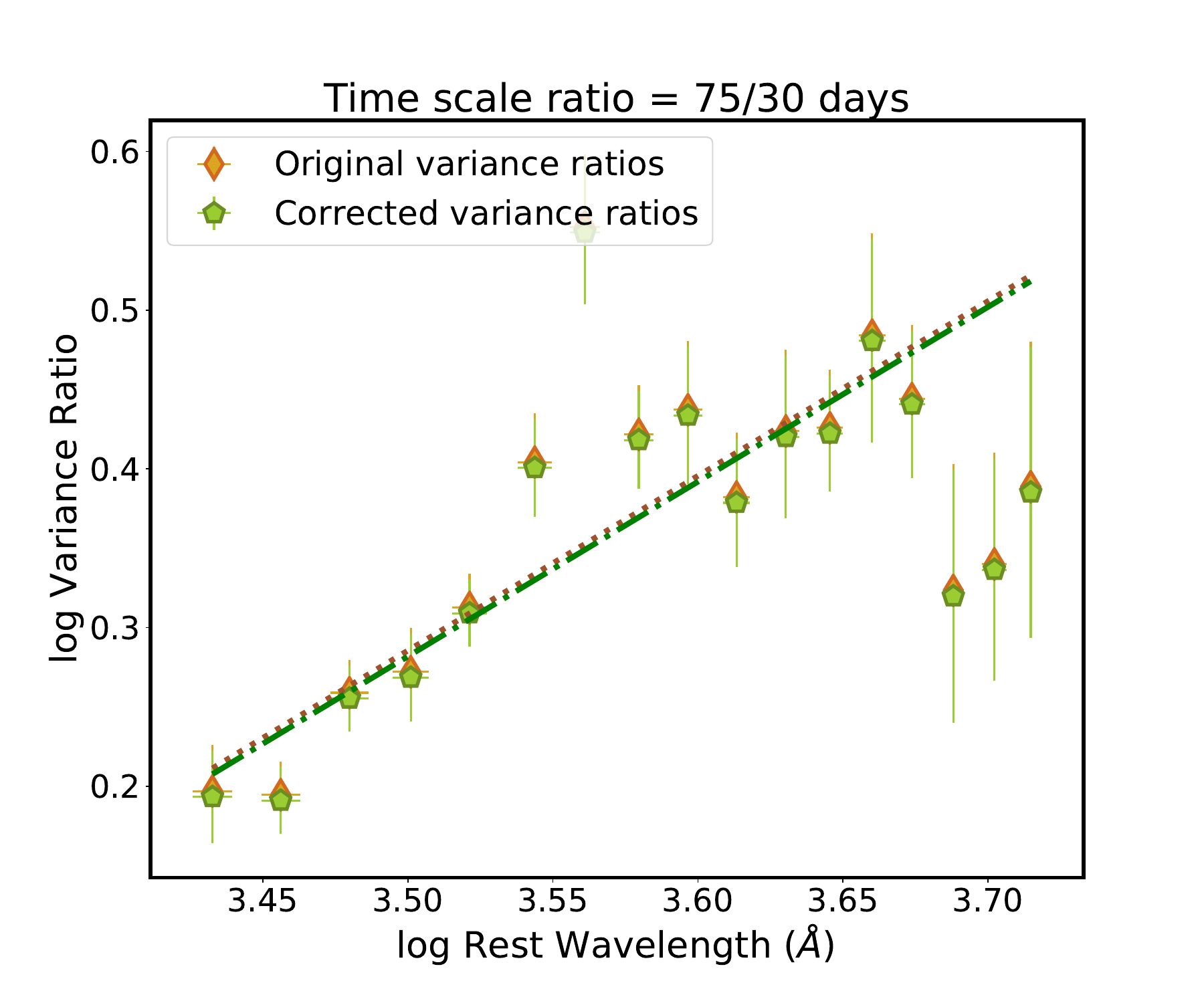}
\caption{Quasar variance ratios as a function of rest-frame wavelength. The median variance ratios are shown with gold diamonds. Green pentagons denote values corrected for spectral suppression. The error bars correspond to the root-mean-squared scatter of the median variance ratio. Linear fits are shown using lines.}
  
\label{fig : variance ratio correction}
\end{figure*}

We carried out a new linear regression as a function of rest-frame wavelength to the corrected variance values determined assuming partial variability given by the spectral components responsivity, with the regression results listed in the second column of Table \ref{tab : linear_fits} (Corrected Data). Essentially, the changes correspond to minor corrections to the slopes and intercepts. Henceforth, the new variance will be referred to as the observed variance. We also examined the new variance ratios, but the results are indistinguishable from the already obtained relations (see Figure \ref{fig : variance ratio correction}).

\section{Testing the CHAR model}\label{sec: CHAR model}

The CHAR model proposes a magnetic coupling between the corona and the underlying cold, thin disk. Fluctuations in the coronal magnetic field (with power $Q_{mc}^{+}$) propagate outward at near $c$ speed and cause changes in the heating rate of the accretion flow. As a result, variations appear in the disk temperature. While calculating the effective temperature of the resulting disk, the model considers vertically integrated thermal-energy conservation (for more details, see Sect. 2 of Sun et al.~\citeyear{2020ApJ...891..178S}). In this section, we examine how well the CHAR model predictions match the observed quasar UV/optical variability. Specifically, we compare the model predictions to the observed relation between variance and rest-frame wavelength and assess the agreement between the two. 

\subsection{Modelling ZTF quasar variability}\label{subsec: Modelling}

Simulated quasar light curves were obtained using the CHAR model in the wavelength range of interest. Fluctuation of $Q_{mc}^{+}$ occurring at different annuli within $\sim 10\, R_s$, move toward the center, resulting in fluctuations in accretion power within the corona. These fluctuations exhibit a power spectral density (PSD), which has been usually presented as the inverse of frequency; that is, $Q_{mc}^{+} \propto 1/f$ (for instance, Lyubarskii ~\citeyear{1997MNRAS.292..679L}; King et al.~\citeyear{2004MNRAS.348..111K}; Ar{\'e}valo et al. ~\citeyear{2008MNRAS.389.1479A}; Lin et al.~\citeyear{2016MNRAS.463..245L}; Noble \& Krolik~\citeyear{2009ApJ...703..964N}). Subsequently, the corona drives outward-propagating magnetohydrodynamic waves into the disk. We further considered alternative PSDs for $Q_{mc}^{+}$, specifically $\propto 1/f^{\beta}$, where $\beta$ ranges from 0.6 to 2, aiming to identify the best fit to the data. It should be noted that only for $\beta = 0$, or “white-noise,” the fluctuations are completely uncorrelated. On the other hand, $\beta > 0$, or “red-noise,” corresponds to a monotonic increase in power at low frequencies, and is a common behavior in astrophysical sources.

After defining the fluctuating characteristics of $Q_{mc}^{+}$, four parameters of the CHAR model that control the relationship between the quasar variability amplitude and the rest-frame wavelength ($\lambda_{\mathrm{eff}}$) need to be defined. The parameters are the dimensionless accretion rate ($\dot{m}$), with a fixed radiative efficiency of $\eta = 0.1$, the black hole mass, $M_{BH}$, the dimensionless viscosity parameter ($\alpha$), and the fractional variability amplitude of the magnetic fluctuations ($\delta_{mc}$). Previous studies indicate an average $\eta$ from 0.05 to 0.1 over cosmic time, depending on the AGN luminosity function and black hole mass density assumed (e.g., Chokshi \& Turner~\citeyear{1992MNRAS.259..421C}; Small \& Blandford~\citeyear{1992MNRAS.259..725S}; Yu \& Tremaine~\citeyear{2002MNRAS.335..965Y}; Marconi et al.~\citeyear{2004MNRAS.351..169M}; Shankar et al.~\citeyear{2004MNRAS.354.1020S}; Cao \& Li~\citeyear{2008MNRAS.390..561C}; Cao~\citeyear{2010ApJ...725..388C}; Li et al.~\citeyear{2012ApJ...749..187L}; Ueda et al.~\citeyear{2014ApJ...786..104U}). We fixed $\dot{m} = 0.1$ and $M_{BH} = 10^{8} M_{\odot}$ to match the properties of our sample of quasars. The typical range of the dimensionless viscosity parameter, $\alpha$, is $\sim 0.1-0.4$ (King et al.~\citeyear{2007MNRAS.376.1740K}).

We constructed models for eight different values ranging from $\alpha=0.1$ to 0.8 in steps of 0.1 to cover this range. The value of $\delta_{mc}$ determines the amplitude of the flux variations but not their power spectra or wavelength dependence, so it can be adjusted a posteriori. This was done for each input $Q_{mc}^{+}$ PSD (where $\beta$ ranges from 0.6 to 2). Then, the equations from Sun et al. (\citeyear{2020ApJ...891..178S}), which govern the variations in disk temperature across radii, were solved. Finally, integrating the black body emission across the entire disk, the simulated light curves at various wavelengths were generated. We simulated 128 light curves of rest-frame wavelengths ranging from 1500 Å to 9000 Å, for each value of $\alpha$. This sample size is large enough to provide robust statistical insights, while remaining computationally manageable. The simulated light curves are in their rest-frame and have the same sampling patterns as the ZTF light curves in the observed frame. Finally, the variance on the four timescales of interest (300, 150, 75, and 30 days) was estimated using the Mexican hat filter and was averaged on wavelength bins. Variance ratios were obtained by dividing the variance of longer timescales by the variance of shorter timescales.  

\subsection{Data—CHAR model comparison}\label{subsec:Model Comparison}

\begin{figure*}[!ht]%
\centering
\includegraphics[width=0.49\textwidth]{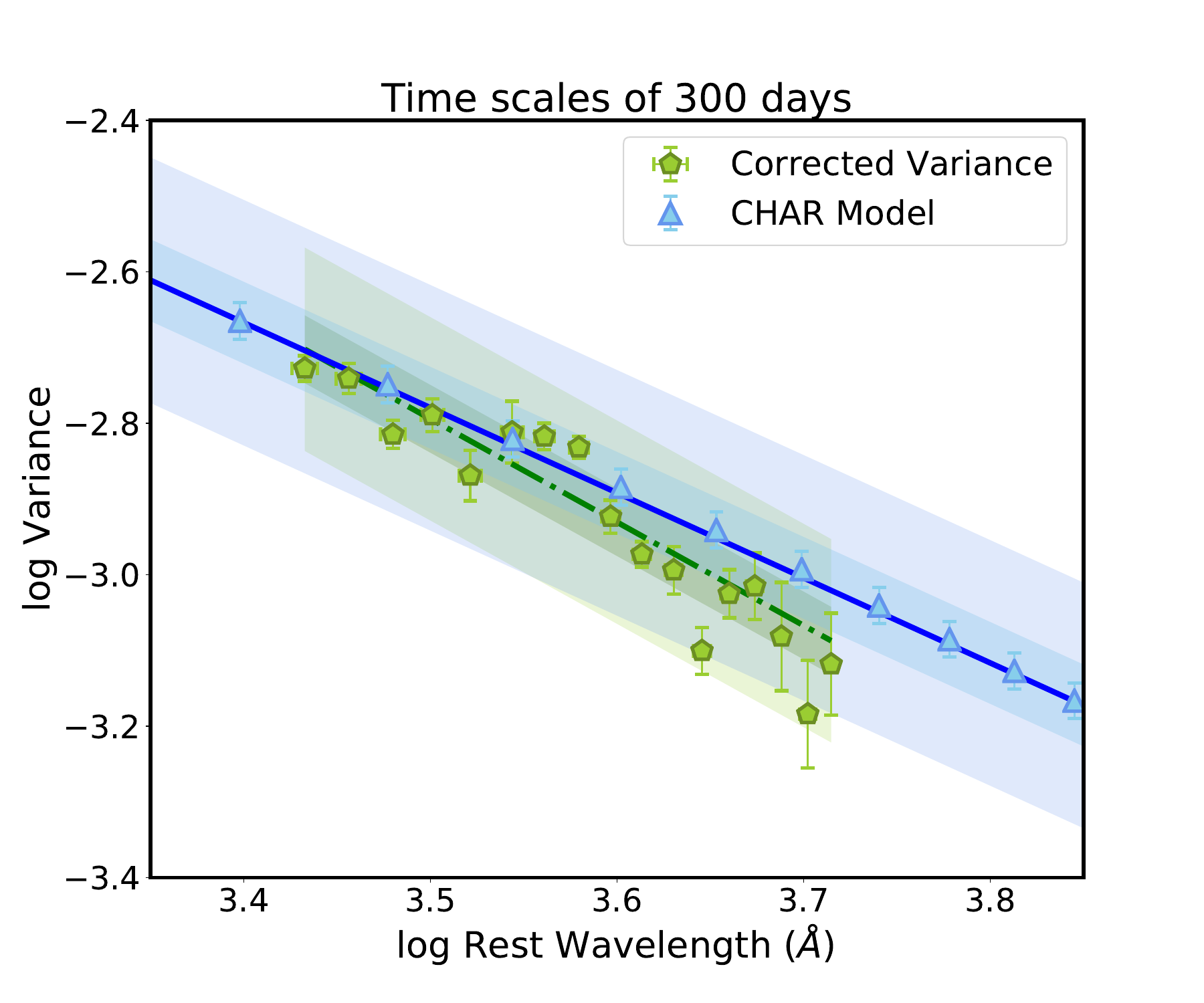}
\includegraphics[width=0.49\textwidth]{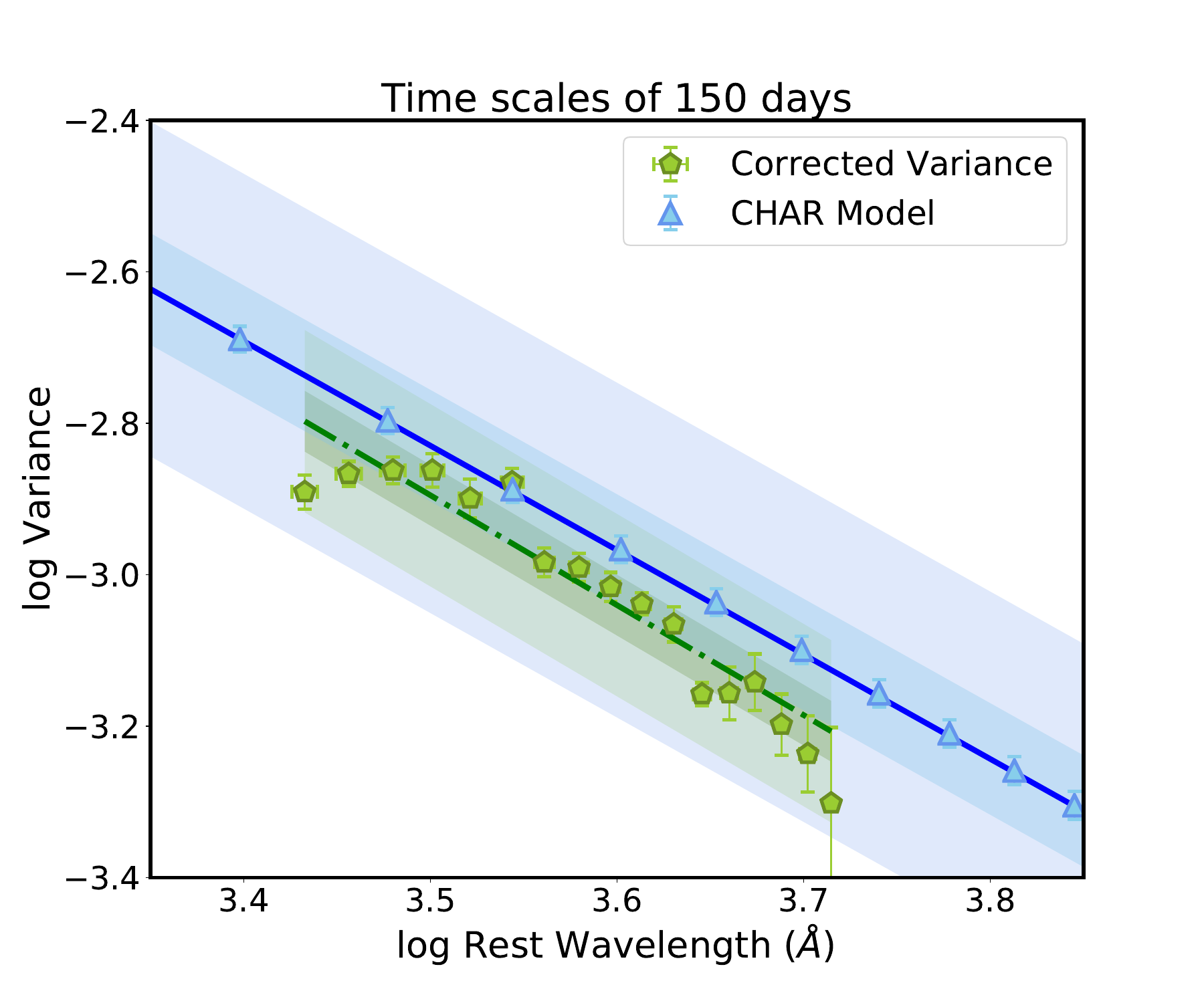}
\includegraphics[width=0.49\textwidth]{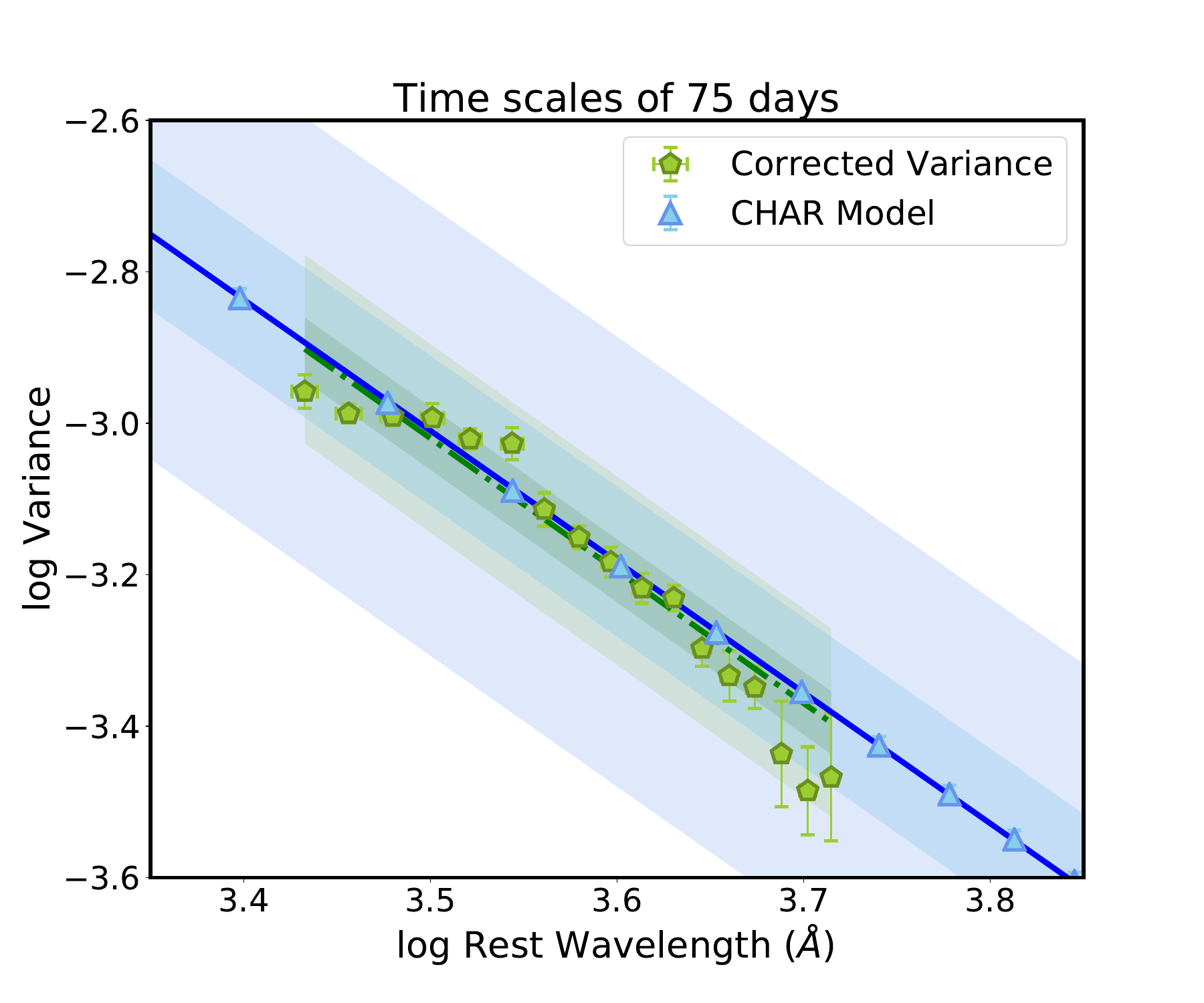}
\includegraphics[width=0.49\textwidth]{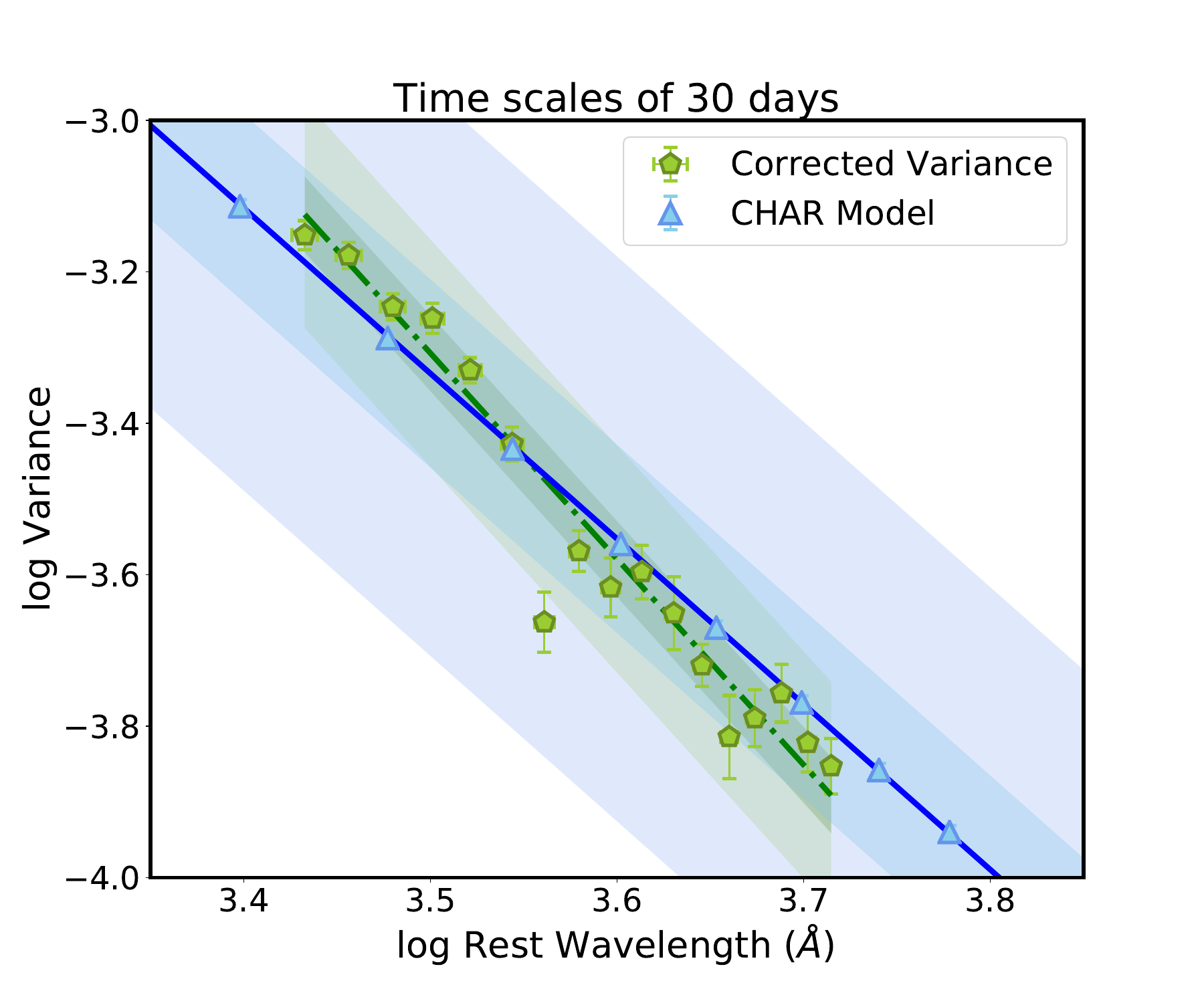}
\caption{Comparison of variance vs.~rest-frame wavelength for ZTF data and CHAR model predictions for four different timescales. The observed median variance is shown with green pentagons, with error bars indicating the root-mean-squared scatter. The linear fit to the observed median variances is shown, while the green shaded areas represent $1 \sigma$ and $3 \sigma$ uncertainties, respectively. The mean CHAR model variance estimated for the corresponding 128 simulated rest-frame wavelength light curves is shown with blue triangles, with error bars representing the standard error of the mean. The linear fit to the mean model variances is shown with the solid blue line. The shaded regions in blue indicate the $1 \sigma$ and $3 \sigma$ uncertainties, respectively.}
\label{fig : CHAR_model_variance}
\end{figure*}

Examination of the variance obtained from the simulated light curves shows that larger values of $\alpha$ produce higher variance and less steep slopes, which can be attributed to the enhanced efficiency of the accretion disk in responding to coronal fluctuations, resulting in a more pronounced variability across a wider temperature range. King et al.~(\citeyear{2004MNRAS.348..111K}) also demonstrate that an increase in $\alpha$ leads to greater normalized root mean square variability in disk luminosity.

We identified that the input $Q_{mc}^{+}$ PSD requires $\propto 1/f^{0.6}$ to provide a good match to the data. In contrast, Sun et al.~\citeyear{2020ApJ...891..178S}, ~\citeyear{2020ApJ...902....7S}, considered $\beta = 1$. A PSD with $\beta=0.6$ shows a less steep decline in power across frequencies compared to $\beta=1$, with more power found in short-term fluctuations. We also found that $\alpha=0.5$ and $\delta_{mc}=1.32$ reproduced most closely the normalization of the observed relations. In Appendix~\ref{sec:appendixC}, plots depicting variance as a function of wavelength for different input $Q_{mc}^{+}$ PSDs, with $\beta$ ranging from 0.6 to 2, and for the four variability timescales studied, are presented. Likewise, figures in Appendix~\ref{sec:appendixD} present variance as a function of wavelength for different values of the viscosity parameter, $\alpha$.

\begin{figure*}[!ht]%
\centering
\includegraphics[width=0.3\textwidth]{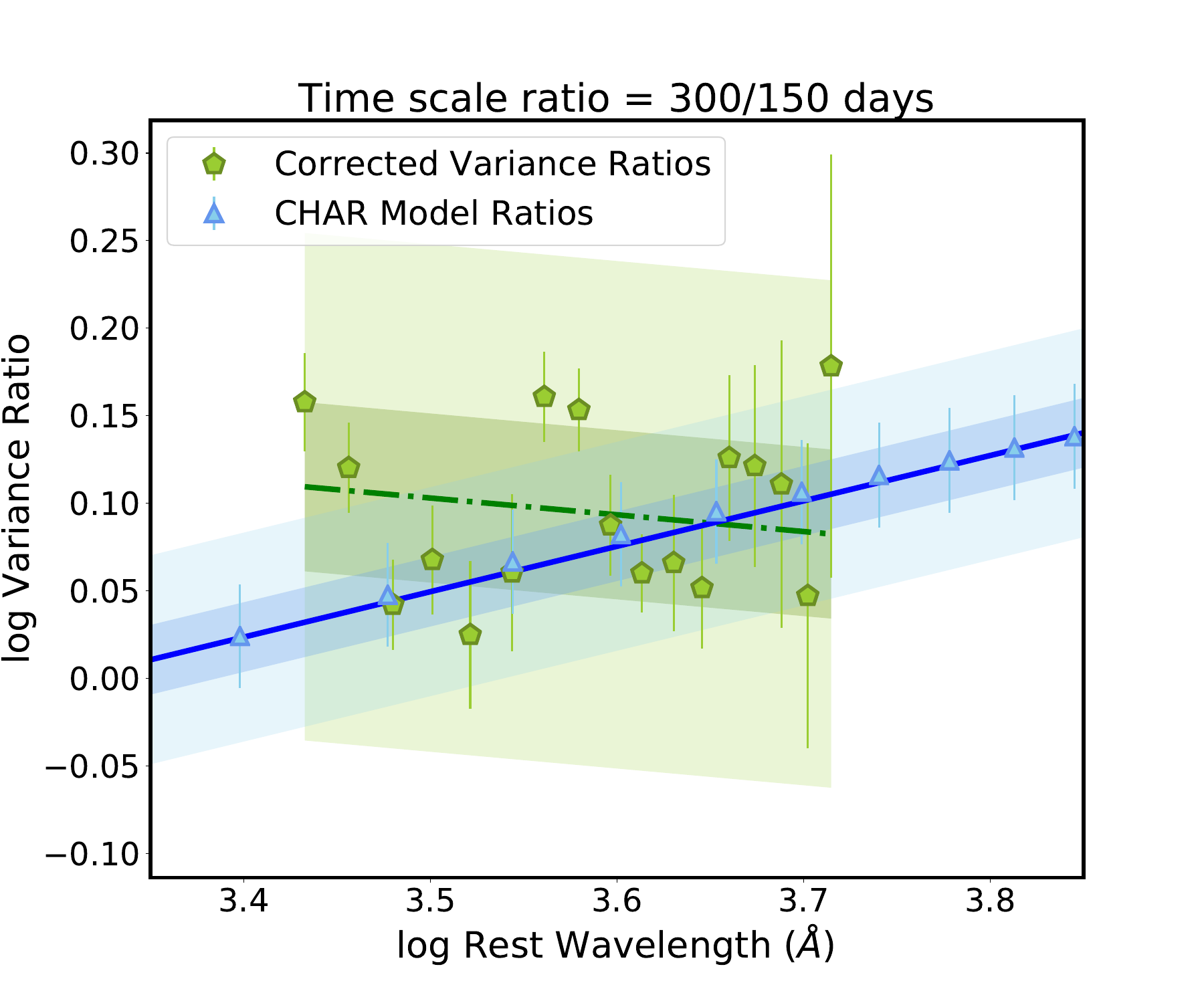}
\includegraphics[width=0.3\textwidth]{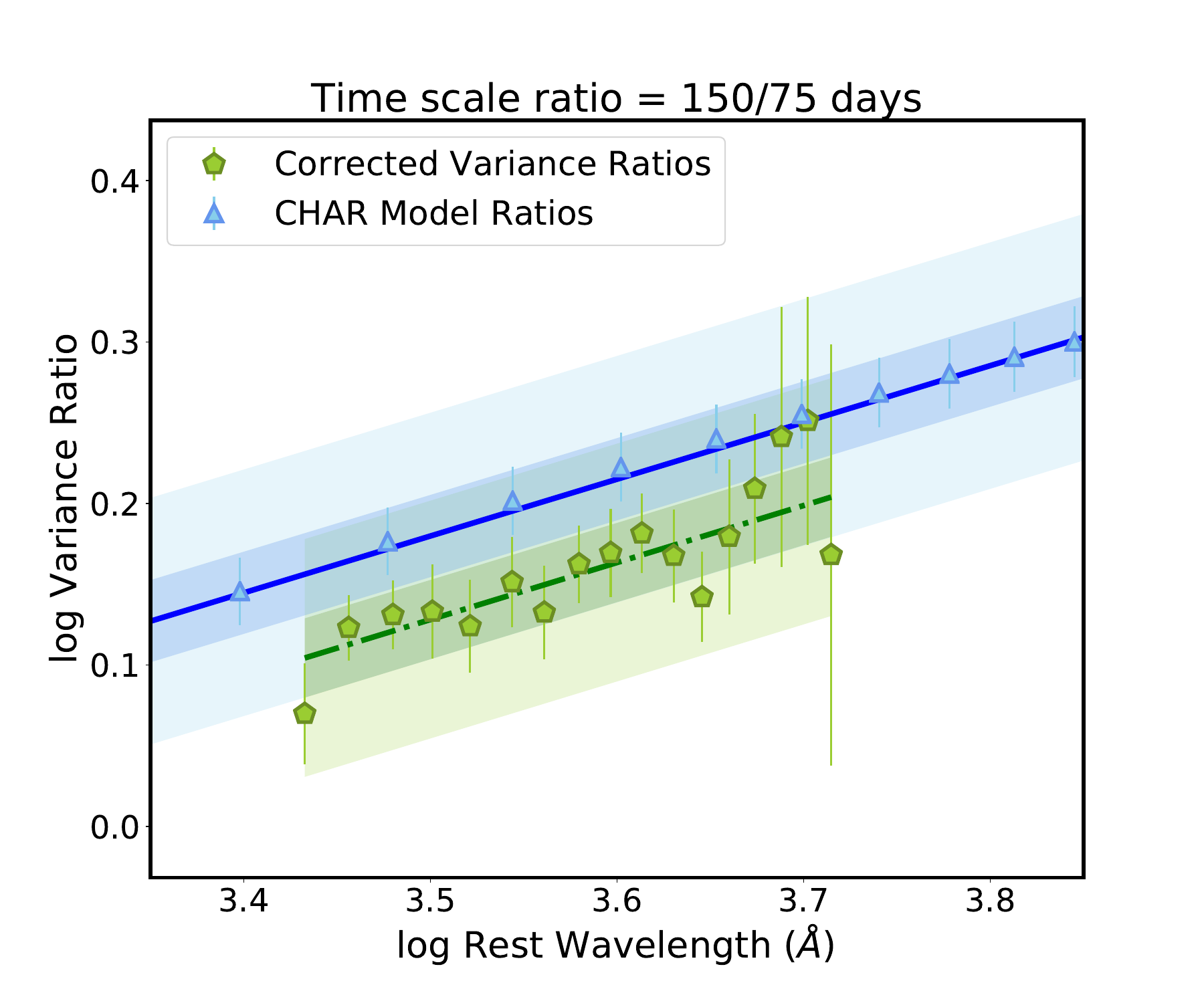}
\includegraphics[width=0.3\textwidth]{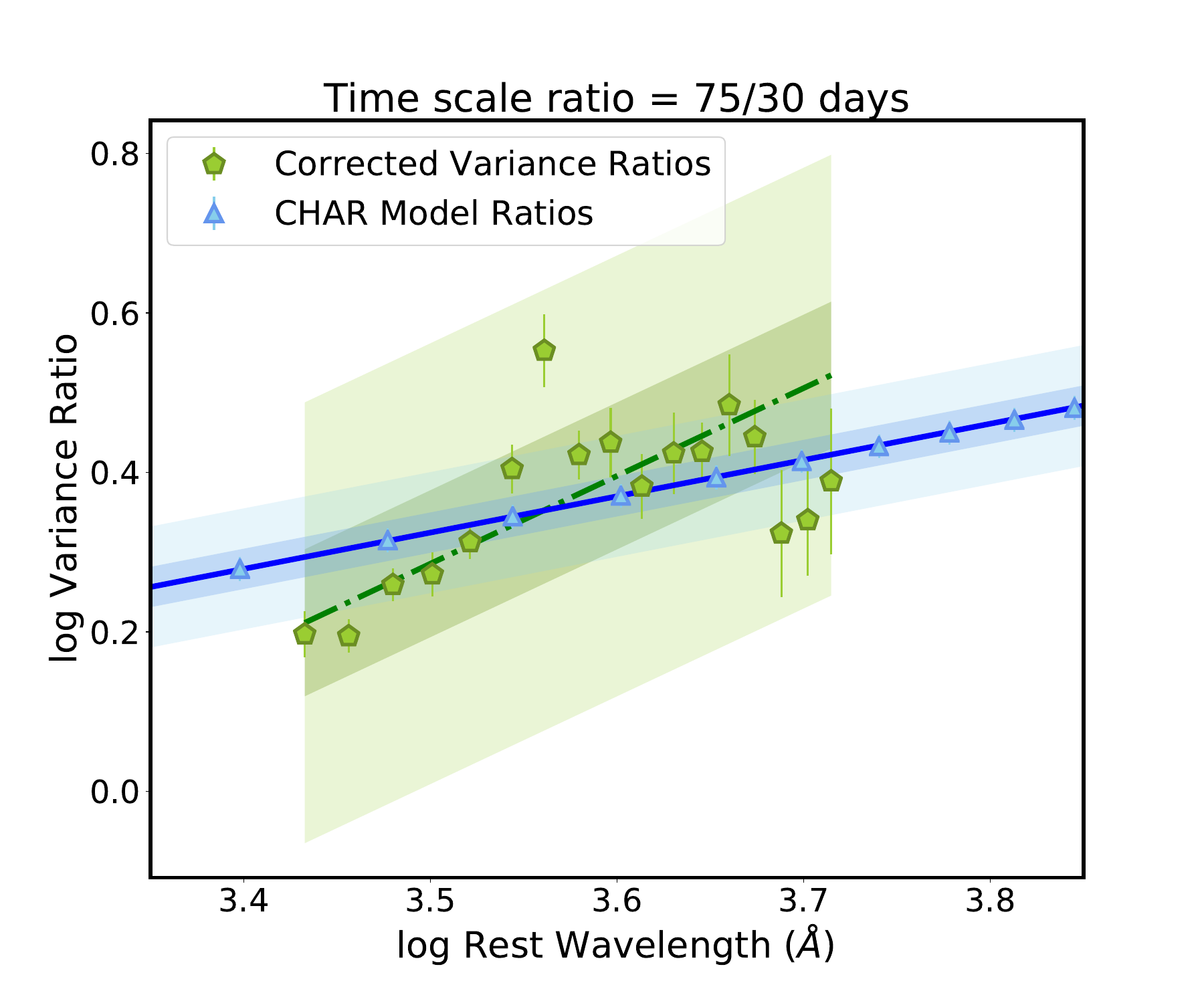}
\caption{Comparison of ZTF data with CHAR model predictions for variance ratios vs.~rest-frame wavelength. Colors and markers have the same meaning as Figure 5.}
\label{fig : CHAR_model_variance_ratio}
\end{figure*}

Figures \ref{fig : CHAR_model_variance} and \ref{fig : CHAR_model_variance_ratio} show the variance and variance ratios, respectively, as a function of wavelength for the data and model together, adopting $\alpha=0.5$ and $\delta_{mc}=1.32$. Notice the different dynamical ranges in the y-axis for all these plots. Table \ref{tab : linear_fits} summarizes the linear regression results for the intercept (a) and the slope (b) for the observed data (corrected and uncorrected) and the CHAR model as a function of the rest-frame wavelength ($\lambda_{RF}$). These results demonstrate that the observed data and the CHAR model variances agree well for most timescales, with consistent slopes within 1 $\sigma$ for the 300, 150, and 75-day timescales. However, at the 30-day timescale, the slopes exhibit a $3 \sigma$ deviation between the observed and CHAR model values, with the observed slope being steeper than that of the model. The intercepts were within $1 \sigma$ of each other for all the timescales studied. 

The investigated timescales exceed the reprocessing timescales estimated for a quasar with a black hole mass of $10^{8}$ solar masses and an Eddington ratio of 0.1. In their study, Kammoun et al.~(\citeyear{2021ApJ...907...20K}) investigate the response of the accretion disk to variable irradiation in order to examine disk thermal reverberation. Their research showed that the time lag between X-rays and UV/optical light curves increases with increasing wavelength, typically remaining under 10 days for rest-frame wavelengths up to 10,000 Å. Reprocessing implies that below this timescale of up to 10 days, fluctuations in the driving signal would get smoothed out by light-travel time effects. Hence, we do not expect significant dampening at the timescales that we are studying (30--300 days). In fact, if variability suppression were present at the 30 day timescales, it would preferentially affect longer wavelengths and give an even steeper observed relation, as this emission comes from more extended regions of the accretion disk, making the mismatch between our observations and the CHAR model even more pronounced.

A positive correlation between the 150/75 and 75/30 variance ratios and the rest-frame wavelength is found for both the CHAR model and the real data (see Fig. \ref{fig : CHAR_model_variance_ratio}). However, for the 300/150 ratio, this correlation was significantly positive for the CHAR model, while consistent with 0 in the observed data. Besides, all variance ratios involving the 30-day timescale show significant discrepancies, as is expected (see above). The CHAR model best reproduces the 150/75 variance ratio.

\begin{table*}[!ht]
\fontsize{8.0pt}{8.0pt}\selectfont
\centering
\caption{Linear regression best-fit values for slope (a) and intercept (b) using $\lambda_{RF}$ as the independent variable.}
\label{tab : linear_fits}
\tablefoot{Columns represent: “Original data” (linear regression fits to the initial variance and variance ratio), “Corrected data” (fits to the new variance and new variance ratio, accounting for estimated variance suppression), and the “CHAR model” (fits to the variance and variance ratio based on the simulated light curves obtained from the CHAR model).}
\end{table*}

\section{Discussion} 
\label{sec:discussion}

\subsection{Comparison with previous work}\label{subsec:Previous work}

This study confirms that the amplitude of the variability is anticorrelated with the rest-frame wavelength, as has already been discussed in previous works (Cutri et al.~\citeyear{1985ApJ...296..423C}; Paltani \& Courvoisier~\citeyear{1994A&A...291...74P}; Vanden Berk et al.~\citeyear{2004AAS...20512002V}; MacLeod et al.~\citeyear{2010ApJ...721.1014M},~\citeyear{2012ApJ...753..106M}; Meusinger et al.~\citeyear{2011A&A...525A..37M}; Zuo et al.~\citeyear{2012ApJ...758..104Z}; Morganson et al.~\citeyear{2014ApJ...784...92M}; Li et al.~\citeyear{2018ApJ...861....6L}; Sánchez-Sáez et al.~\citeyear{2018ApJ...864...87S}). Moreover, this study shows a negative correlation between variance and rest-frame wavelength for the four timescales considered in this work (300, 150, 75, and 30 days). This anticorrelation is stronger for shorter timescale variations (see Table \ref{tab : linear_fits}) and shows that if the rest-frame wavelength represents the radius of the accretion disc, optical fluctuations at all timescales are less pronounced as the radius increases. This pattern is stronger on short timescales. This study also shows that the amplitude of variability at long timescales is larger than at short timescales for rest-frame wavelengths from 2700 Å to 5200 Å. Figure \ref{fig: variance_all_timescale} summarizes the normalized variance as a function of $\lambda_{RF}$ for all timescales considered here.

\subsection{Power spectrum and the variance--$\lambda_{RF}$ correlation}\label{subsec:PSD} 

We can understand the correlation obtained between quasar variability and the rest-frame wavelength by plotting the PSD for two different rest-frame wavelengths. The power spectrum ($P(f)$) quantifies the level of variation observed across various timescales ($t$) or frequencies ($f$), where $f = 1/t$. In previous studies, the quasar optical power spectra have been described using a damped random walk (DRW) model, with its power spectra similar to a random walk at high frequencies $(P(f) \propto 1/f^{2}$) but flattening out for frequencies below a break ($f < f_{b}$). The break frequency ($f_{b}$) corresponds to a damping timescale, $\tau_{damp} = 1/f_b$. Besides, Burke et al.~\citeyear{2021Sci...373..789B} have shown that $\tau_{damp} \propto 107^{+11}_{-12} \times (M_{\rm BH} / 10^8 M_{\odot})^{0.38^{+0.05}_{-0.04}}$ days for $\lambda_{RF} = 2500$ \AA, which is close to the starting wavelength in our data (2700 \AA).

\begin{figure}[!htbp]%
\centering
\includegraphics[width=0.48\textwidth,trim= 30 20 30 10]{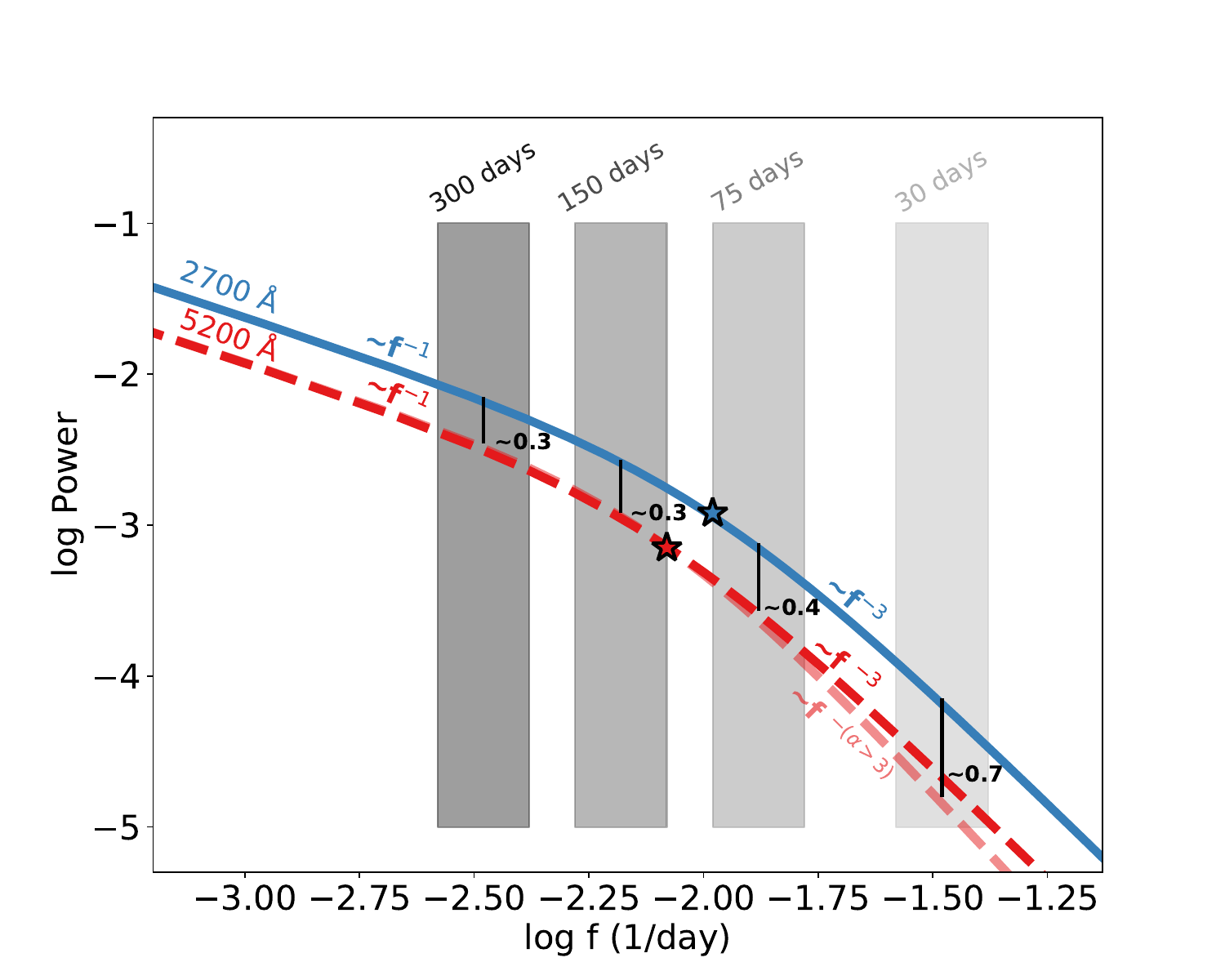}
\caption{Illustration of the PSD for two different rest-frame wavelengths (2700 and 5200 \AA), for a fixed black hole mass and Eddington ratio. It shows that longer rest-frame wavelengths correspond to lower normalization in the power spectrum. The break frequency ($f_b$) varies inversely with $\lambda_{RF}$ and predicts that the bending will occur at higher frequencies for bluer wavelengths. The predicted values of $f_b$ for 2700 and 5200 \AA\ are shown with a blue and red star, respectively. However, the observed variance trends present a steeper decrease at 5200 \AA\ than predicted, as is depicted by the dashed line in the figure, suggesting that the high-frequency slope of the PSD is wavelength-dependent.}
\label{fig : PSD_wavelength}
\end{figure}

The change in the behavior of the PSD below and above $f_b$ can be more generally modeled as a bending power-law model. This model describes the PSD with a high-frequency ($\alpha_{H}$) and low-frequency ($\alpha_{L}$) slope (see Fig. \ref{fig : PSD_wavelength}). Ar\'evalo et al.~\citeyear{2024A&A...684A.133A} recently found that bending power laws with combinations of $\alpha_{H} = -3$ and $\alpha_{L} = -1$, as well as $\alpha_{H} = -2.5$ and $\alpha_{L} = -0.5$, correctly describe the PSD of quasars with $7.5 < \log (M_{BH}/M_{\odot}) < 9.5$ and $-2 < R_{Edd} < 0$ for a fixed $\lambda_{RF} \sim 2900$ \AA. In contrast, the DRW model ($\alpha_{H} = -2$, $\alpha_{L} = 0$) shows a significantly poor fit. Furthermore, their findings predict that, specifically for the model with $\alpha_{H} = -3$ and $\alpha_{L} = -1$, the break frequency for our adopted mass and accretion rate is $96^{+9}_{-20}$ days at this wavelength, which we can adopt as the break frequency for the blue-most wavelength of our ZTF data ($\sim 2700$ \AA). A further bending of the power law at even shorter frequencies is suggested by the recent analysis of Stone et al.~(\citeyear{2022MNRAS.514..164S}), but it needs to be confirmed with data of longer temporal baselines.

\begin{figure}
\centering
\includegraphics[scale=0.32]{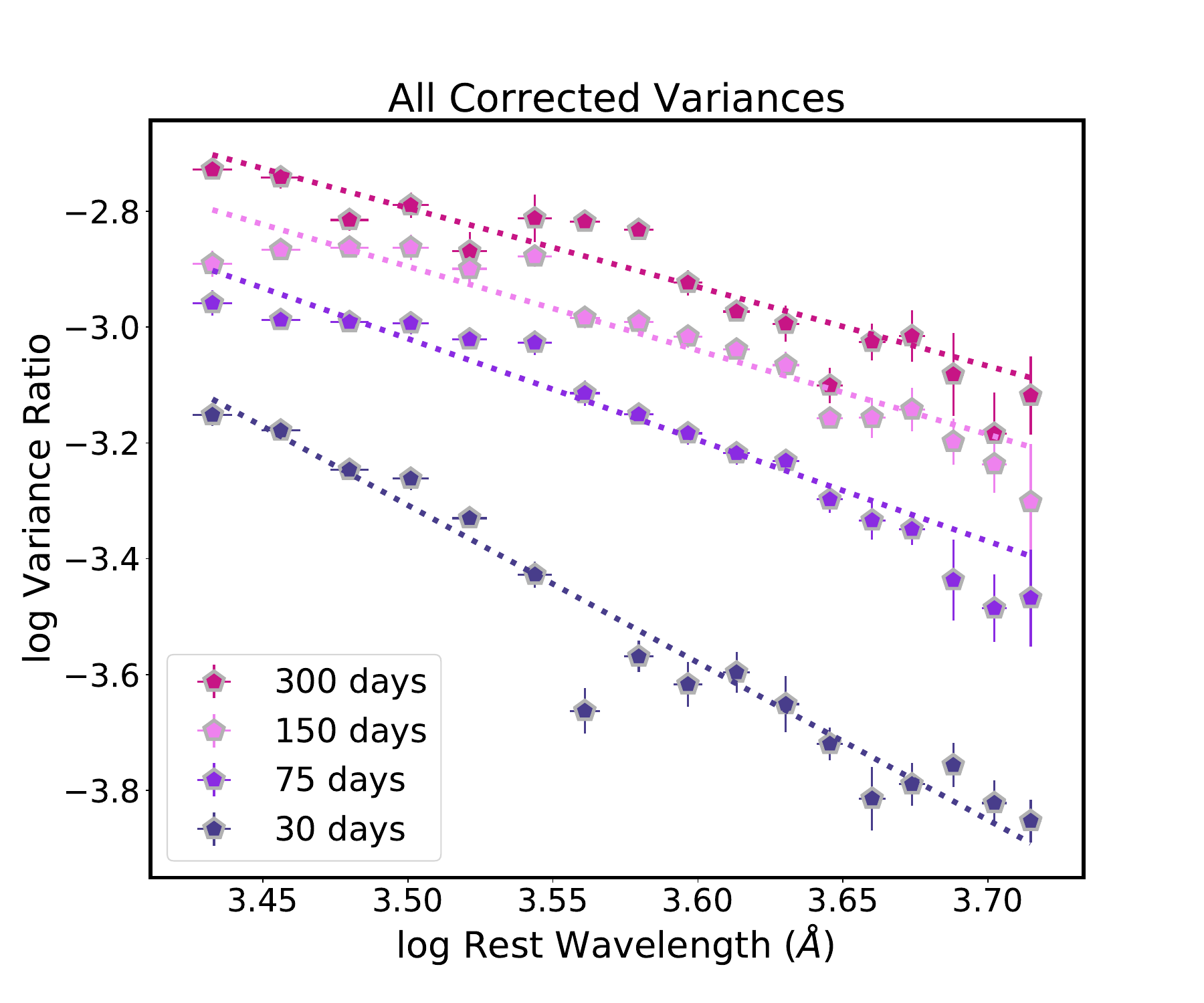} 
\caption{Normalized variance as a function of $\lambda_{RF}$ for different timescales. Observed median variance values are shown with triangles. Straight lines represent the linear regression fits, while different colors represent different timescales.}
\label{fig: variance_all_timescale} 
\end{figure}

\begin{figure}
\centering
\includegraphics[scale=0.32]{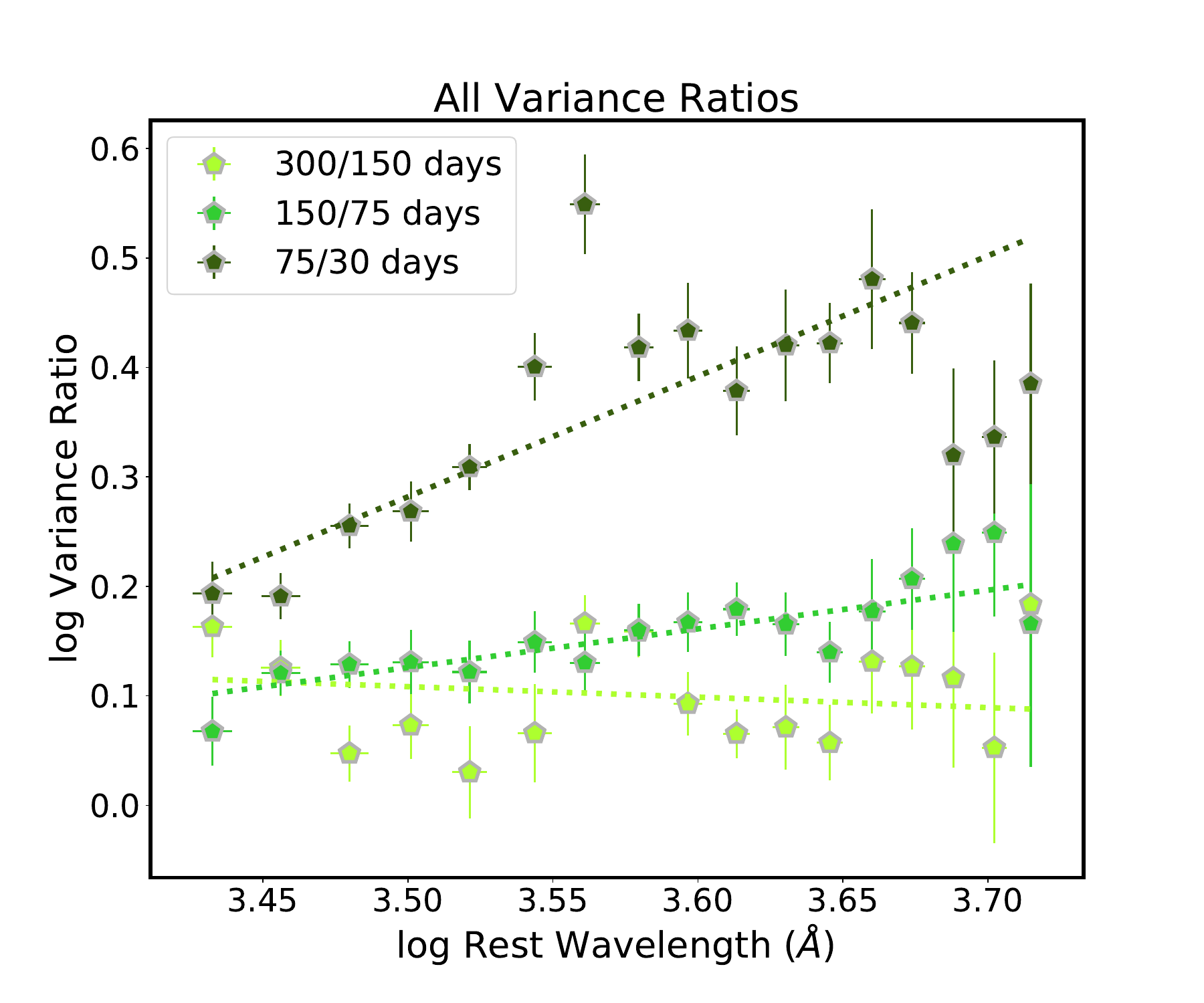} 
\caption{Variance ratio is plotted against $\lambda_{RF}$. Triangles represent the observed median variance ratios with associated errors. Linear regression fits are represented by straight lines, and different colors indicate different variance ratios.}
\label{fig: variance_ratios_all_timescale} 
\end{figure}

Our analysis finds a negative correlation between observed variance and the rest-frame wavelength, which can be represented by the equation log(variance) $= a \times \log(\lambda_{RF}) + b$ (see Fig. \ref{fig:variance plots} and Table \ref{tab : linear_fits}). The simplest interpretation is that as the value of \(\lambda_{RF}\) decreases, the power spectrum normalization increases. However, this scaling alone does not provide a correct description of the different dependence of variance with \(\lambda_{RF}\) across timescales. If this were the scenario, the curves presented in Fig. \ref{fig: variance_all_timescale} would only show a vertical shift, while it is found that the variance at shorter timescales presents steeper slopes compared to longer timescales. In fact, Fig. \ref{fig: variance_ratios_all_timescale} shows a nearly constant value for the dependency of the variance 300/150 ratio as a function of \(\lambda_{RF}\), suggesting a flat dependence on wavelength at these timescales, while the 75/30 ratio shows a strong and significant positive correlation. We conducted linear fits as $\log(\mathrm{variance\ ratio})= a \times \log(\lambda_{RF}) + b$  (see Table \ref{tab : linear_fits}). The linear fits for the binned data are shown alongside the data points in Fig. \ref{fig: variance_ratios_all_timescale}. The fact that the slope of the relation between variance and wavelength becomes steeper for shorter-timescale fluctuations shows that the power spectral shape, not just its normalization, must depend on wavelength.

MacLeod et al.~(\citeyear{2010ApJ...721.1014M}) have reported a dependence of the damping timescale on $\lambda_{RF}$ for quasars, which follows $\tau_{damp} \propto \lambda_{RF}^{B}$, where $B = 0.17 \pm 0.02$. Stone et al.~(\citeyear{2022MNRAS.514..164S,2023MNRAS.521..836S}), on the other hand, found a higher value of the exponent, with $B = 0.30 \pm 0.13$.\footnote{Stone et al.~(\citeyear{2022MNRAS.514..164S,2023MNRAS.521..836S}) report larger slopes for $B$ but these values were later revised to $0.30 \pm 0.13$ (see Zhou et al.~(\citeyear{2024ApJ...966....8Z}).} The slow rolling of the PSD around $f_b$, coupled with its shift as a function of wavelength found by MacLeod et al.~(\citeyear{2010ApJ...721.1014M}) and Stone et al.~(\citeyear{2023MNRAS.521..836S}) could introduce the different dependence of variance on wavelength as a function of timescale (see Fig. \ref{fig : PSD_wavelength}, where the position of $f_b$ is shown with a blue star for $\lambda_{RF} = 2700$ \AA\ and a red star for $\lambda_{RF} = 5200$ \AA). We can test this hypothesis. 

Our results show that the 300-day variance drops in power by $\sim 0.33$ dex from the blue to the red end of the wavelengths available to us ($\sim 2700 - 5200$ \AA), while drops of $\sim 0.33, 0.44$, and 0.68 dex are observed for timescales of 150, 75, and 30 days between the same two wavelengths (see Fig. \ref{fig: variance_all_timescale}). We shall test whether these variance drops agree with a unique bending power-law PSD and a wavelength-dependent $f_b$. The drop of 0.33 dex on a 300-day timescale sets the normalization ratio of the PSDs for $\lambda_{RF} = 2700 $\AA\ and $\lambda_{RF} = 5200 $\AA\ (see Fig. \ref{fig : PSD_wavelength}), as this is the lowest frequency available to us and corresponds to a timescale long enough to suffer little impact from the break, which is supported by the constant slope seen in the 300/150 variance ratio. Using the expression of a bending power law for the PSD (see Eq. 1 in Arévalo et al.~(\citeyear{2024A&A...684A.133A})), with $\alpha_{H} = -3$ and $\alpha_{L} = -1$, the model-predicted variance as a function of frequency is $\propto (f (1 + (f/f_b)^2))^{-1}$, while a variance ratio at a given frequency for red over blue wavelengths is $A^{\rm red}(1 + (f/f^{\rm blue}_b)^2)/A^{\rm blue}(1 + (f/f^{\rm red}_b)^2)$, where $A^{\rm red}$ and $A^{\rm blue}$ correspond to the normalizations of the PSDs. We have already shown that $f_b = 1/96$ days$^{-1}$ at the blue end ($\lambda_{RF} = 2700$ \AA), and this is predicted to be $\sim 1/(107\pm4)$ and $1/(117\pm18$) days$^{-1}$ at the red end ($\lambda_{RF} = 5200$ \AA), following MacLeod et al.~(\citeyear{2010ApJ...721.1014M}) and Stone et al.~(\citeyear{2023MNRAS.521..836S}). Inserting the normalization difference (0.33 dex) for the PSDs corresponding to the blue-most and red-most wavelength ranges on 300-day timescales, we obtained a model-predicted variance difference at $\lambda_{RF} = 5200$ \AA\ of $\sim 0.50$ dex for the 75-day timescale and $\sim 0.56$ dex for the 30-day timescale, adopting $f_b = 1/135$ days$^{-1}$, maximizing the variance changes with the highest allowed value of $117 + 18$ days. Therefore, the dependence of $f_b$ on wavelength is able to explain the variance change seen on 75-day timescales, but it is not sufficient to yield the necessary drop in variance observed on 30-day timescales. Simulations show that scenarios with a wavelength-dependent normalization and bend frequency fail to capture the stronger steepening on the shortest timescales (see Appendix~\ref{sec:appendixE}).

In order to explain the observed trends in variance and variance ratios (as a proxy for the PSD slopes), the dependence of $f_b$ on wavelength should be considered, but we also suggest that a change of slope in the PSD for $f > f_b$ might also be necessary to explain the behavior observed at the highest frequencies. A more accurate determination of the dependency of $f_b$ on wavelength is therefore necessary, particularly since the works of MacLeod et al.~(\citeyear{2010ApJ...721.1014M}) and Stone et al.~(\citeyear{2023MNRAS.521..836S}) assumed a DRW model for their studies. Another significant finding from our results is the evidence that below the break frequency disk variability at different wavelengths only requires a different normalization factor (at least for timescales up to 300 days), while a chromatic response that changes the shape of the PSD as a function of wavelength is seen above $f_b$.

\subsection{Insights from CHAR model—data comparison}\label{subsec:comparison}

In Sect.~\ref{subsec:Model Comparison}, we compared variance and variance ratios as a function of wavelength between observations and CHAR model predictions. As a result, it was found that a viscosity parameter of $\alpha = 0.5$ provides the best match. The CHAR model predicts an inverse relationship between variance and wavelength, in agreement with the observations.

In standard accretion disk theory, the viscosity parameter, $\alpha$, sets the value for the thermal timescale, $\tau_{\rm TH} = \tau_{\rm DYN} / \alpha$, where $\tau_{\rm DYN}$ is the dynamical or orbital timescale, which increases with radius to the power of 3/2. $\tau_{\rm TH}$ corresponds to the timescale beyond which a region can respond in thermal equilibrium to any internal or external perturbation. A region responding in thermal equilibrium will act coherently and the timescale and amplitude of its response will be inversely proportional and proportional to its size, respectively. A region outside equilibrium will act as a collection of smaller regions, and the sum of their emissions will present more rapid fluctuations but with smaller amplitudes, as they will add incoherently. This dampening of short timescale variations (i.e., shorter than the thermal timescales) produces a drop in the power spectrum toward high frequencies, modifying the power spectrum of the original driving fluctuations. If the disk surface temperature decreases outward so that longer wavelengths are emitted by larger regions where the average thermal timescale is longer, it is expected that longer wavelengths will lose more high-frequency power. 

This behavior is indeed reproduced by the CHAR model, which predicts an inverse relation between variance and wavelength that gets steeper for shorter wavelengths. As can be seen in Fig. \ref{fig : CHAR_model_alpha}, for any given value of $\alpha$, the model shows steeper slopes for shorter timescale fluctuations, similarly to the behavior of the data. In fact, it is clear from these figures that the model reproduces very closely the relation of variance with wavelengths for the different timescales, in particular for the three longest timescales of variability, only showing discrepancies on the shortest timescale. This agreement is remarkable considering the simplifications of the model and its few free parameters. It is possible that further refinements, such as considering a nonconstant value of $\alpha$ in further modeling, will describe the data even better. 

\section{Conclusions}
\label{sec:conclusion}

In this study, we investigated the dependence of the quasar ensemble variability on the rest-frame wavelength, specifically for a black hole mass of $10^{8} M_{\odot}$ and an Eddington ratio of $10^{-1}$. By selecting this constrained bin in the black hole mass and Eddington ratio, we could distinctly separate the dependence of variance on wavelength from those on mass and the accretion rate, allowing for a more detailed comparison with theoretical models. Our study examines the variability of quasars across four distinct timescales (300, 150, 75, and 30 days) and a rest-frame wavelength of $\sim 2700 - 5200$ \AA. The main findings of this research are as follows:

\begin{enumerate}
\item  The quasar variance anticorrelates with the rest-frame wavelength for a fixed black hole mass of $10^{8} M_{\odot}$ and an Eddington ratio of $10^{-1}$. These results support previous findings (e.g., Li et al.~\citeyear{2018ApJ...861....6L}; Sánchez-Sáez et al.~\citeyear{2018ApJ...864...87S}). 

\item We provide new linear equations between variance and rest-frame wavelength for four different timescales in the range of 30 to 300 days in the quasar rest-frame while controlling for black hole mass and Eddington ratio. The significant decreasing trend for the variance with increasing rest-frame wavelength becomes steeper for shorter time scale fluctuations.

\item We found a positive correlation between variance ratios and rest-frame wavelength, except for the 300/150-day timescale, which shows a flat dependence. This result confirms previous findings by Li et al. (\citeyear{2018ApJ...861....6L}).

\item Our results show that shifting the break frequency with wavelength alone is insufficient and suggest that the slope of the power spectrum above the break frequency increases in steepness at longer wavelengths.

\item In our comparison of observed variability with the CHAR model, as is detailed in Sun et al.~(\citeyear{2020ApJ...891..178S}), we observe that the CHAR model, for adopted parameters of $\alpha = 0.5$ and $\delta_{mc} = 1.32$, aligns well with the observed data for the 300, 150, and 75-day timescales, both in terms of slope and intercept. This consistency is evident in the similar slopes of the CHAR model and measured variance ratios at 300/150 and 150/75 days. However, on the 30-day timescale, we note a significant deviation in the slope between the CHAR model and the observed data, particularly affecting the 75/30 day variance ratio. This discrepancy indicates a need for a further refinement in the CHAR model's ability to accurately represent shorter-term quasar variability.

\end{enumerate}

This paper, along with the findings by Sun et al.~\citeyear{2020ApJ...891..178S},
~\citeyear{2020ApJ...902....7S}, demonstrates the ability of the CHAR model to quantitatively describe quasar variability. In this work, we used ZTF DR15 light curves, which have a cadence of approximately four days and extend up to 4 years and 8 months, including yearly gaps. Consequently, obtaining variance on timescales of less than 30 days or more than 300 days was not feasible. This baseline, however, can be further extended by the Legacy Survey of Space and Time (LSST, Ivezić et al.~\citeyear{2019ApJ...873..111I}). LSST will provide high-cadence and up to 10-year light curves from blue to near-infrared wavelengths, covering wavelengths of 3200–10500 Å. Extending this project to include the deep drilling fields, which will have a higher cadence, has the potential to enhance our understanding of AGN short-timescale variability. In particular, such a dataset would enable variance estimation across various timescales and more wavebands for each individual object, providing better constraints to fully understand the disk temperature profile and the mechanism behind accretion disk variability.

\begin{acknowledgements}
We thank the anonymous referee for their constructive feedback, which helped improve the presentation of our results. The authors acknowledge support from the National Agency for Research and Development (ANID) grants: Programa de Becas/Doctorado Nacional 21222298 (PP), 21212344 (SB), Millenium Nucleus NCN19\_058 (PA, PL, MLMA) (TITANs); FONDECYT Regular 1201748 (PL), Millennium Science Initiative AIM23-0001 (MLMA),and from the Max-Planck Society through a Partner Group grant (PA).

Based on observations obtained with the Samuel Oschin Telescope 48-inch and the 60-inch Telescope at the Palomar Observatory as part of the \textit{Zwicky} Transient Facility project. ZTF is supported by the National Science Foundation under Grant No. AST-2034437 and a collaboration including Caltech, IPAC, the Weizmann Institute for Science, the Oskar Klein Center at Stockholm University, the University of Maryland, Deutsches Elektronen-Synchrotron and Humboldt University, the TANGO Consortium of Taiwan, the University of Wisconsin at Milwaukee, Trinity College Dublin, Lawrence Livermore National Laboratories, and IN2P3, France. Operations are conducted by COO, IPAC, and UW.

This research has made use of the SVO Filter Profile Service "Carlos Rodrigo" (\url{http://svo2.cab.inta-csic.es/theory/fps/}), funded by MCIN/AEI/10.13039/501100011033/ through grant PID2020-112949GB-I00.
\end{acknowledgements}

{\bf Data Availability Statement:} The ZTF photometric data (Masci et al.~\citeyear{2019PASP..131a8003M}) are accessible to the public at \url{https://www.ztf.caltech.edu/ztf-public-releases.html}.
SDSS DR14 spectra are publicly available in the public SDSS archives at \url{https://dr14.sdss.org/optical/spectrum/search}. 

{\hspace*{0.3cm}\bf Code Availability Statement:} The Mexican Hat filter code (Arévalo et al.~\citeyear{2012MNRAS.426.1793A}) can be requested to the corresponding author.

%
\bibliographystyle{aa} 
\bibliography{references.bib} 
%

\appendix

\section{Filtered light curves and variability}\label{sec:appendixA}

\begin{figure}[!htbp]
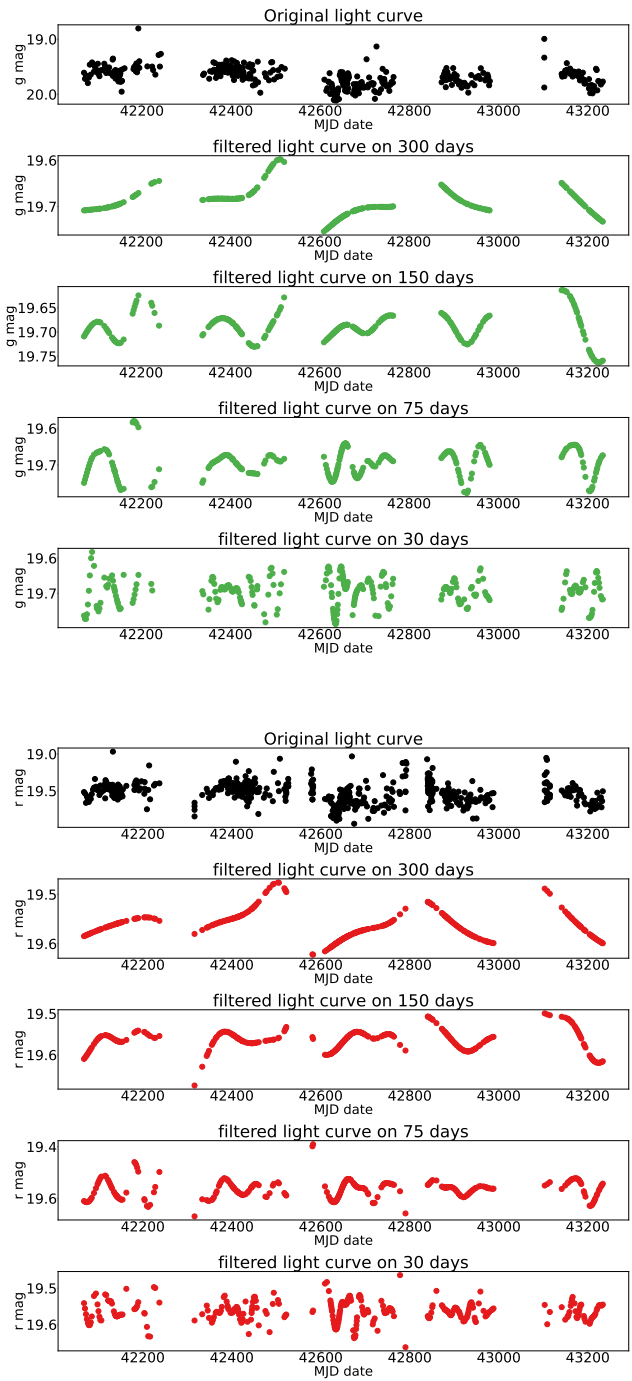

\centering
\includegraphics[width=0.49\textwidth,trim= 320 300 30 270]{lc_gband_filtered.pdf}
\includegraphics[width=0.49\textwidth,trim= 320 300 30 130]{lc_rband_filtered.pdf}
\caption{The top panel displays the original $g$-band light curve, while the filtered light curves at the four studied timescales are shown beneath it. Similarly, the bottom panel shows the original $r$-band light curve, with filtered light curves at the same timescales arranged from top to bottom.}
\label{fig: Filtered lcs}%
\end{figure}

Figure \ref{fig: Filtered lcs} presents both the original and filtered light curves for four timescales as obtained using the Mexican Hat filter (Ar\'evalo et al.~\citeyear{2012MNRAS.426.1793A}). Normalized variance values were obtained after filtering the light curves for a specific timescale.

\section{Emission line variability} \label{sec:appendixB}

The sensitivity of emission lines to continuum fluctuations, quantified by line responsivity ($\eta$), varies across different line types and plays a significant role in understanding quasar variability (Goad et al.~\citeyear{Goad_1999}). Line responsivity, denoted by $\eta$, is a linear coefficient that relates line emissivity to the ionization parameter (Goad et al.~\citeyear{1993MNRAS.263..149G}). High-ionization emission lines (CIV, HeII, SiIV) typically have $\eta \sim 1$, Balmer lines ($H\alpha, H\beta, H\gamma$) have $\eta \sim 0.6$, and Mg II emission lines have $\eta \sim 0.2$ (Goad et al.~\citeyear{Goad_1999}; Korista \& Goad \citeyear{2004ApJ...606..749K}). Mg II and Fe II lines exhibit weak variability due to low responsivity (Kokubo et al.~\citeyear{2014ApJ...783...46K}). The Balmer pseudo-continuum in the Small Blue Bump region ($2200-3646$ Å) is a key component generated by the recombination of hydrogen atoms (Yip et al.~\citeyear{2004AJ....128.2603Y}). The Small Blue Bump region includes the Fe II and Balmer pseudo-continua and the Mg II emission line (Wills et al.~\citeyear{1985ApJ...288...94W}). As described, all these features, particularly the Mg II emission line and Fe II emission, show low responsivity.

\begin{figure}[!htbp]
\centering
\includegraphics[width=0.5\textwidth]{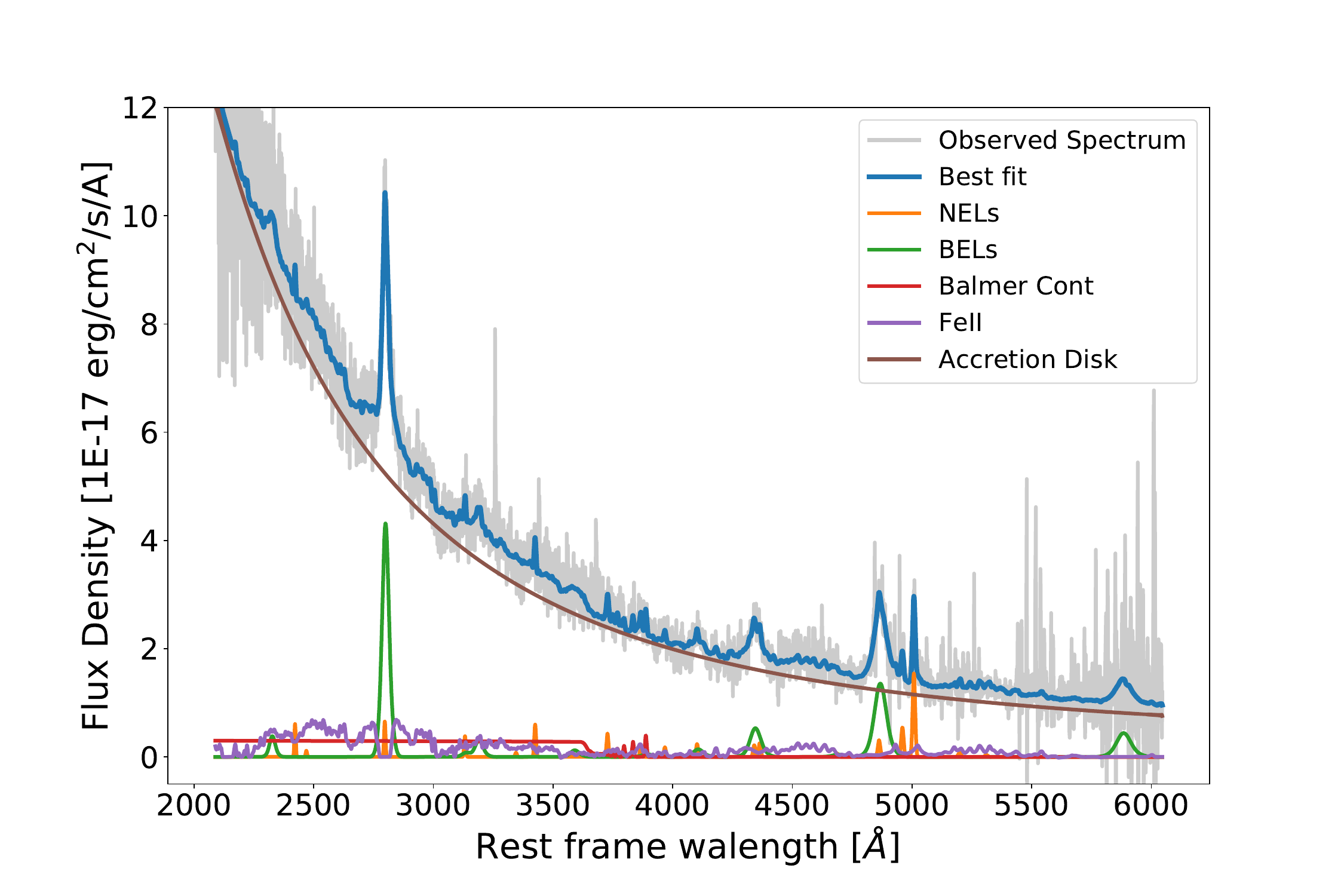}
\includegraphics[width=0.5\textwidth]{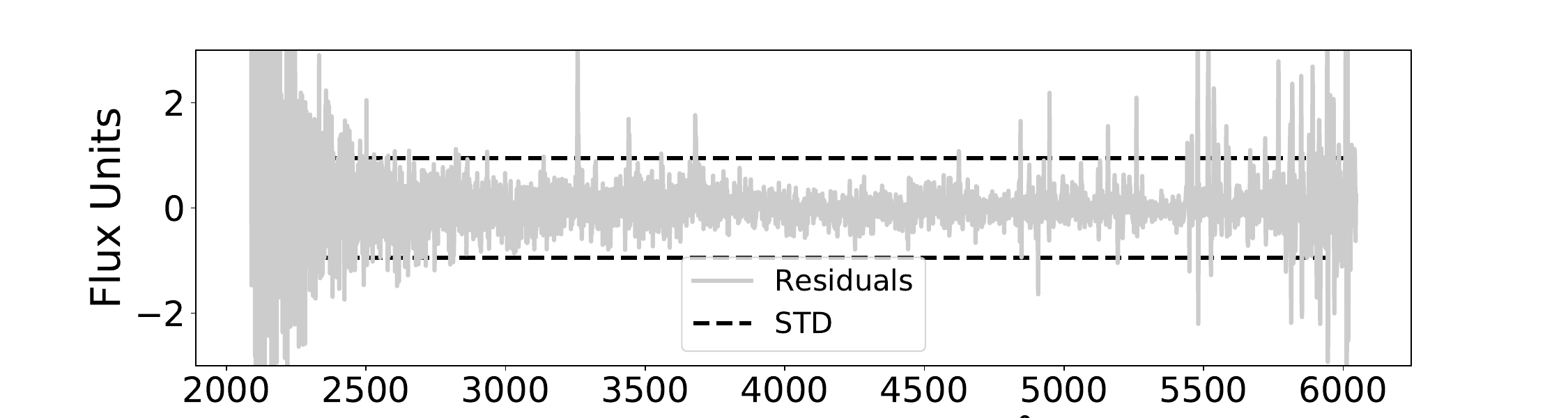}
\caption{Spectral decomposition of a quasar spectrum found at $z = 0.71$ with undetected host features. Upper subplot: Observed Spectrum (gray), Best Fit (blue), Narrow Emission Lines (orange), Broad Emission Lines (green), Balmer pseudo-Continuum (red), FeII pseudo-continuum (purple), and Accretion Disk power-law emission (brown). Bottom plot: residuals (gray).}
\label{fig : spectra_3767-55214-0738}
\end{figure}

Figure \ref{fig : spectra_3767-55214-0738} displays the spectral decomposition of the SDSS DR14 spectrum spec-3767-55214-0738, one of the quasars in the sample, characterized by signal-to-noise values of 10 at 5100 Å (continuum level) and 14 at 3000 Å. This modeling is used to quantify the flux associated to the accretion disk continuum, Fe II and Balmer pseudo-continua, and emission lines. 

\section{Predicted CHAR Model Variance versus wavelength for different input PSDs}\label{sec:appendixC}

\begin{figure*}[!htbp]
\centering
\includegraphics[width=0.49\textwidth]{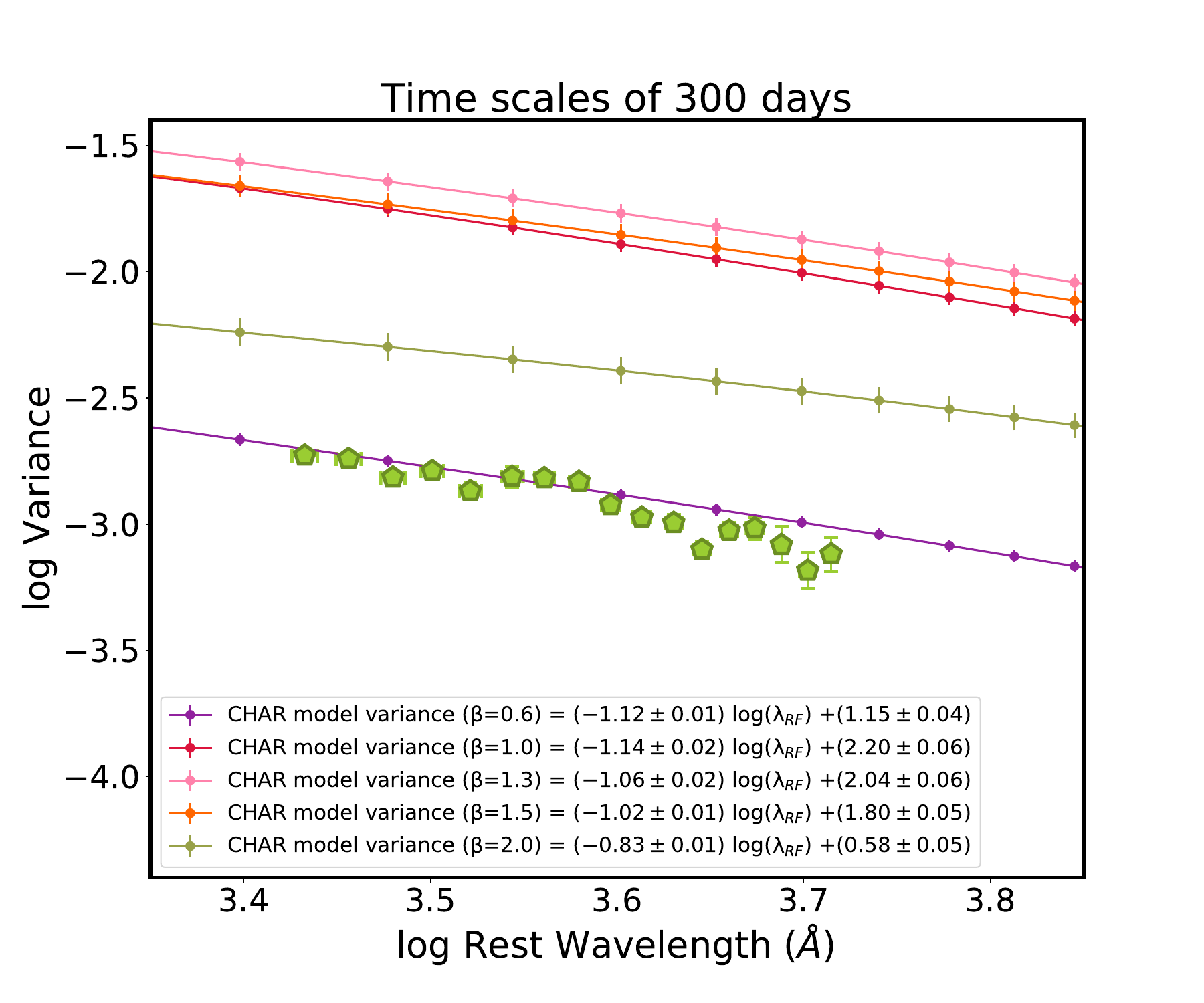}
\includegraphics[width=0.49\textwidth]{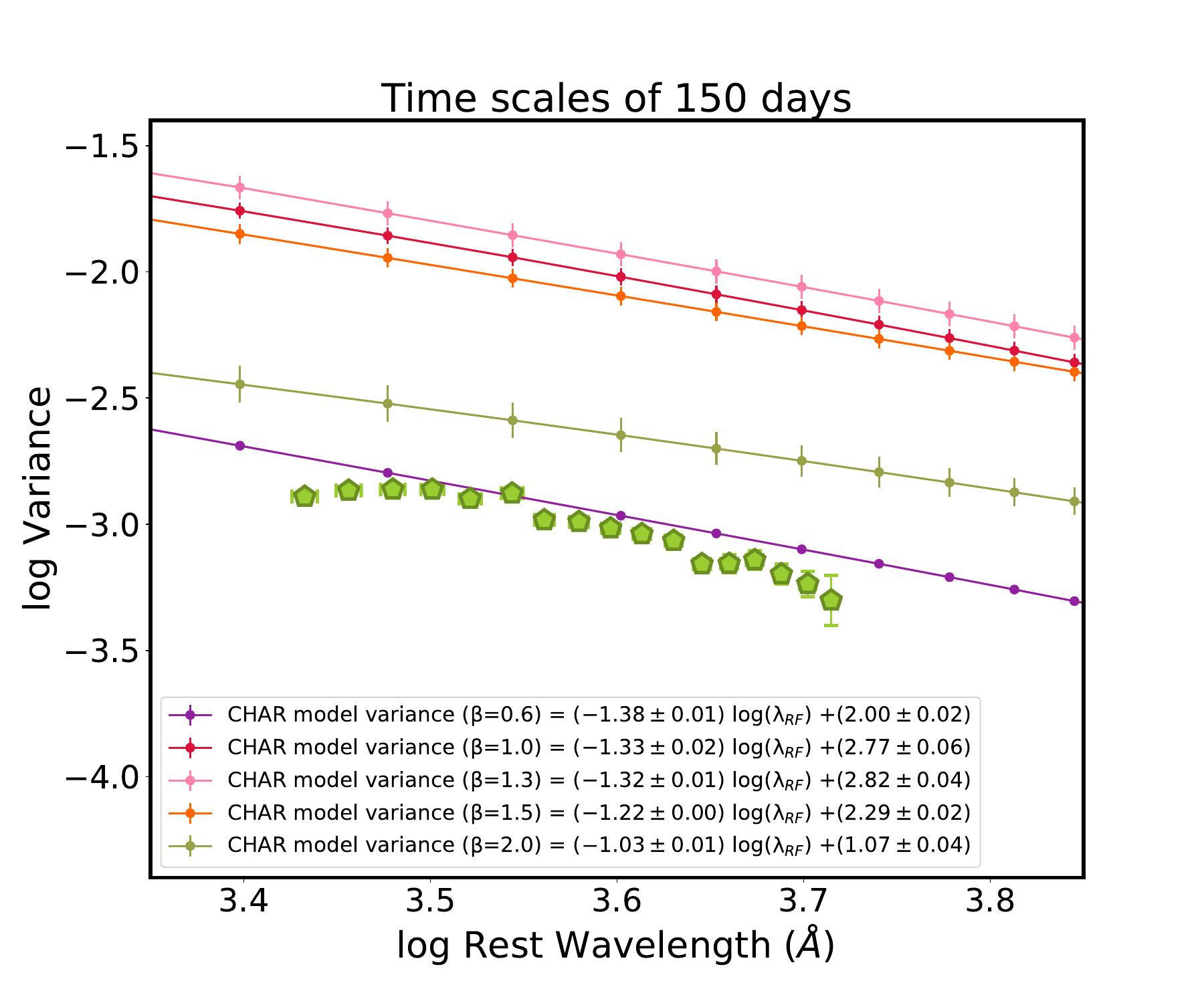}
\includegraphics[width=0.49\textwidth]{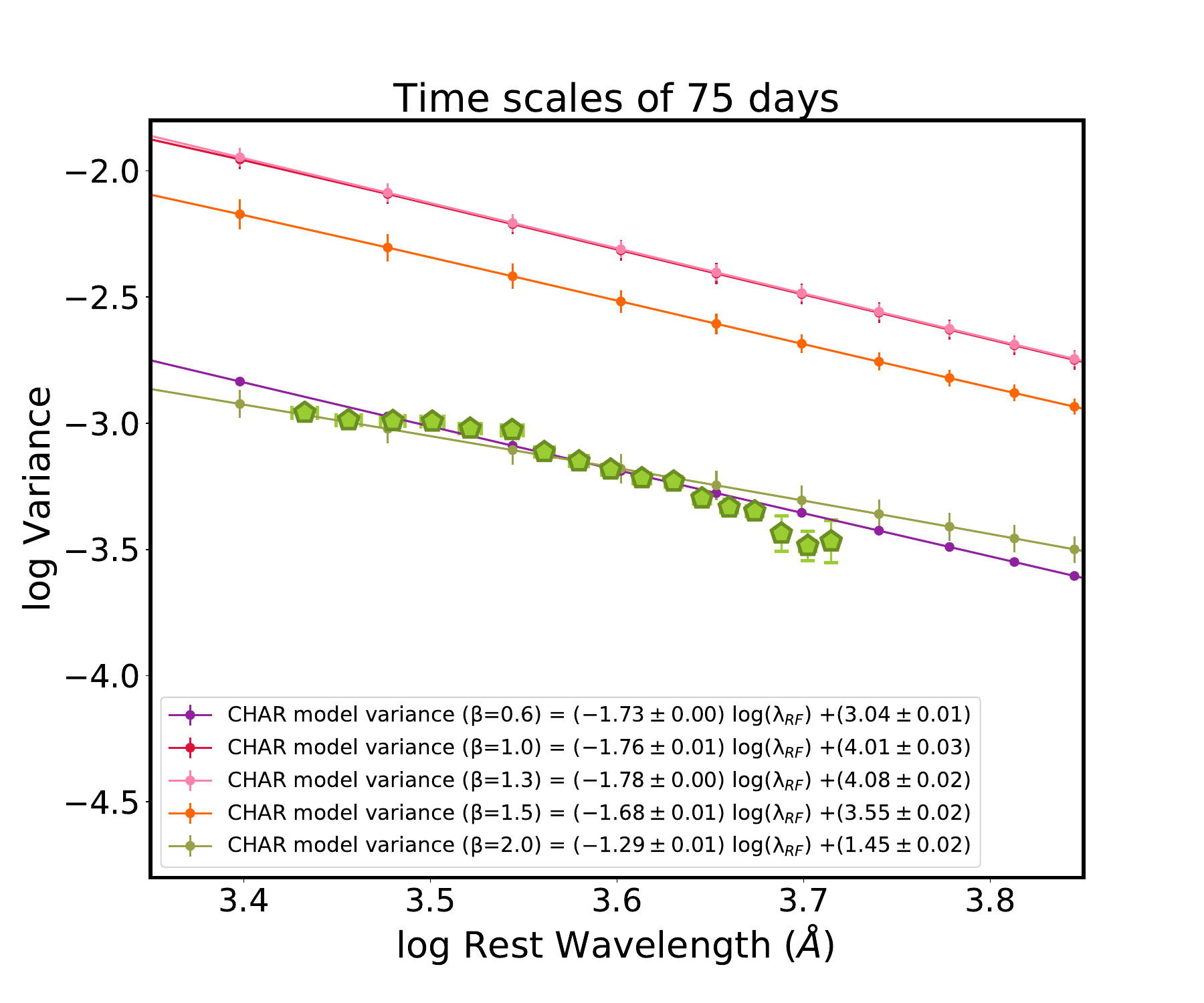}
\includegraphics[width=0.49\textwidth]{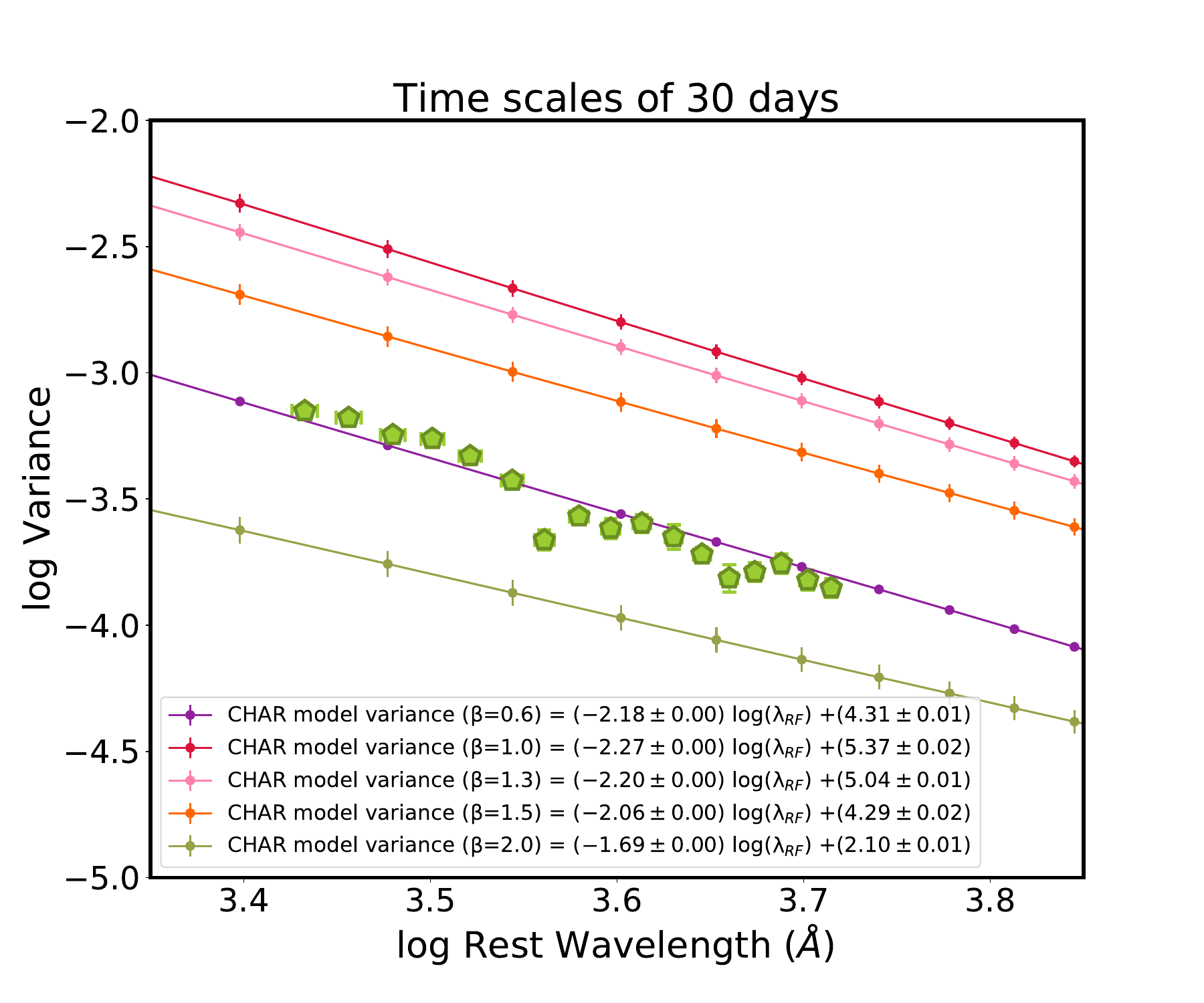}
\caption{The variance-wavelength relationship for the CHAR model for four timescales exploring a PSD $\propto 1/f^{\beta}$ with $\beta$ ranging from 0.6 to 2. The plots display multiple linear regressions, as indicated in the legends, showing the influence of different PSDs on the model behavior. Green pentagons represent observed median ZTF variance values, with error bars indicating root-mean-squared scatter.}
\label{fig : CHAR_model_PSD}
\end{figure*}

Figure \ref{fig : CHAR_model_PSD} presents variance vs. wavelength results for the CHAR model across four distinct timescales together with the observed variance values. Model parameters are: $\dot{m}$ = 0.1, $M_{BH}$ = $10^{8} M_{\odot}$, $\alpha$ = 0.5 and $\delta_{mc}$ = 1.32. A $Q_{mc}$ PSD $\propto 1/f^{\beta}$ is adopted, where $\beta$ ranges from 0.6 to 2.0. A PSD with $\beta = 0.6$ better represents the observed variance versus wavelength across the four studied timescales.

\section{Impact of dimensionless viscosity parameter on variance-wavelength patterns in the CHAR model} \label{sec:appendixD}

\begin{figure*}[!htbp]
\centering
\includegraphics[width=0.49\textwidth]{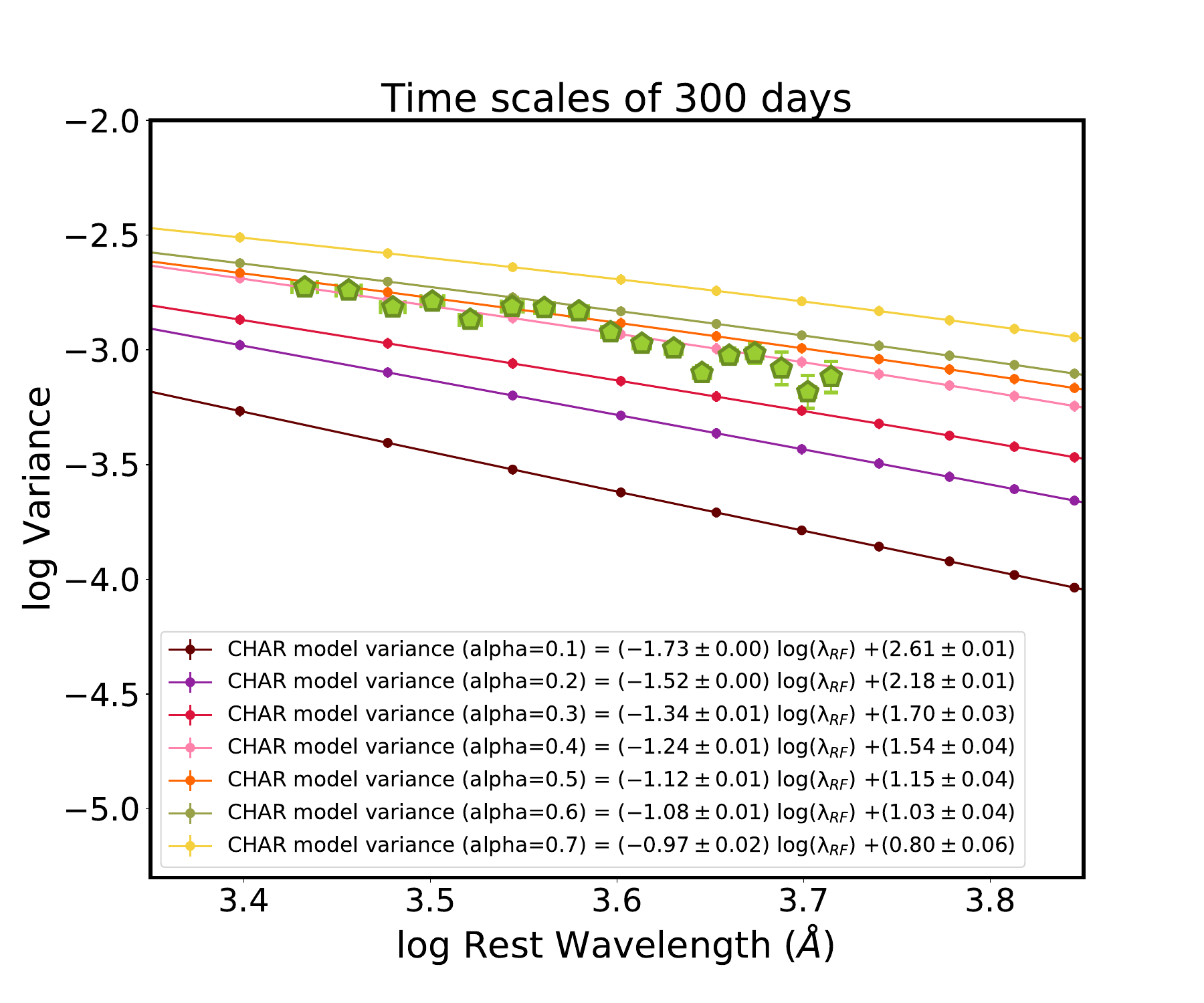}
\includegraphics[width=0.49\textwidth]{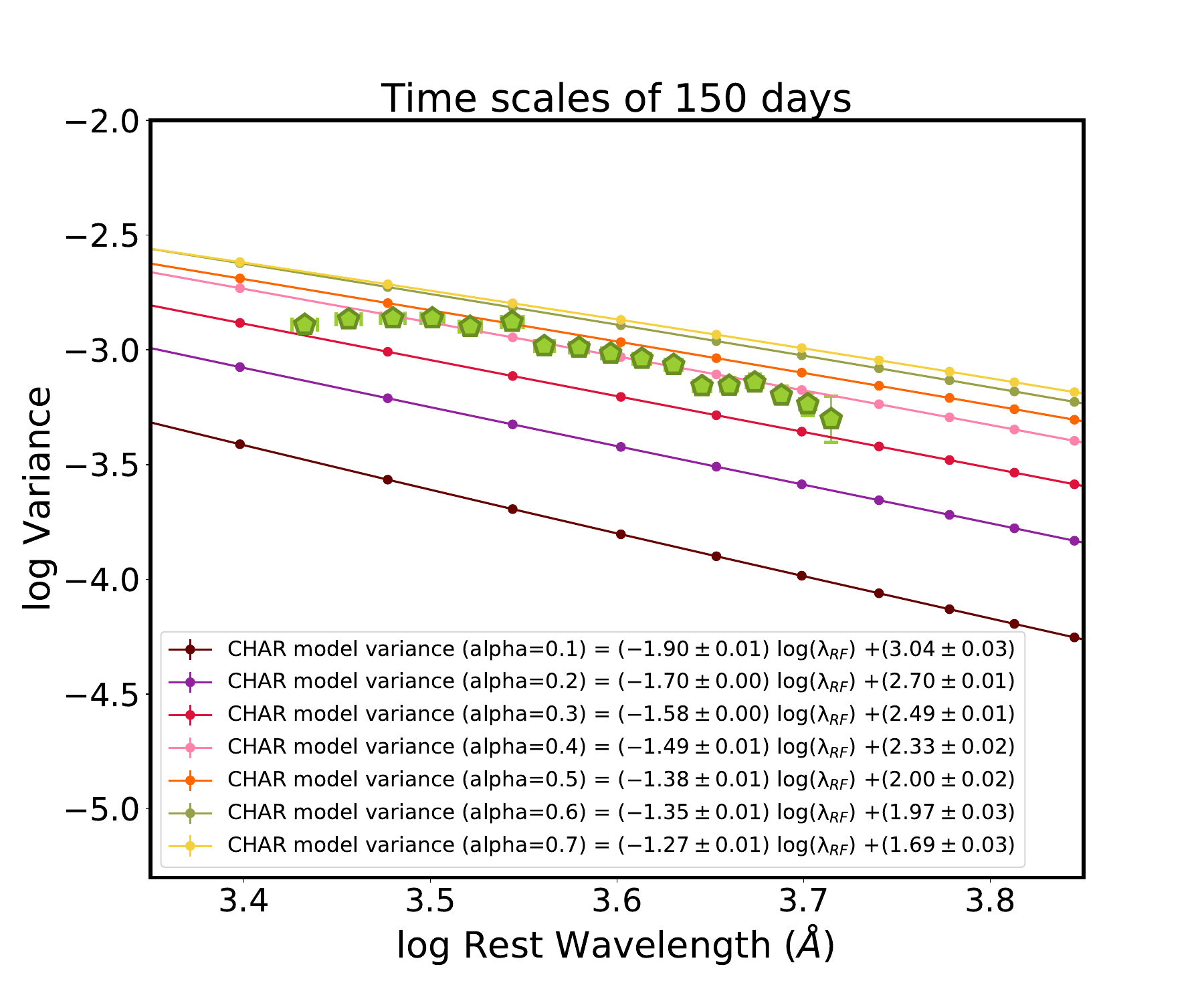}
\includegraphics[width=0.49\textwidth]{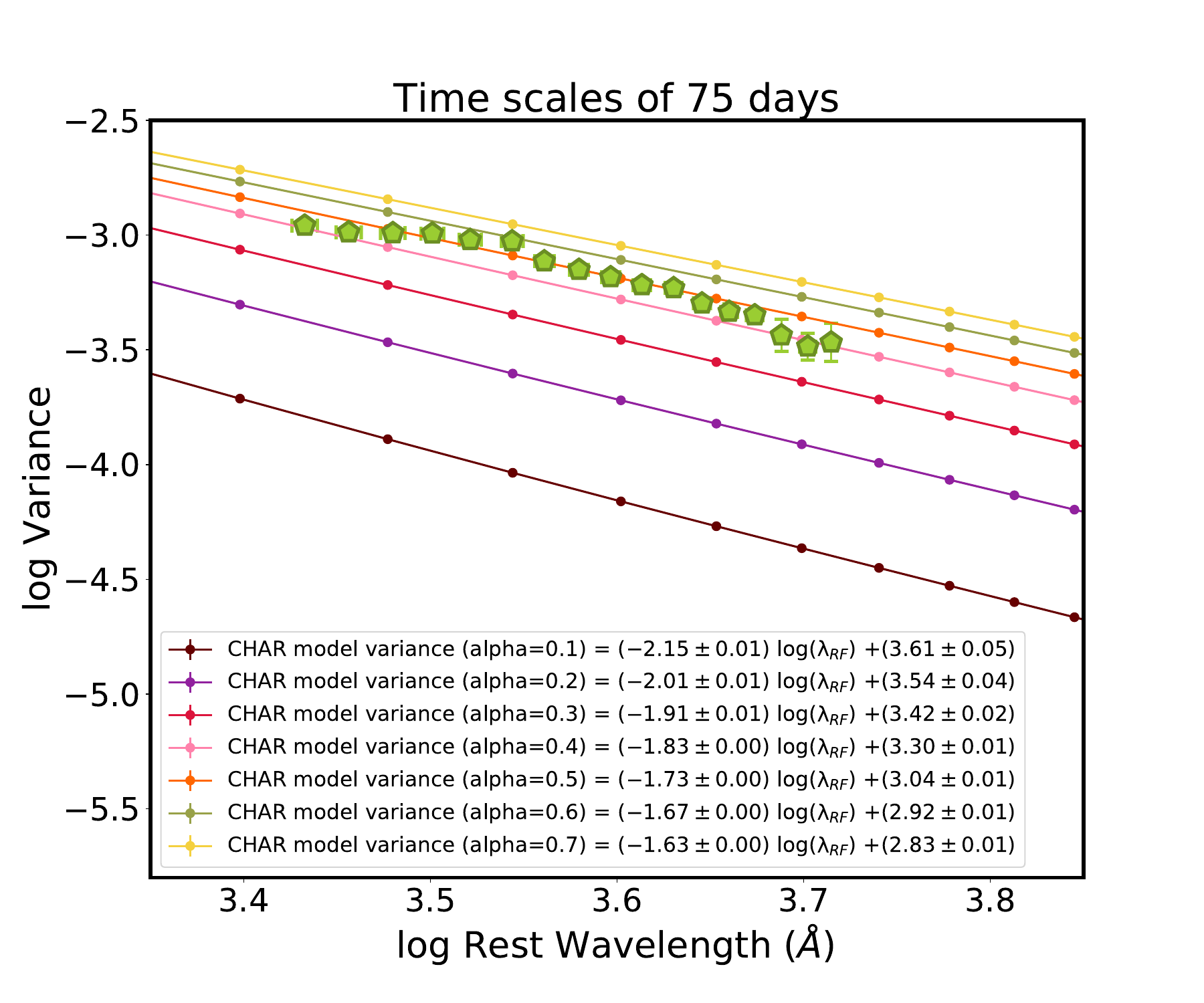}
\includegraphics[width=0.49\textwidth]{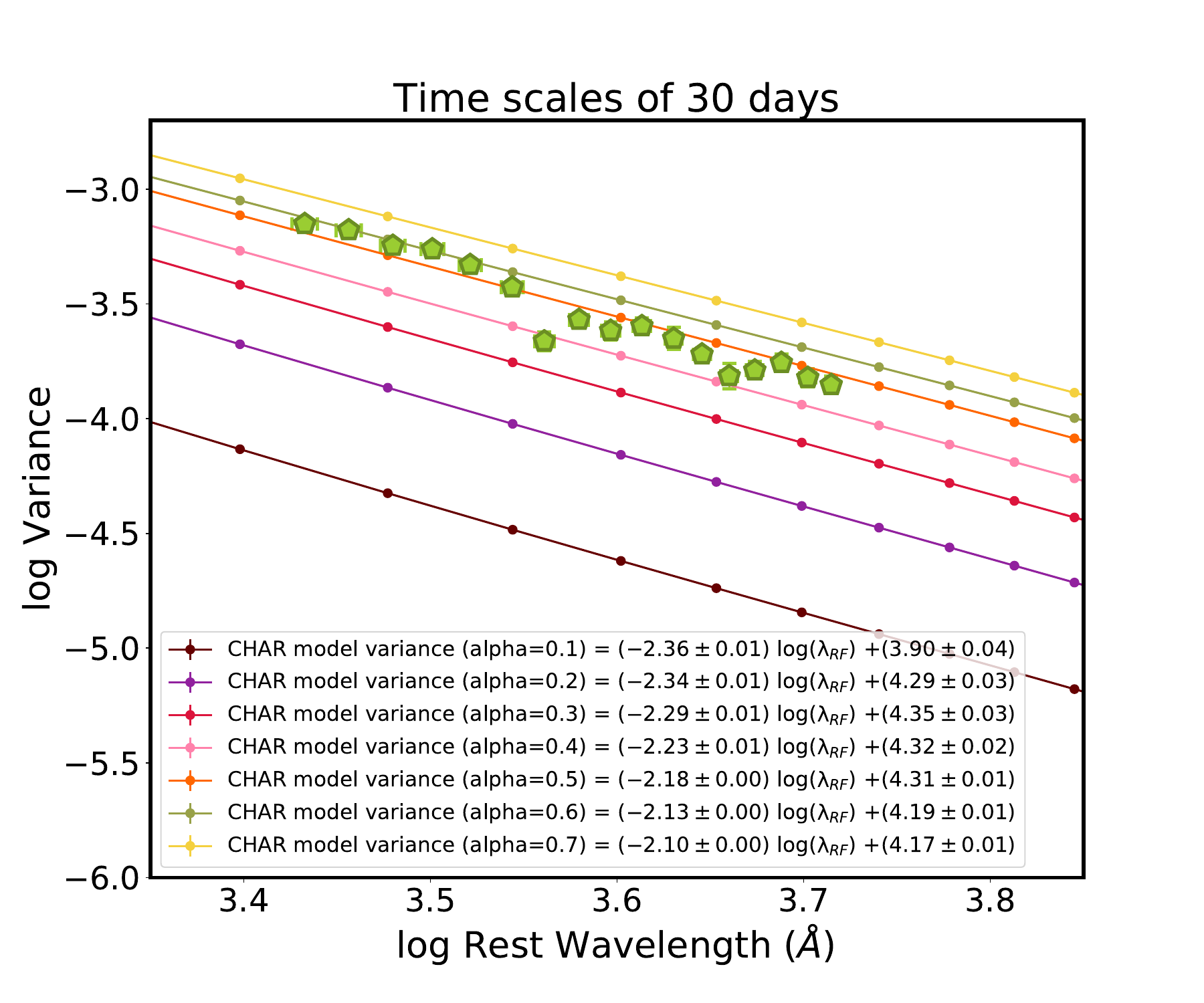}
\caption{Variance-wavelength relationship for the CHAR model examining varying $\alpha$ values from 0.1 to 0.8. The displayed multiple linear regression models indicated in the legend, show how the model responds to different $\alpha$ values. Additionally, green pentagons represent observed ZTF median variance, with error bars indicating root-mean-squared scatter.}
\label{fig : CHAR_model_alpha}
\end{figure*}

Figure \ref{fig : CHAR_model_alpha} presents variance as a function of rest-frame wavelength for the CHAR model for values of the dimensionless viscosity parameter ($\alpha$) ranging from 0.1 to 0.8, for all timescale studied. Model parameters are: $\dot{m}$ = 0.1, and  $M_{BH}$ = $10^{8} M_{\odot}$, $\beta = 0.6$ and $\delta_{mc}$ = 1.32. Observed variances are also included. It can be seen that as $\alpha$ increases, the slope of the variance-wavelength relationship becomes flatter. Among these, $\alpha = 0.5$ was identified as the best value, providing the closest match to the observed variance across all timescales studied.

\begin{figure}[!htbp]%
\centering
\includegraphics[width=0.45\textwidth]{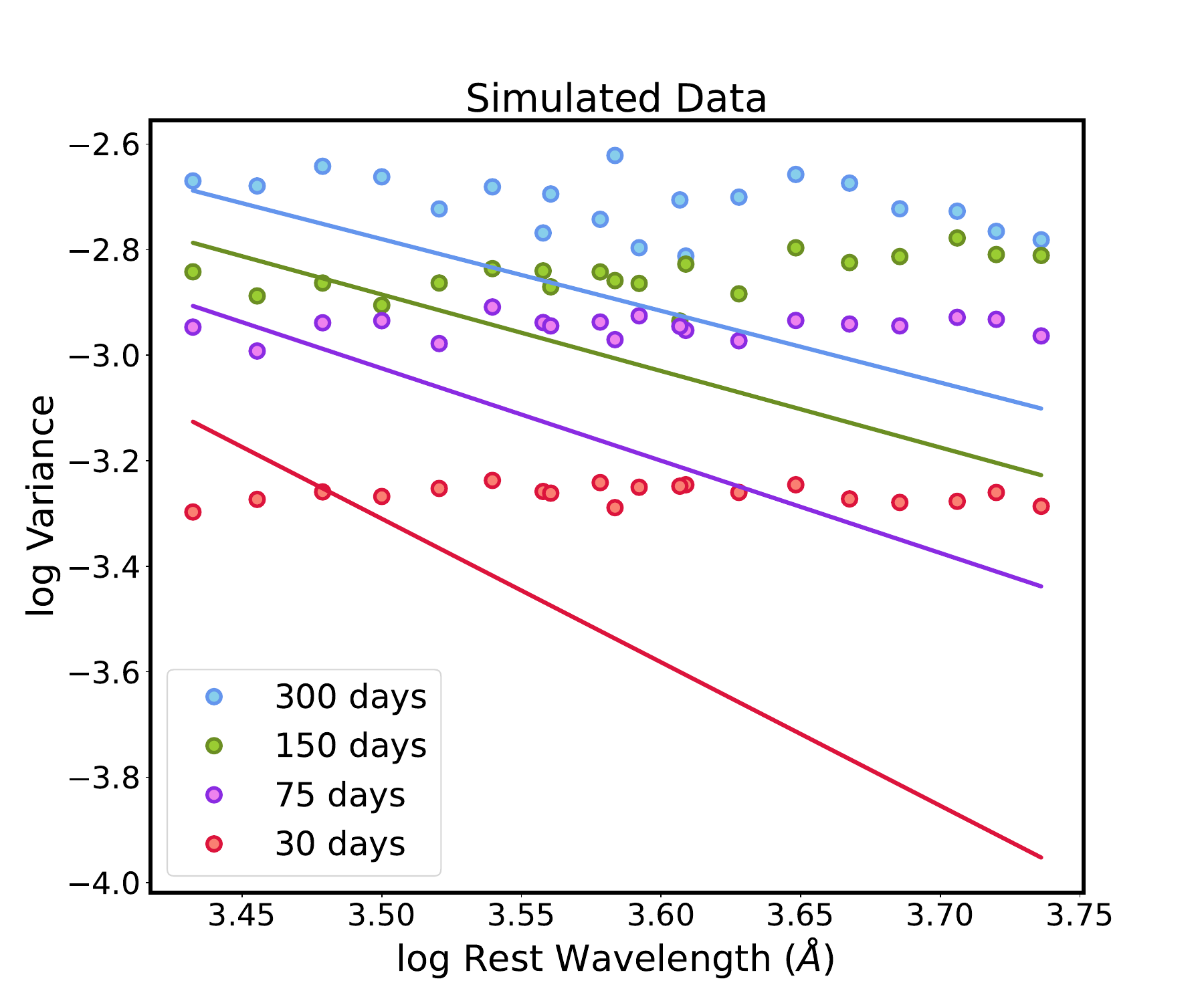}\\
\includegraphics[width=0.45\textwidth]{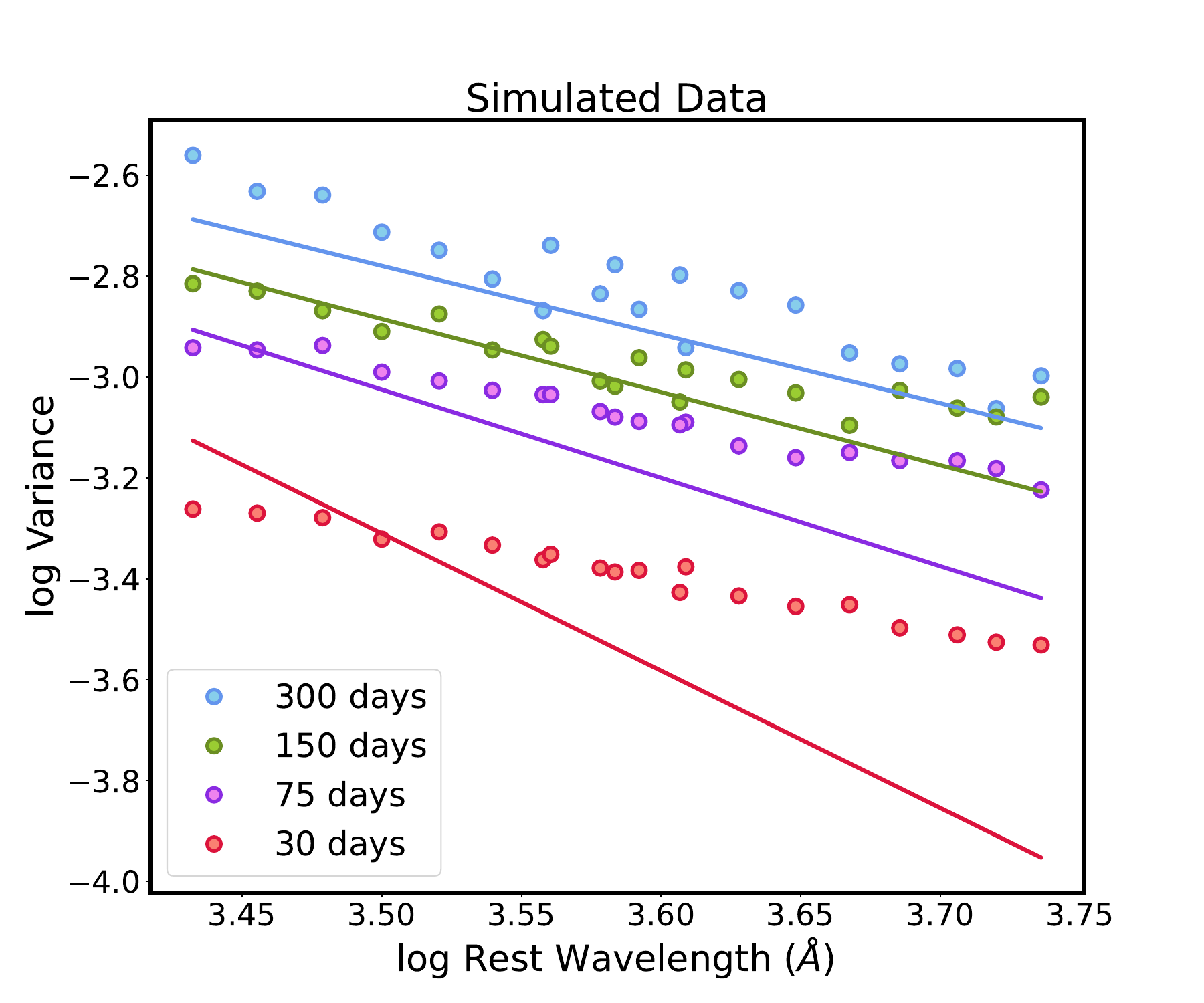}\\
\includegraphics[width=0.45\textwidth]{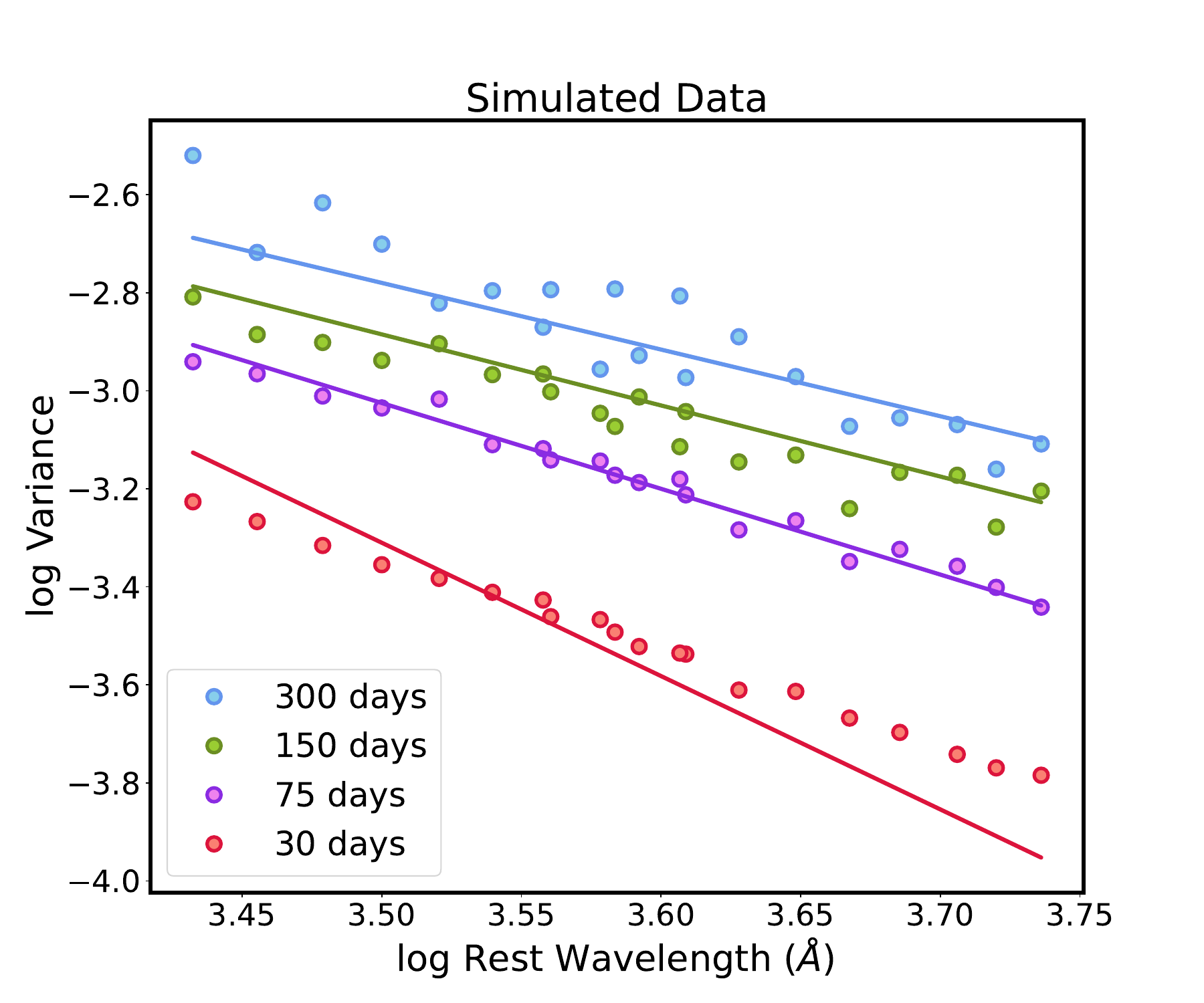}
\caption{ZTF Data-Simulation Comparison. This figure compares ZTF data with simulations from the bending power law model for variance versus rest-frame wavelength at four different timescales: 300, 150, 75, and 30 days. The circles represent the mean variance from simulations, while the solid lines show the linear fit to the ZTF data. The first subplot (left) shows results assuming no wavelength dependence of the power spectrum or its normalization. The second subplot (middle) includes wavelength dependence only in the normalization of the power spectrum. The third subplot (right) incorporates wavelength dependence in both the normalization and the break frequency. Despite these adjustments, the model does not fully account for the steepening at 30 days, indicating that additional factors, such as a wavelength-dependent slope in the power spectrum, might be needed to explain the observed trends.}
\label{fig:variance_simulations}
\end{figure}

\section{Light curve simulation and variance analysis}\label{sec:appendixE}

\subsection{Simulating the light curves}

The measurements of variance in light curves are inevitably affected by the sampling pattern and the methods used for variance estimation. To reduce these biases, we employed asimulation approach based on the model proposed by \citet{Timmer95}. This model was adapted to incorporate dependencies on key physical properties, including the black hole mass ($M_{BH}$) and the Eddington ratio ($R_{\text{Edd}}$), and rest-frame wavelength ($\lambda_{RF}$). 

\[
\text{P}(f) = A \times \frac{f^{\alpha_{\text{L}}}}{1 + \left(\frac{f}{f_b}\right)^{\alpha_{\text{L}} - \alpha_{\text{H}}}}
\]

where $f$ is the frequency, $A$ is the amplitude normalization, $f_b$ is the break frequency, and $\alpha_{\text{L}}$ and $\alpha_{\text{H}}$ are the low and high-frequency slopes, respectively. This model was adapted to incorporate dependencies on key physical properties, i.e., black hole mass ($M_{BH}$) and the Eddington ratio ($R_{\text{Edd}}$), and for this test also the rest-frame wavelength ($\lambda_{RF}$). 

The model incorporates $\alpha_{L} = -0.5$ for low frequencies and $\alpha_{H} = -2.5$ for high frequencies, with a break frequency \(f_b\). This approach is based on the study by Ar\'evalo et al.~(\citeyear{2024A&A...684A.133A}) where this pair of parameters showed a good match to the power spectra of objects observed in the $g$-band and located at $0.6<z<0.7$. In addition to the indices, the model includes the dependence of the normalization (\(\alpha_{A}\)) and the break frequency (\(\alpha_{f_{b}}\)) on the rest-frame wavelength \(\lambda_{\text{RF}}\). According to the fits in Ar\'evalo et al.~(\citeyear{2024A&A...684A.133A}), the break frequency \( f_b \) is determined by the black hole mass \( M_{BH} \) and the Eddington ratio \( R_{\text{Edd}} \):

\begin{equation}
    \log(f_b) = \log(B) + C \times (\log(M_{BH}) - 8.0) + D \times (\log(R_{\text{Edd}}) + 1.0),
\label{eq:fb}
\end{equation}

and the overall normalization \( A \) is dependent on  \( R_{\text{Edd}} \) as :
\[
A = 10^{F \times (\log(R_{\text{Edd}}) + 1.0)}
\]
where \( B = 0.007 \), \( C = -0.55 \), \( D = -0.35 \), and \( F = -0.40 \). 

As in the present work, the black hole masses and Eddington ratios are about the same for the whole sample (and those that deviate slightly from the reference value have had the variance already scaled by this model), we use a single value of these parameters for the modeling below. 

To incorporate possible wavelength dependencies we introduced a power law dependence of the amplitude $A$ on the rest-frame wavelength $\lambda_{RF}$, normalized by a reference wavelength similar to the characteristic wavelength of all light curves used in \citet{2024A&A...684A.133A}, i.e., 2900\AA. Therefore the normalization of the power spectrum will have the following form:
\begin{equation*}
    A = 10^{F \times (\log(R_{\text{Edd}}) + 1.0)}\left( \frac{\lambda_{RF}}{2900.0} \right)^{\alpha_{A}}
\end{equation*}

Similarly, to allow a wavelength dependence of the break frequency, we replaced the value of $B$ in Eq.~\ref{eq:fb}  with 
\begin{equation}
    B = 0.007 \times \left( \frac{\lambda_{RF}}{2900.0} \right)^{\alpha_{f_{b}}} 
\end{equation}

For each set of model parameters, we generated synthetic light curves by constructing the time series from Fourier components that follow a Gaussian distribution, as described by \citet{Timmer95}. Specifically, the real and imaginary parts of the Fourier components were derived by multiplying Gaussian random values by the square root of half the power at each frequency. This method introduces the necessary stochasticity into the simulated light curves, ensuring that they capture the inherent variability observed in real astronomical data. After generating the frequency-domain representation, we applied an inverse Fourier transform to produce the simulated light curve in the time domain.

Each set of simulations generated 100 light curves for specific rest-frame wavelength bins, ranging from 2706.41 Å to 5446.39 Å. These simulated light curves were resampled to match the actual observation times using 35 real sampling patterns already rescaled to the rest-frame time of the quasars. In the lower redshift bins, which correspond to higher rest-frame wavelengths, fewer real sampling patterns were used. The variance of each simulated light curve was measured using the Mexican hat filter method across four different timescales. The logarithms of these variances were then averaged over all simulations to provide a robust measure of variance at each timescale, following the same methodology applied to the real observational data.

\subsection{Analysis of variance versus wavelength across different cases}

To understand the effects of different dependencies of the power spectrum on wavelength, we simulated three scenarios, each with a distinct dependence of the power spectrum on wavelength i.e., exponents \(\alpha_{A}\) and \(\alpha_{f_{b}}\), with the power spectrum slopes fixed at $\alpha_{L} = -0.5$ and $\alpha_{H} = -2.5$. These simulations were then analyzed to produce variance as a function of rest-frame wavelength across four rest-frame timescales: 300, 150, 75, and 30 days. Below, we discuss each case and its implications. Figure~\ref{fig:variance_simulations} presents the variance versus wavelength for these simulations, in markers, and compares them with linear fits obtained from the ZTF data (see Table~\ref{tab : linear_fits}), in solid lines.

In the first scenario, we assumed no dependence of the power spectrum or its normalization on wavelength. Specifically, we set the exponents \(\alpha_{A}\) and \(\alpha_{f_{b}}\) to $0.0$. This simulation will simply reflect the biases introduced by the sampling patterns and by the method used to estimate the variance. If the method was perfect, there would be no wavelength dependence of the variances. We note that the difference in the variance levels for different timescales (different colors in these plots) is given by the power spectral shape and the value of the break frequency, described above. 

The effect of cosmological time dilation shifts the periodicity of the gaps in the light curves across different timescales, sometimes approaching the timescales where we measure the variance. For this reason, the biases in the variance estimates are different for the different redshifts and timescales, and the simulated data points are not exactly at the same level but instead show wiggles. We also note that the wavelength range is composed of $g$ and $r$ band light curves over the same redshift range so that the effects of time dilation on the sampling pattern are repeated (i.e., the points in the middle of the wavelength range have the same redshift and therefore time dilation as the points at the far right). Therefore, the wiggles in the variance appear repeated.

Notably, the trend in the simulated data shows an increase in variance with wavelength, which is opposite to the anticorrelation between variance and \(\lambda_{\text{RF}}\) found in the ZTF data. This suggests that the anticorrelation found in our study is real and not biased by the sampling artifacts present in the simulations as well as in the data.

In the second scenario, we introduced a wavelength dependence for the normalization of the power spectrum by setting \(\alpha_{A}\) to $-1.0$ while keeping \(\alpha_{f_{b}}\) at $0.0$. This configuration explains the variance observed at timescales before the break frequency, particularly at 300 and 150 days. The introduction of this particular wavelength dependence in the normalization aligns with the observed variance trends, indicating that the normalization alone can account for some of the observed features. However, this model does not explain the steepening observed in the variance as a function of wavelength at shorter timescales, such as 75 days, and especially at 30 days. Repeating this experiment for slightly higher or lower values of \(\alpha_{A}\) produces similar results, i.e., the wavelength dependence can be reproduced for some but never for all the timescales simultaneously. We note that the discrepancy between data and model is timescale-dependent, therefore it is unlikely that a different wavelength dependence of the normalization $A$ will solve it.

In the third scenario, we considered a more complex model where both the normalization and the break frequency depend on wavelength, with \(\alpha_{A}\) set to $-1.0$ and \(\alpha_{f_{b}}\) set to $-0.43$. This value is taken from Stone et al.~(\citeyear{2023MNRAS.521..836S}), who reported the dependence of the damping timescale on $\lambda_{RF}$ on rest-frame wavelength for quasars, which follows $\tau_{damp} \propto \lambda_{RF}^{0.30 \pm 0.13}$ and keep the largest possible value at $1 -\sigma$ level (see the footnote in Sect. \ref{subsec:PSD}).

In this case, the model successfully explains the variance vs. wavelength observed at 300, 150, and 75 days, where both the normalization and break frequency dependencies play a role. However, despite these adjustments, the model still cannot account for the steepening of the variance vs. wavelength at 30 days. This suggests that additional factors, potentially including a wavelength dependence in the slope of the power spectrum itself, are necessary to fully explain the observed variance trends. These results support the arguments we presented in Sect. \ref{sec:discussion}, where we concluded that the slope of the power spectrum should also depend on wavelength.

\end{document}